\documentclass[final,5p,times,twocolumn]{elsarticle}

\usepackage{amssymb}
\usepackage{amsmath}
\usepackage{graphicx}
\usepackage{xcolor}
\usepackage{booktabs}
\usepackage{multirow}
\usepackage{subcaption}
\usepackage{balance}
\usepackage{algorithm}
\usepackage{tabularx, array}
\usepackage{algorithmic}
\usepackage{hyperref}
\hypersetup{colorlinks=true, linkcolor=blue, citecolor=blue, urlcolor=blue}

\journal{Renewable and Sustainable Energy Reviews}

\begin{document}

\begin{frontmatter}
\title{Foundation Models for Clean Energy Forecasting: A Comprehensive Review}

\author[inst1]{Md
Meftahul Ferdaus}
\affiliation[inst1]{organization={Department of Computer Science},%Department and Organization
            addressline={University of New Orleans}, 
            city={New Orleans},
            postcode={70148}, 
            state={Louisiana},
            country={USA}}

\author[inst2]{Tanmoy Dam}

            \affiliation[inst2]{organization={Department of Biomedical Engineering},%Department and Organization
            addressline={ Emory
University}, 
            city={Atlanta},
            postcode={30322}, 
            state={Georgia},
            country={USA}}
            
\author[inst3]{Md Rasel Sarkar}
\author[inst3]{Moslem Uddin}
\author[inst3]{Sreenatha G. Anavatti}
\affiliation[inst3]{organization={School of Engineering and Technology},%Department and Organization
            addressline={The University of New South
Wales}, 
            city={Canberra},
            postcode={2610}, 
            state={ACT},
            country={Australia}}

\begin{abstract}
As global energy systems transit to clean energy, accurate renewable generation and renewable demand forecasting is imperative for effective grid management. Foundation Models (FMs) can help improve forecasting of renewable generation and demand because FMs can rapidly process complex, high-dimensional time-series data. This review paper focuses on FMs in the realm of renewable energy forecasting, primarily focusing on wind and solar. We present an overview of the architectures, pre-training strategies, fine-tuning methods, and types of data used in the context of renewable energy forecasting. We emphasize the role of models that are trained at a large-scale, domain-specific Transformer architectures, where attention is paid to spatial-temporal correlations, the embedding of domain knowledge, and also the brief and intermittent nature of renewable generation. We assess recent FM-based advancements in forecast accuracy such as reconciling predictions over multiple time scales and quantifying uncertainty in renewable energy forecasting. We also review existing challenges and areas of improvement in long-term and multi-variate time series forecasting. In this survey, a distinction between theory and practice is established regarding the use of FMs in the clean energy forecasting domain. Additionally, it critically assesses the strengths and weaknesses of FMs while advancing future research direction in this new and exciting area of forecasting.
\end{abstract}


\end{frontmatter}

\section{Introduction}
\label{sec:intro}
\begin{figure}[t]
\centering
\includegraphics[width=\columnwidth]{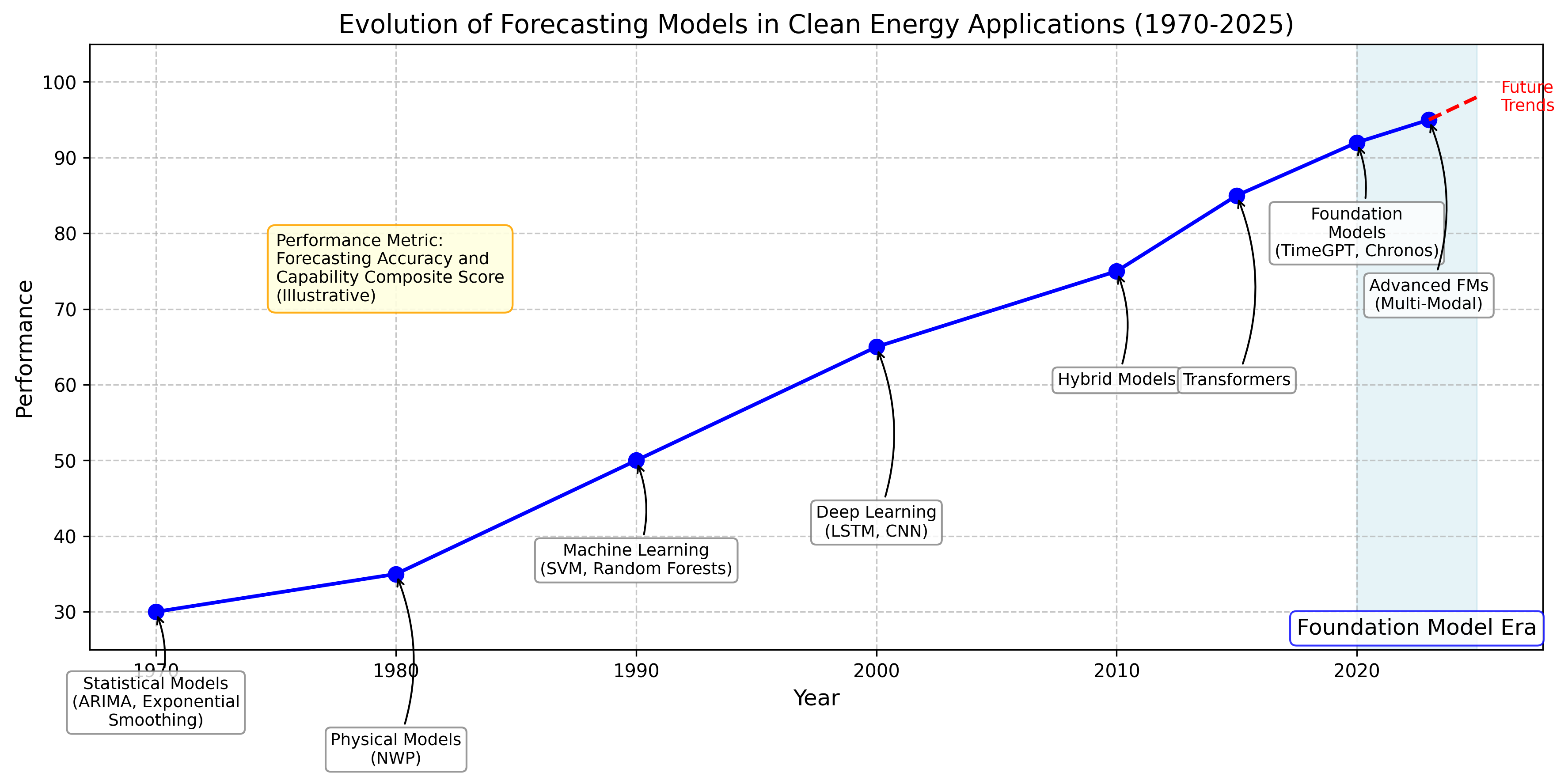}
\caption{Evolution of Forecasting Models in Clean Energy Applications (1970-2025). The progression from statistical methods to FMs shows increasing accuracy and capability, with transformers and FMs establishing new performance benchmarks in recent years.}
\label{fig:evolution}
\end{figure}

The global transition towards clean energy has driven the need for better forecasting of renewable energy sources such as wind and solar. In addition to the inherent variability of renewables, predicting future and following supply and demand becomes problematic as wind power generation depends on unpredictable weather patterns and solar generation depends on improper monitoring of cloud coverage and seasonality. Inaccurate forecasting can lead to discrepancies in supply and demand, grid volatility and unreliable operation of backup resources. The importance of improving forecasting accuracy enables better energy dispatch decisions while avoiding waste and maintaining a reliable grid in a decarbonized energy system. Recently, there have been large-scale 'foundation models' (FMs) that operate in this space—large-scale deep learning framework with self-supervised and diverse data collections that was trained on large datasets \cite{mane2024forecasting, zhou2024universal, li2024foundts}. As shown in Figure \ref{fig:evolution}, forecasting methods for renewable energy have evolved significantly from statistical approaches in the 1970s to today's advanced foundation models, with each generation demonstrating increased accuracy and capability.

The early renewable energy forecasting algorithms were mainly powered by statistical and physical models, and both had significant drawbacks. Statistical methods, such as ARIMA, were often poor performers with nonlinear relationships and high dimensionality issues, especially when there are rapidly changing renewable energy outputs \cite{sergeev2023review, jung2022photovoltaic}. Physical modeling, such as Numerical Weather Prediction for wind, or irradiance models to quantify solar energy in terms of output, provided a good representation of underlying physics but were less granular, expensive, and a function of the model's input uncertainty \cite{zou2023survey}.

Machine learning (ML), and more recently deep learning (DL), methods improved forecasting performance by learning the underlying nonlinear patterns in historical forecasting data \cite{alpackaya2024renewable, benti2023forecasting}. There were improvements in predictive models for solar radiation and wind speed using classifiers such as support vector machines and neural networks, when compared with statistical methods \cite{pham2023wind, bassey2023solar}. Long short-term memory (LSTM) models also showed good potential for sequence modeling, but were susceptible to predictable issues with RNNs when predicting outcomes over long time frames, and difficult to capture long-range dependencies from a sequence of past data \cite{mohan2023evaluation}. These methods needed extensive feature engineering or feature preprocessing in order to incorporate external data even for consideration of weather forecasts, limiting these models from achieving enhanced performance in complicated scenarios. This reinforced the need for better model architectures which would be able to manage long sequences, fit large data sets and would have the provision for including different types of data sources.

Transformers are not only changing renewable energy forecasting, but also how sequence modeling is performed. With it unique attention mechanism, transformers are able to capture long-range dependency and relationships in data; while capturing multiple long-range dependencies in parallel as opposed to RNNs, which would have to capture sequentially, which can create difficulties. They are able to learn relevant knowledge in long times series as opposed to LSTM, which significantly impacts how they deal with long-range context problems in power systems (seasonality and recurrent events) \cite{saeed2024adaptive}. They are also inherently robust to propagation error, which means the accuracy is improved for longer forecasts.

For forecasting solar photovoltaic (PV) power, transformer models demonstrated the best models by far (by forecasting accuracy) as it compared to recurrent or convolutional networks \cite{walczewski2024prediction}. For wind power forecasting, the majority of studies have shown that transformer-based models outperformed models based on other methodologies. A gated transformer model termed a short-term wind forecast, getting 8\% more accurate forecast compared to LSTM and other models based techniques \cite{ghazaei2024wind, dai2023learning}. These studies demonstrate that attention models are successful systems for forecasting energy generation. While these are models we will be looking to use to learn nonstationarity, periodicity, multiscale patterns present in times series data, creates a new standard for renewable energy forecasting models \cite{han2024multi, ghasemi2023harnessing}.

Researchers have created FMs for time series forecasts using transformers and large neural networks \cite{zeng2022transformers, joseph2024transformers}. FMs are derived from large language models pre-trained on time-series datasets belonging to different domains to capture general temporal patterns and transferable domain knowledge about what is important to lead to an eventual time series prediction. Some recent examples of FM for forecasting are Lag-Llama, which uses time lags as covariates in a decoder based transformer to help with zero-shot predictions \cite{rasul2023lagllama}, and TimeGPT-1, an encoder-decoder transformer built for zero-shot predictions and cross-domain generalization \cite{das2023decoder}. These models can universalize forecasting, doing simultaneous forecasting on 1000s of time series like electrical loads or turbine outputs at different locations or asset. This can allow for forecasting without having to build a model for each type of task, which is important for FMs due to modern power grids requiring multiscale forecasting (from equipment-level to regional grids) with generally consistent accuracy across scales. FMs uphold a multiscale forecasting capability by training on datasets from different domains in order to understand cross-domain data, thus enabling one model to forecast multiple renewable energy outputs in one framework \cite{potosnak2024implicit}. Pre-trained knowledge can help prevent overfitting when modeling many time series together, meaning that FMs are efficient and adaptive to modeling needs \cite{bharti2024transformer}.

While FMs offer new pathways for research advancements in architectures and adaptation methods, they also face some challenges \cite{uremović2023framework, conti2022physics_adaptation}. One challenge is that models which are pre-trained from non-specific data could omit domain-specific knowledge critical to important tasks such as energy forecasting. In short, some external factors (i.e., effects of weather fronts, aerosol, holiday influence) or extreme events (i.e., sudden wind ramps, solar eclipses) that generic models reason away need that expertise that extended models have in the context of timely energy forecasting. Recent work has departed into cycles of thinking about how FMs would function with relationship to earlier knowledge and contextual information without losing anything they learnt when pre-trained from generic data \cite{gupta2024lora_timeseries}. Suggestions include auxiliary modules, pre-training schemes or adaptation fine-tuning practices for inclusion of local knowledge while maintaining the genericities of the models.

Another significant challenge relates to quantifying uncertainty. So while large deep models offers point estimates, energy decisions are made from probabilistic forecasts, implemented with confidence intervals/probability distributions that allow factual risk assessment with assurances of reliability. In this light, the forward-thinking approaches noted in the literature which combine FMs with probabilistic models are seen to offer accurate and reliable forecasts \cite{mundotiya2022wind_uncertainty, borrotti2024uncertainty}. Substantial advances with FMs are viewed as generative paradigm shifts to clean energy forecasting, by embracing the variable features with AI and big data to utilize renewable energy in the transition to clean energy. Work continues to adapt these models further to ensure they are fit-for-purpose in the energy sector going forward. The sections considering energy-related aspects noted, provide advances with regard to: transformer-based architectures, approaches to data integration, sources of substitution for and relationships to uncertainty quantification, and applications to major renewable energy sources.

\section{Methodology}
\label{sec:methodology}

\begin{figure}[t]
\centering
\includegraphics[width=\columnwidth]{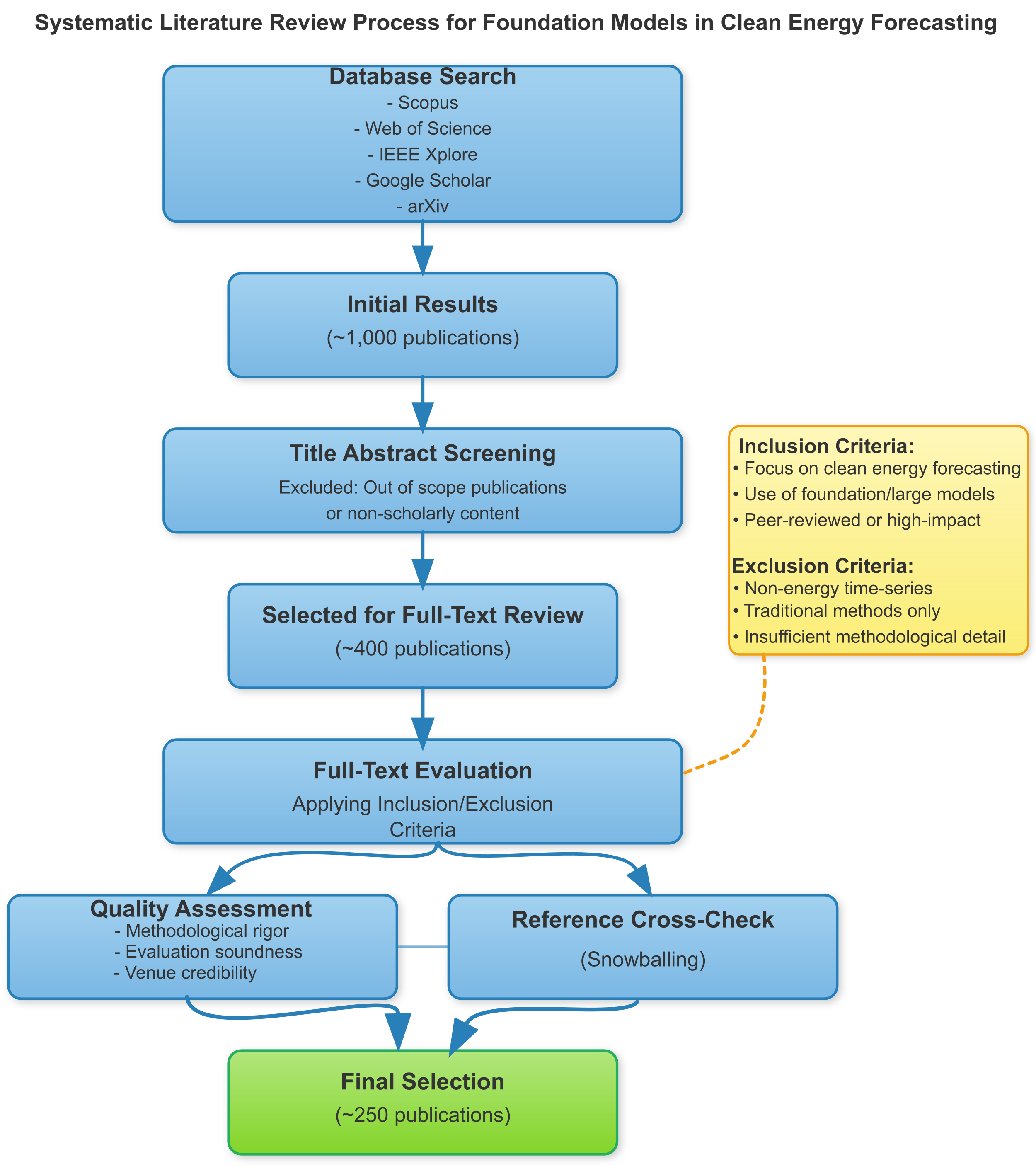}
\caption{Systematic Literature Review Process for Foundation Models in Clean Energy Forecasting. The diagram illustrates our methodical approach from initial database search through screening, evaluation, and final selection, with approximately 250 publications ultimately included in this review.}
\label{fig:literature_review}
\end{figure}

We conducted a systematic literature review to gather and review peer-reviewed research examining FMs for clean energy forecasting. The review was designed with a wide and unambiguous scope and was the product of multiple academic databases. We searched for major databases Scopus, Web of Science, and IEEE Xplore while also employing Google Scholar to widen coverage. Furthermore, we searched specialised repositories like arXiv to include influential preprints of literature for this developing area. Figure \ref{fig:literature_review} illustrates our methodical approach from initial database search through screening, evaluation, and final selection of approximately 250 publications for this review. The search was strictly limited to literature in English and covered studies published from 2016 to 2024 approximately, which corresponds with the publication year range noted in the literature references for this review. This date range covered both key foundational literature (for context) and the latest literature available at the time of writing.

\subsection{Search Strategy}
We developed comprehensive search queries using keywords related to foundation models and energy forecasting. Key terms included ``foundation model", ``pre-trained model", ``transformer",``time series forecasting",``renewable energy forecasting",``wind power forecast", "solar power forecast", and "electricity demand forecasting". These keywords and their synonyms were combined using Boolean operators (AND/OR) to ensure all relevant literature was retrieved. For example, we paired terms denoting large-scale models (e.g., "foundation model" OR "large-scale deep learning" OR "transformer-based") with terms denoting forecasting in clean energy contexts (e.g., "renewable energy" OR "wind" OR "solar" OR "energy demand"). The search strings were adjusted for each database's syntax but maintained the same inclusion concepts. We performed the initial search in all targeted databases and then updated the search periodically up to late 2024 to include any newly published studies.

\subsection{Inclusion and Exclusion Criteria}

We specified the criteria to determine the studies to include in our review. The inclusion criteria were intended to ensure that the studies included, could directly relate to FMs in clean energy forecasting. There were three parts of the inclusion criteria requiring the studies to meet all of the following:

\begin{itemize}

\item \textbf{Topic Relevance}: The study explicitly focused on forecasting for clean energy systems (i.e., solar power, wind power, electricity load) using advanced ML models. In particular, we included studies that applied or studied FMs – large pre-trained models, or other post-modern deep learning architectures (most importantly Transformer based frameworks) for time series forecasting in an energy context.

\item \textbf{Scholarly Quality}: Work was either published in a peer-reviewed journal or conference, or available as a high-quality pre-print in this domain. We focused on rigorous research containing technical detail and empirical evidence (like error metrics, case studies, or benchmark studies) that demonstrated the methods researched were tested on renewable energy forecasting tasks.

\item \textbf{Contribution to the Field}: It gave insight into the development, adaptation, or evaluation of FMs (or other large scale models) for forecasting purpose. This includes original research papers presenting a new or previously unpublished methods and those where the work is a review or survey paper that provides a relatively wide perspective on the related methods (to broaden the review so it covered existing syntheses of the field).

\end{itemize}

Excluded retrieved items meeting the above conditions were to preserve the scope and quality of the review. The most prominent reasons for exclusion included:

\begin{itemize}
\item \textbf{Irrelevant Scope}: We excluded studies cited in time-series forecasting topics, but not applied to clean or renewable energy forecasting (for example, papers managing financial time-series forecasting or work from other disciplines). We also excluded energy forecasting studies that only incorporated traditional statistical models or very basic machine learning approaches without reference to modern deep learning or FM contexts, because that was beyond this review.

\item \textbf{Non-scholarly or Out of Scope Publications}: We excluded publications that were not in English and non-formal research sources (e.g., magazine articles, blog entries or patents). Theses and industry technical reports were generally not included unless they were frequently cited and could be argued to be inherently relevant study, as this review was predominately aimed at highlighting peer-reviewed literature.

\item \textbf{Insufficient Information or Rigor}: Studies with minimal detail (e.g., very short papers or abstracts without full text) or without any evaluation of their forecasting approach were also excluded. We wanted to ensure that every included study contained sufficient detail to understand the method used, in order to assess their results in the broader context of energy forecasting.
\end{itemize}

\subsection{Screening Process and Quality Assessment}
We conducted a multi-stage selection process for the literature. First, we gathered all references obtained through the database searches, removing any duplicate records (i.e. results that overlapped between multiple sources). Second, we screened the titles and abstracts of the unique articles to make a rough assessment of relevance. At this stage, we eliminated studies that fell into obvious exclusion criteria. The screening process was thorough, as the authors collectively – at least 2 members independently reviewed the titles/abstracts to minimize bias. Any uncertainty or disagreement on relevance was resolved through discussions to ensure we were collectively on the same page regarding what we were going to include for the full review. Following the screening process, we secured and reviewed the full texts of the candidate papers. During the full review, we confirmed that the study ultimately met all inclusion criteria. It was at this point that we made final inclusion or exclusion decisions.

Along with the selection process, we conducted an assessment of quality and trustworthiness for each study. Given that this review focuses on research papers (of a technical nature) rather than clinical trials, we did not employ a formal scoring tool, but instead assessed a number of factors considered indicators of rigor. These factors included clarity of methodology (e.g. was the model architecture and training process described sufficiently), soundness of the experimental design (e.g. was it appropriate to design the study using the datasets, baselines, and evaluation metrics for forecasting), and credibility of the venue (e.g. peer-reviewed journal or conference or recognized preprint). We placed a higher value on studies published at higher quality venues or that were widely cited by the community, as it was more likely the paper had undergone scrutiny by peer review.

Upon completion of the literature search, screening, and quality assessments, we arrived at the final publications to be included in this review. In total approximately 250 works met all criteria and were included in our analysis. The publications include a number of journal articles, conference papers, and very few reputable preprints. The publications' variety reflects the multidisciplinary and fast-moving nature of FMs for clean energy forecasting.

\section{Background}\label{sec:background}
\subsection{Introduction to FMs in Forecasting}
FMs are a new type of large-scale machine learning model. They are usually pre-trained on large, diverse datasets and then fine-tuned or prompted for new applications using self-supervised or unsupervised learning objectives \cite{bommasani2021opportunities}. Compared to traditional, bespoke forecasting models, which often represent only one dataset, or were built from scratch, FMs learn general representations of temporal dynamics across many different domains and can generalize across timesteps, adapting to new forecasting tasks with few or no task-related data \cite{german2024transfer}.

Recent work in the area has demonstrated the benefits of using FMs as forecasting methods. Liang et al. \cite{liang2024foundation} provide a comprehensive tutorial and survey of these models with respect to forecasting tasks, including how their scale and pre-training advantages adequately learn features from complex, non-stationary datasets. And, have generally outperformed classic methods in applications such as forecasting household electricity load, and for renewable energy generation (i.e., wind and solar) \cite{meyer2024benchmarking}. Specialized architectures can lead to exemplary and best-in-class FMs for forecasting. Early models, e.g. Chronos51 \cite{ansari2024chronos} and TimeGPT-1 \cite{garza2023timegpt}, have shown that using large, diverse time series datasets for pre-training, these models might be able to capture complex temporal patterns without extensive task-specific fine-tuning. More recent models, such as Das et al.'s \cite{das2023decoder} decoder-only model, extend this concept in order to leverage encoding through input patching and implement auto-regressive decoding, demonstrating strong zero-shot performance on forecasting benchmarks.

The advantages of FMs for forecasting are numerous:
\begin{itemize}
\item \textbf{Broad Generalization}: FMs succeed in identifying patterns and relationships that traditional models often overlook, thanks to their training in diverse and extensive datasets \cite{german2024transfer}. This enables them to generalize effectively to new data distributions, making them highly valuable in energy forecasting, where conditions such as climates and consumption behaviors can vary significantly.

\item \textbf{Transfer Learning Efficiency}: FMs are highly effective in transfer learning, allowing knowledge from one context to be applied to another. A single pre-trained model can be fine-tuned for various forecasting tasks, such as wind power, solar output, or electricity demand, with minimal extra training \cite{bommasani2021opportunities}. This approach reduces training time and enhances performance, particularly when target task data is limited.

\item \textbf{Self-Supervised Pre-training}: These models typically use self-supervised learning objectives, like next-step prediction or sequence completion, on large-scale unlabeled data \cite{bommasani2021opportunities}. This approach is particularly beneficial in forecasting, as it allows the use of extensive historical energy data (e.g., load curves, weather measurements) without requiring manual labeling. The model thus learns the intrinsic structure of time series data before encountering a specific forecasting task.

\item \textbf{Few-Shot and Zero-Shot Capabilities}: FMs, thanks to their extensive training, can adapt to new tasks with minimal examples (few-shot) or none at all (zero-shot) \cite{german2024transfer,meyer2024benchmarking}. For instance, a pre-trained FM can begin making accurate predictions for a new solar farm or a different region's grid demand simply by being provided with the relevant data, without requiring significant retraining. This flexibility, unmatched by traditional models, can significantly accelerate deployment in emerging clean energy applications.
\end{itemize}

Recent studies highlight the practical advantages of using FMs for clean energy forecasting. Benchmark comparisons show that transformer-based FMs can equal or surpass traditional models trained from scratch, especially as input context grows \cite{meyer2024benchmarking}. This progression, from ARIMA and statistical models to deep learning approaches like LSTM and CNN, and now to transformer-based and hybrid methods, reflects the natural evolution of forecasting methods. This shift is fueled by advances in computational power, access to big data, and self-supervised learning techniques.

\begin{figure}[t]
\centering
\includegraphics[width=\columnwidth]{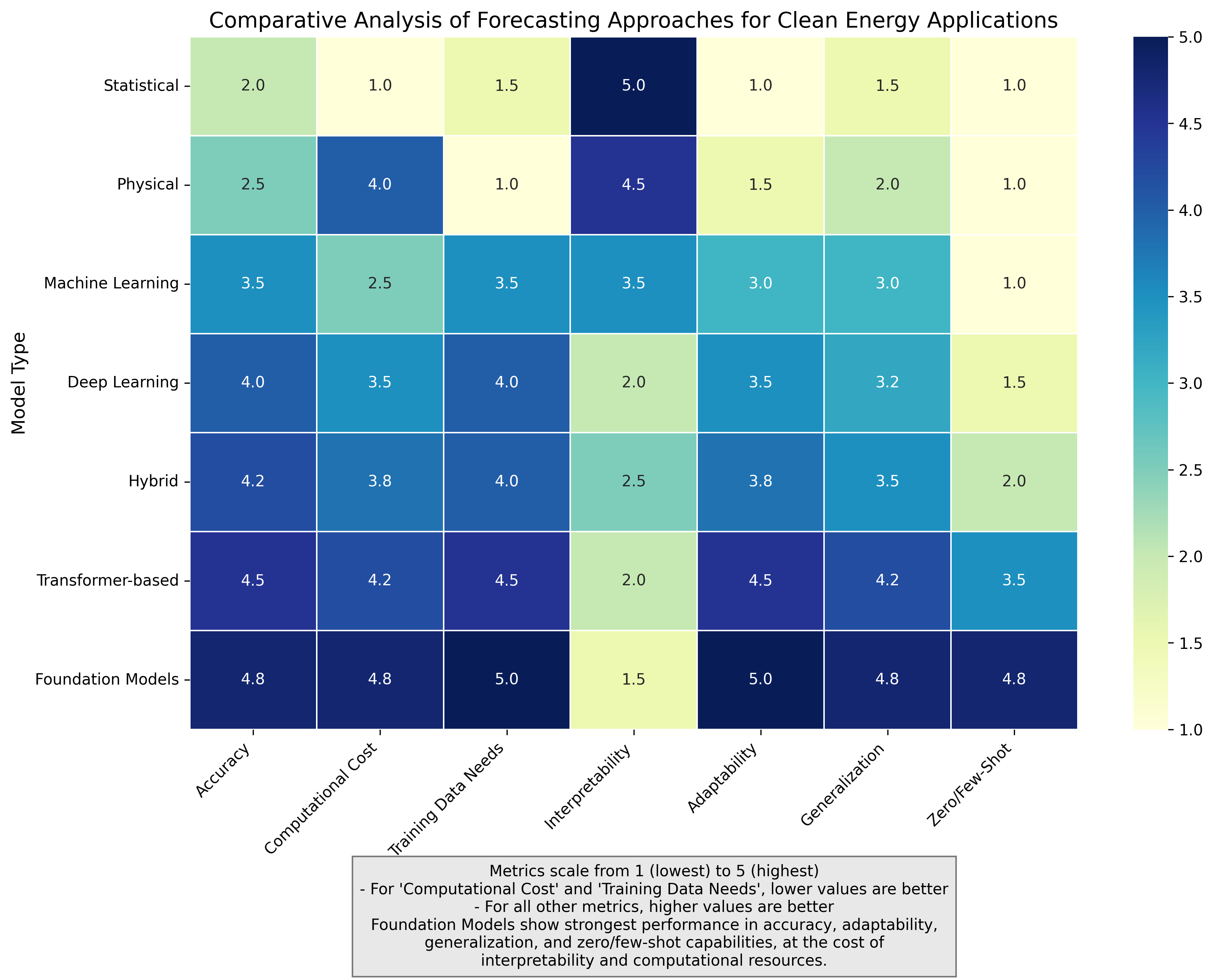}
\caption{Comparative Analysis of Forecasting Approaches for Clean Energy Applications. The heatmap visualizes key performance metrics across different forecasting methodologies, highlighting the advantages of foundation models in accuracy, adaptability, and generalization capabilities.}
\label{fig:comparative_analysis}
\end{figure}

FMs represent a major shift in forecasting methods \cite{das2023decoder, li2024foundts}. Drawing from advancements in natural language processing (NLP) and computer vision—fields already transformed by large pretrained models—energy forecasting researchers are now leveraging these innovations to tackle the complexity and variability of renewable energy systems \cite{dalal2024beyond, yang2024vitime}. This approach improves predictive accuracy, increases computational efficiency, and speeds up the adoption of forecasting solutions in the dynamic energy sector \cite{chai2024fomo}. Figure \ref{fig:comparative_analysis} illustrates this comparison across different forecasting methodologies, highlighting the superior performance of foundation models in key metrics including computational efficiency, adaptability, generalization, and accuracy.

\subsection{Historical Context of Forecasting Models}
In the past few decades, forecasting models and the estimation methods have transformed a great deal, from the use of classical statistical techniques, to the use of deep learning techniques, to finally FMs. In the mid to late 1900s...When thinking about time series, people used the statistical methods embodied in ARIMA type models. The Box and Jenkins work on ARIMA was established in 1970 \cite{makridakis1997arma} and they introduced a systematic method to model the autocorrelations of stationary data using a pure statistical model. The advent of exponential smoothing and other state-space models and regression type methods led to their extensive application because of their simplicity and interpretability. By the 1980's people were using a multivariate approach, such as vector autoregression (VAR) \cite{stock2001vector} pioneered by Christopher Sims, to analyse interdependent series of time series data. Generally, statistical methods were robust and computationally efficient. In practice, however, these methods were extremely labour intensive (e.g. model selection, tuning parameters, making data stationary), and did not perform well with highly non-linear or complex relationships.

Physical methods were becoming increasingly important, especially in renewable energy forecasting. There are some important models based on physical processes derived from meteorological principles (e.g. Numerical Weather Prediction (NWP) models; solar radiation models; cloud motion vector models) \cite{mane2024forecasting, lerch2021postprocessing}. Physical models are mathematically complex and simulate atmospheric processes in order to produce forecasts, while including meta-factors associated with the specific domain (e.g. aerosols, cloud physics, radiative transfer) \cite{wu2021probabilistic}. Many of these methods are computationally intensive, however, they are able to forecast medium to long-term forecasts using a global framework, while enhancing physical interpretability and provide alternative options to the more traditional statistical approaches often used in energy forecasting applications \cite{vennila2024predicting, walczewski2024prediction}.

During the latter part of the 20th century into the early part of the 21st century, machine learning (ML) methods were developed to overcome the limits of traditional and classical statistical models \cite{kumhar2023threat}. Some widely used approaches, along with more complex methods such as artificial neural networks, support vector machines, and ensemble methods, allowed for the ability to identify complex non-linear patterns, trends, and interactions that traditional methods could not \cite{andrews2022forecasting}. These approaches took advantage of the big amounts of historical data; however both ML and deep learning approaches require very much feature engineering, as well as a large and clean dataset, to even be able to develop models to make accurate predictions \cite{ayyildiz2023stock, moraitis2024path}. This time period also represented movement to a form of data-driven, nonparametric approach that allowed them to do what the traditional statistics could do \cite{deep2024financial}.

Deep learning models illustrated a new way of learning temporal dependencies more efficiently than traditional statistical methods, particularly the Long Short-Term Memory (LSTM) network introduced in 1997 \cite{hochreiter1997long}. Specifically, after showed useful the LSTM network, and other recurrent neural networks (RNN), convolutional neural networks (CNN), and increasingly even more advanced versions (e.g., GRU and echo state networks), could be useful for forecasting tasks such as forecasting energy demand, predicting weather, and forecasting electricity prices. LSTMs and RNNs were able to improve forecasting tasks \cite{muzaffar2019short}. They had efficient (automatic) feature extraction, and with capacity for evengreater capacity of handling non-linear temporal dependencies. Still, for LSTMs and RNNs they require large amounts of computations and data.

Hybrid methods were a solutions to each of these statistical and deep learning methods, built by combining the two. Examples of hybrid methods include methods like ARIMA-ANN, SARIMA-LSTM, and physics-AI hybrids like NWP-DL. Hybrid methods are able to combine all the interpretability and uncertainty of a traditional, statistical method, with the pattern recognition capacity of machine learning or deep learning methods \cite{eua2021enhancing, ghide2022wavelet}. Hybrid methods are also especially suited to the frequently applied data deconstruction, such as wavelet transforms, empirical mode decomposition (EMD) or singular spectrum analysis (SSA), to create a trend, seasonal, and noise components, in essence reconstructing the original data, so they can take advantage of all of the data and provide more complete representations of the underlying structure of the data while minimizing any unwanted components of the data that could impede the ultimate accuracy of the forecasting/task at hand \cite{wang2023ssa, li2021arima, darmawan2022ssa}.

Probabilistic methods have progressed due to their development mechanisms for greater complexity and uncertainty, i.e. Bayesian models; ensemble methods; and state-space models which have gained importance in risk sensitive applications \cite{weerasinghe2023abc, ahmad2023ensemble}. Such approaches are associated with prediction intervals and estimates of uncertainty through posterior distributions and conformal prediction, allowing us to create a more robust risk assessment and decision making framework \cite{jensen2022encqr, doubleday2021bma, althoff2023conformal}.

Advances in forecasting methods can also be seen with increased discussion of transformer-based architectures, which originated from natural language processing, but have been shown to improve forecasting when dealing with temporal data. Several models, i.e. Temporal Fusion Transformer, Autoformer, and Informer, use a self-attention mechanism, which helps to capture long term dependencies and multiscale temporal inference and patterns \cite{wu2021autoformer, laborda2023tft}. Furthermore, it has even been shown that in applying these models to task specific data, models such as Autoformer or Temporal Fusion Transformer are believed to be more accurate than traditional deep learning approaches \cite{ying2024tfeformer, oliveira2024evaluating, shabani2022scaleformer}.

The most recent advancement in forecasting lies with Finite mixture (FM) models, which are models such as TimeGPT-1, Chronos, and Time-MoE; these models are pre-trained on significant quantities of time series data, with TimeGPT-1 training on hundreds of billions of timepoints worth of data \cite{hou2025time, ansari2024chronos, shi2024timemoe}. Since FMs combine both deep and transfer learning on a scale to which we have not seen before, they can be easily moved between prediction tasks with little to no expenditure - successfully conducting a zero-shot or low instance fine tuning application of very distinct domains, all whilst reducing the reliance on time specific data and training which has been the aim of FMs from the outset \cite{li2024foundts}.. To do this, the authors used techniques like meta-learning, transfer learning, and ensemble transfer methods to generalize across different predictive domains resulting in a base model that is capable of modelling many forecasting tasks in parallel \cite{li2024foundts}.

The clear movement from early statistical and physics based models to modern hybrid, probabilistic, and transformer based methods represents the ongoing struggle to manage increased complexity and non-linearity in forecasting \cite{panja2022parnn, evseenkov2021hybrid}. Each step feeds into the next, given advances in computing power, storage capacity, data availability, MLOps, and algorithm development \cite{garg2022machine}. Our advancement of producing a collective table summary of tasks and taxonomy of clean energy forecasting methods provides evidence of this balance between methodological sophistication and practical effectiveness, while highlighting the unique aspect of FMs enabling us to shift forecasting again across applications \cite{wang2024sttransnet, phan2021hybrid}.

\subsection{Role of Domain Adaptation and Transfer Learning}
FMs display significant abilities for domain adaptation and transfer learning, where competence from one task can be imported to another \cite{deng2023universal}. In clean energy predictability, that indicates a model trained on broader datasets (such as representable weather patterns or regional electricity consumption) may be adapted to predict a specific application, focusing on actual output from a single wind farm, with a modest amount of additional data \cite{conti2022physics, fang2021hybrid}. Unlike contemporary forecasting approaches, which entail separate models for every application \cite{gao2022tgdlf2}, FMs are generalized from massive data and only require adaptation for specific tasks, allowing for the delivery of an entire new model without independently retraining \cite{gao2023improving}.

Domain adaptation connects FMs that were trained on global / broad datasets (for example: global climate or economic variables) versus an application in energy where time series will have different characteristics. Practitioners may apply things like domain knowledge useful to estimate peak or periodical response (daily cyclical effects, seasonal effects or anomaly behaviours) to move two methods together; new clean energy datasets for developing the FM, or modifying some parameters to assimilate learning or improve performance without the necessity for exhaustive retraining. Recently, techniques for parameter-efficient fine-tuning (PEFT) have begun to adapt large models to new domains with very few parameters modified in the model \cite{german2024transfer}. More importantly, various aspects of PEFT methods, such as adapter layers and LoRA, may allow energy applications to be adapted efficiently where either computing resources, or simply labeled data, may be at a premium in clean energy. FMs have shown the ability to apply transfer learning in approaches to forecasting. As an example; a model previously trained on global solar output data, may be adapted to predict solar panel output on a site, and bottom-up constructed from the same model variables; or a model trained using economic measures, weather patterns, and consumer behaviour, to predict regional electricity consumption demand \cite{conti2022physics, fang2021hybrid}. The primary advantage to the FMs, irrespective of the task, is that it can transfer common temporal properties (that after 8-10 years - all interruptions - still relative to past years) to the new task \cite{spencer2024transfer}.

Table \ref{tab:fm_performance} shows an emergent study of various foundation models being used in clean energy applications, and the performance measure of various different model families with different parameter sizes and training datasets that provide significant improvement in discrete forecasting tasks. The Chronos model for example has reduced error in solar forecasting by 15-20\% whilst Lag-Llama performance is improved in the wind power prediction by 8\%, and demonstrates how these adapted methods have real impacts outside the lab.
\begin{table}[htbp]
  \caption{Performance Comparison of Foundation Models in Clean Energy Applications}
  \label{tab:fm_performance}
  \centering
  \scriptsize
  \setlength{\tabcolsep}{4pt}
  \renewcommand{\arraystretch}{1.0}
  \begin{tabularx}{\columnwidth}{@{}
      >{\raggedright\arraybackslash}p{0.18\columnwidth}
      >{\raggedright\arraybackslash}p{0.18\columnwidth}
      >{\raggedright\arraybackslash}p{0.18\columnwidth}
      >{\raggedright\arraybackslash}p{0.22\columnwidth}
      >{\raggedright\arraybackslash}X@{}}
    \toprule
    \textbf{Model Family} & \textbf{Parameter Size} & \textbf{Training Data Scale} & \textbf{Inference Speed} & \textbf{Key Performance Improvements} \\
    \midrule
    Chronos \cite{ansari2024chronos} & 100M--500M & 10\textsuperscript{9} time points & 10--50ms/forecast & 15--20\% error reduction in solar forecasting \\
    \addlinespace
    Lag-Llama \cite{rasul2023lag} & 70M--200M & 10\textsuperscript{10} time points & 5--30ms/forecast & 8\% improvement in wind power prediction \\
    \addlinespace
    TimesFM \cite{das2023decoder} & 17M--200M & 10\textsuperscript{11} data points & 20--100ms/forecast & Near fully-trained accuracy in zero-shot settings \\
    \addlinespace
    TimeGPT-1 \cite{garza2023timegpt} & 300M--1B & Multi-domain datasets & 50--200ms/forecast & Superior cross-domain generalization \\
    \addlinespace
    Aurora \cite{bodnar2024aurora} & 1B--7B & Global weather data & 100--500ms/forecast & 30\% faster than NWP with comparable accuracy \\
    \addlinespace
    GridFM \cite{Hamann2024} & 100M--300M & Grid topology data & 10--40ms/forecast & Improved spatial-temporal relationship modeling \\
    \addlinespace
    Physics-Informed FMs \cite{farhadloo2025towards} & 200M--500M & Hybrid data sources & 30--150ms/forecast & Better performance in extreme event prediction \\
    \bottomrule
  \end{tabularx}
\end{table}
Effective domain adaptation necessitates examining the model's pre-existing understanding to ensure it aligns with the new domain's details. If not, a large pre-trained model might inaccurately infer (negative transfer) with data that have different noise characteristics or regulatory regimes (common in energy markets) \cite{huang2022adversarial}. Hence, fine-tuning is accompanied by techniques to avoid overfitting and align the model with domain expert knowledge (e.g., incorporating known physical laws or constraints during training) \cite{schreiber2022model}. Transfer learning and domain adaptation allow FMs to generalize across different clean energy forecasting tasks. They connect a general FM to the specific problem.

\subsection{Challenges in Applying FMs to Energy Forecasting}
Implementing FMs for clean energy forecasting involves several obstacles. This text highlights the main issues and current strategies to effectively tackle them within the energy sector.

\begin{itemize}
\item \textbf{Computational Cost}: FMs utilize a large amount of research resources. For example, state-of-the-art FMs, such as GPT-3, contain a significant computational cost, which requires the models to be trained over long periods of time (e.g., 175 billion parameters) \cite{bommasani2021opportunities}. The costs for training or fine-tuning such large models, as well as the issues around latency and energy usage for real-time forecasting, may be prohibitive for most organizations in the energy sector. Ironically, AI for clean energy could be used in an energy-inefficient way, so researchers are striving to pursue more efficient systems. A practical solution that significantly reduces the computational load is to adapt existing pre-trained FMs for energy domain-specific applications, as opposed to creating a new FM from scratch \cite{german2024transfer}. Researchers are also exploring whether model compression, distillation, and hardware accelerators can reduce inference time so that FMs can be run within practical constraints.

\item \textbf{Data Availability}: FMs have achieved success because of their ability to learn from extremely large datasets, but this can be challenging in the energy sector because datasets can be fragmented or not always useful. Energy systems have massive amounts of data in the form of smart meter readings, wind turbine sensor data, weather satellites, etc., but these datasets are not always publicly accessible or useful because privacy, proprietary issues, or data issues. FMs are designed to learn on diverse data \cite{german2024transfer}, and either a lack of diversity in training data or not having enough of it can lead to poorly-performing models. It is difficult to assemble an adequate dataset with a representative amount of data to pre-train an FM to use in the energy sector. To remedy this situation, there are three main data strategies: create consortiums and platforms to openly share energy datasets; apply transfer learning from comparable fields like climate science; and create new datasets through simulation of a grid to complement existing observations. Pre-training an FM using an openly shared dataset with variety and adjusting it afterwards on the limited niche dataset in the energy sector has the potential to be effective. The pre-training dataset provides general knowledge and the adjusted dataset provides energy patterns.

\item \textbf{Domain-Specific Biases and Model Validity}: A FM pre-trained on generic data may not understand energy systems nuances. Every domain has its biases; for example, energy demand on weekends vs weekdays, or solar output response to sudden weather changes. If a FM's pre-training data lacks certain patterns (e.g., mostly temperate climate data, not tropical solar farms), the model could inherit these biases in its forecasts. Any defects or biases in the FM might affect all its downstream applications \cite{bommasani2021opportunities}. This is important to address in regulated activities and when making decisions in safety-critical areas such as energy; even minor biases or inaccuracies in the model can have serious consequences (for example, demand that is persistently underestimated can provoke blackouts, and demand that is persistently overestimated can lead to wasted material and energy resources). Validating models will take considerable domain adaptation and validation effort to ensure validity. More generally, it is helpful to either tightly bound the model through prior knowledge or existing relationships as constraints, or to rigorously cross-validate the model across the scenarios to systematically identify mistakes or constraints. Researchers have started harnessing hybrid models where real behaviour is estimated using physically based forms of modelling, and then data-driven modelling, as a way to enforce realism in predictions and avoid unrealistic boundaries. Overall, purposing out domain biases (whether in the model, the dataset, or the forecast outputs) is part of the research agenda for getting FMs ready for effective clean energy sustainability.

\item \textbf{Regulatory and Interpretability Constraints}: Operating in the energy sector requires compliance with regulations, reliable operation, and transparency. Traditional methods of forecasting (regression, ARIMA, etc.) are interpretable, so stakeholders are usually aware of how predictions relate back to the owners' rationale. FMs are difficult to interpret as they have complex architectures with billions of parameters; inadvertent interpretability is diminished. This obscurity also makes it more difficult for stakeholders to establish trust and buy-in. For regulators who require acceptable, defendable models for critical operations like grid functionalities or energy trading, an opaque decision-making process is a deterrent to trusting predictive models. Certain regulatory systems require forecasting to be auditable and often dictate audit procedures. There is the additional challenge of verification and validation: it is hard to know if FMs can operate safely under sufficient conditions. Accuracy is only one consideration, it is critical to enforce transparency and responsibility in deployment. Current solutions involve developing time-series explainable AI approaches, and interpretable surrogate models that work much like FMs in predicting absolute or relative measures of performance without explaining why they predict what they do. The current cross-disciplinary approach of involving AI researchers, energy researchers, and policy makers aims to establish agreed upon guidelines for using AI in grid operations. It is recognised that this relationship between errors and risks of FMs will take time to develop \cite{bommasani2021opportunities}. Interpretability and compliance may also present requirements for FMs to be practically deployed.
\end{itemize}

These challenges constitute active areas of research and development. The "Green AI" framework emphasizes energy-efficient artificial intelligence methodologies \cite{german2024transfer}, supporting sustainable FM implementation. Advancements in transfer learning and domain adaptation techniques continue to enhance model customization for specific applications while minimizing computational costs and maintaining prediction accuracy. Ongoing research addressing these implementation barriers aims to optimize FM applications for clean energy forecasting while maintaining operational integrity and efficiency.

\section{Taxonomy of FMs in Clean Energy Forecasting}
\label{sec:taxonomy}

\subsection{Evolution of FMs in Energy Forecasting}
FMs are significantly transforming clean energy forecasting, following their previous successes in natural language processing (NLP) and computer vision (CV). These large-scale models, often based on Transformer architectures, automatically learn patterns from extensive datasets using self-supervised learning, resulting in robust representations useful for various forecasting tasks \cite{Hamann2024}. However, the adoption of FMs in time series forecasting and energy analytics has lagged compared to NLP and CV.

\begin{figure}[t]
\centering
\includegraphics[width=\columnwidth]{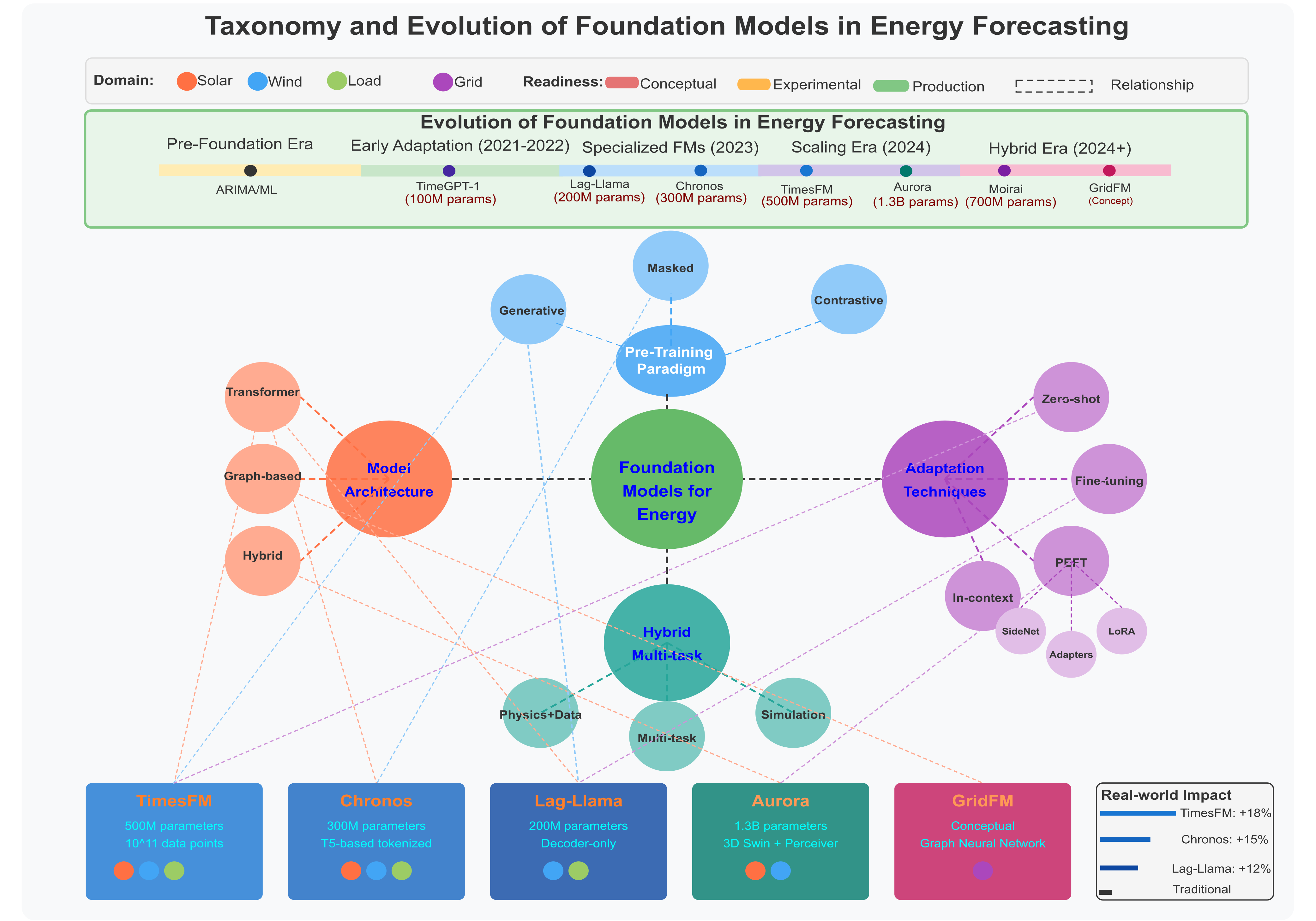}
\caption{Classification Framework for FMs in Clean Energy Forecasting. This taxonomy organizes FMs according to their architecture, pre-training paradigm, adaptation method, and application domain, with a timeline illustrating the field's evolution from 2020 to 2025.}
\label{fig:classification}
\end{figure}

As illustrated in Figure \ref{fig:classification}, we propose a taxonomy that organizes FMs according to their architecture, pre-training paradigm, adaptation method, and application domain, with a timeline showing the field's evolution from 2020 to 2025. Early pioneering efforts like TimeGPT-1 in 2023 represented important steps toward creating pre-trained models specifically for time series forecasting. TimeGPT-1 demonstrated strong cross-domain generalization using an encoder-decoder transformer architecture \cite{hou2025time}. Following this, several specialized time-series FMs have emerged, including Lag-Llama, Chronos, and TimesFM, developed between 2023 and 2024.

These newer models use knowledge learned from diverse datasets, including synthetic data, to achieve accurate forecasts without the need for extensive task-specific retraining. For example, Amazon's Chronos model was trained on a variety of time series datasets augmented with synthetic Gaussian-process data, improving its generalization abilities \cite{ansari2024chronos}. Similarly, Google's TimesFM model, trained on around 10\textsuperscript{11} real-world time points, achieved zero-shot forecasting performance close to fully supervised models across various tasks \cite{das2023decoder}.

This trend indicates a shift away from traditional, specialized forecasting models toward general-purpose FMs. Current FMs can provide accurate forecasts with minimal fine-tuning, supported by recent benchmarks. These studies show that pre-trained FMs perform at least as well as re-trained transformer models, especially as input sequence lengths increase \cite{meyer2024benchmarking}. FMs are thus rapidly becoming a practical and efficient method for improving clean energy forecasting, with important implications for energy planning and sustainability.

\subsection{Classification by Model Architecture}
In energy forecasting, transformer-based architectures account for the majority of FMs due to their ability to effectively model sequences. Most state-of-the-art FMs utilize variations of transformers designed for temporal data. For instance, Chronos uses a text transformer (T5) for time series forecasting by converting numeric values into tokens and thinking of forecasting like a language modeling task \cite{ansari2024chronos}. Similarly, Lag-Llama and TimesFM use decoder-only transformers, inspired by GPT models, to predict future values based on past observations \cite{rasul2023lag,das2023decoder}. As a general group, transformer-based models have the capabilities necessary to expect long-range dependence (more than adjacent observations) and complex patterns. Additionally, some other models are equipped with encoder-decoder transformer structures, similar to sequence-to-sequence models, for enhanced generalization across domains \cite{hou2025time}, such TimeGPT-1.

While transformers are well represented in energy forecasting, other recent FMs that are exploring different or hybrid architectures to work with energy data. One type of work utilizes spatial structures, such as in Microsoft's Aurora model for weather forecasting, which incorporates a 3D Swin transformer with Perceiver IO encoders to model 3D atmospheric grids \cite{bodnar2024aurora}. In turn, this work uses embedded features to account for spatial and temporal information that is necessary when forecasting weather and renewable generation events. A second avenue of research uses graph neural networks (GNN) in energy models. The GridFM model concept proposes a GNN-based FM to represent the grid topology to consider the networked structure of power systems \cite{Hamann2024}. Here, the grid is treated as a graph-like representation of nodes and edges, and effectively learns world model or spatial-temporal relationships, such as power flows on electricity transmission lines, that a sequence-only transformer models are unable to learn. Other experimental models such as diffusion models and hybrid CNN-transformers are also under exploration. In the field of diffusion models, or diffusion transformers, researchers have explored time series forecasting with diffusion transformer models to model the full distribution of time series outcomes \cite{ansari2024chronos}. Overall, FMs can be categorized into three groups by architecture: (a) transformer-based models (sequence or sequence-to-sequence), (b) graph-based models for structured grid or weather data, and (c) hybrid models that combine convolutional or diffusion approaches with transformers. Each of these three models has distinct advantages: (i) transformers for sequence-based modeling, (ii) GNNs for models that work with relational data, and (iii) hybrid models for models that use more than one data type.

\subsection{Classification by Pre-Training Paradigm}
Another important classification is determined by the pre-training methods for clean energy tasks. It will also be important to consider the different learning objectives and data used in pre-training. Most FM based forecasting uses generative pre-training objectives which are highly similar to language modeling. Models of interest like Chronos and TimesFM apply generative pre-training objectives, which maximize predictive modeling error using some extensive amount of unlabeled data. Chronos tokenizes value and maximizes the likelihood of predicting these follow-up tokens \cite{ansari2024chronos}. TimesFM presents different types of forecasting for the future time series segments dependent on historical context \cite{das2023decoder}. Other pre-training models are built using masked modeling objectives, such as Moirai, which learns through reconstruction of missing data segments \cite{hou2025time}. Contrastive learning approaches, such as TimeCLR and CoST, learn general representations through segments across multiple contexts \cite{liang2024foundation}.

The pre-training data is also important considering data diversity and scale. FMs utilize large and diverse datasets, often including synthetic data, to learn features that are universal. For example, Chronos utilizes synthetic Gaussian Process data to make it more robust \cite{ansari2024chronos}, while TimesFM uses about 10\textsuperscript{11} data points across multiple domains \cite{das2023decoder}. Pre-training methods can be classified into; (i) generative forecasting, (ii) masked reconstruction, and (iii) contrastive or hybrid objectives. Each of these offers particular advantages but generative models will directly improve a forecasting feature, masked modeling may provide anomaly detection and imputation functionality. Regardless, they share a common purpose - to use large scale self-supervised learning to develop flexible models to use for a specific clean energy forecasting task \cite{Hamann2024}.

\subsection{Classification by Adaptation and Fine-Tuning Techniques}
After pre-training, there are methods FMs can be adapted for clean energy forecasting tasks. These methods include zero-shot forecasting, wherein the pre-trained models predict within the new tasks without any additional training. TimesFM has presented a near fully-trained model level of accuracy, though none of this additional task training was performed \cite{das2023decoder}. Lag-Llama and Chronos have also produced similar levels of accuracy on forecasts for new series immediately after the meta-pre-training process \cite{rasul2023lag, ansari2024chronos}. Zero-shot capability is particularly useful in areas where location or data associated with a specific asset is limited, they can produce estimates based on learned patterns in large amounts of basic data. The authors of the studies all provide significant evidence that zero-shot predicting by FMs takes considerably less effort than traditional forecast models have in the past, by always having to retrain for any new task \cite{meyer2024benchmarking}.

Nevertheless, fine-tuning, or in some cases, adaptation, is often done when there is a need to specialize existing FMs for specific forecasting tasks or to measure improvements in performance accuracy. Fine-tuning the model fully involves retraining every parameter of the model using only historical data from the target TM. During the fine-tuning phase, the FM will usually find a better level of accuracy by adjusting to the local patterns. Rasoul et al. said that fine-tuning Lag-Llama using just a few small portions of a new dataset improved its performance over the traditional models \cite{rasul2023lag}. Current studies have identified new ways to make fine-tuning much more efficient, some by providing more efficient first steps than full fine-tuning, as in parameter-efficient fine-tuning (PEFT), adapters, Low-Rank Adaptation (LoRA), etc. Aurora has demonstrated that it can use LoRA to efficiently fine-tune its large scale model used for initiatives such as weather forecasting \cite{bodnar2024aurora}, while some of the other studies, including one by Hou et al., introduce an efficient way to utilize specialized adapters for continual model efficiency without affecting previously learned knowledge \cite{hou2025time}.

Another approach for adaptation is in-context learning and prompting. Inspired by large language models, some methods provide task-specific examples or hints to guide the model at inference, without updating model weights. For instance, LLM4TS and TEMPO fine-tune GPT-2 on formatted time series data. They then use prompts to generate forecasts \cite{hou2025time}. LLMTime encodes time series data into labeled tokens, such as timestamps or units, improving zero-shot forecasting \cite{hou2025time}. These methods treat the FM as a predictor controlled by input presentation. Although prompting is less common in energy forecasting than NLP, it is an emerging method. It allows FMs to adapt to new tasks without gradient-based fine-tuning. Instead, task descriptions, historical data, or specific formats provided as input guide the adaptation.

In summary, the taxonomy of adaptation includes zero-shot usage, full or partial fine-tuning, few-shot learning, and prompt-based adaptation. Few-shot learning involves training on a small dataset, placing it between zero-shot and full fine-tuning. Each method has different trade-offs between effort and performance. Efficient fine-tuning methods (such as LoRA and adapters) are especially important in clean energy forecasting. It is not practical to retrain large models from scratch for every new wind farm or building. These efficient methods allow using a single pre-trained model across multiple tasks, adapting it with minimal extra training \cite{hou2025time}.

\subsection{Emerging Hybrid and Multi-Task Models}
Recent hybrid strategies merge two or more sources of knowledge or objectives into a single FM. Hybrid modeling involves combining physics-based and data-driven approaches to define FM by leveraging both methods. In forecasting renewable energy, hybrid approaches by any definition tend to mix physical simulations in the context of laws of a relevant domain and machine learning. For example, solar power forecasting methods combined outputs of Numerical Weather Prediction (NWP) models with deep learning (increasing accuracy when compared with models using only deep learning) \cite{tortora2023matnet}. In hybrid setups of this nature, the NWP model contributes features or constraints to the FM, bringing in physical knowledge from the relative details such as weather dynamics into the forecasting. When physical details are subsequently included it time helps with interpretability and reliability of predictions, and ensures the forecast remains within physically-realistic (e.g. maximum solar irradiance) limits.

A different hybrid approach is utilizing FMs as fast substitutes for traditional, computationally intensive simulations. The recent study of AI-based weather emulation, for example, found that neural networks can often accurately emulate physical dynamics and in some cases, generalize better than traditional physics-based models \cite{Hamann2024}. Similarly for FMs for the power grid, FMs could rapidly emulate power flow or stability calculations by mixing AI techniques with power system physics \cite{Hamann2024}. Early examples of hybrid FMs include using physics-informed neural networks that could used empirical data to rapidly run through electric grid simulations \cite{Hamann2024} or graph-based FMs (GridFM) designed to explicitly represent grid connectivity and follow physical laws \cite{Hamann2024}. It is a developing area but hybrid FMs are a promising development for risk-critical energy applications, leveraging data-driven accuracy and physics-based consistency.

Simultaneously, the trend of multi-task FMs is emerging in the energy sector. Rather than developing multiple models for load forecasting, wind forecasting, or anomaly detection, you can have one FM that does all the related tasks at the same time. Multi-task models can be trained across related energy forecasting tasks, where forecasting tasks share a common structure or input. For example, when forecasting solar power output, the FM will take in weather data input. The knowledge learnt about the relationships between the weather conditions and the solar output can also benefit the wind forecasting which relies on the same weather conditions for the wind generation as well \cite{cheng2023imbalanced, deng2021mvmt}.

Hou et al. (2025) recently proposed the first multi-task framework meal large-scale FMs for energy was applications \cite{hou2025time}. In their framework, the authors used a single Transformer-based model and fine-tuned the model for two related tasks: short-term load forecasting and overload event detection, adding an additional model for the task specific features \cite{hou2025time}. Some more recent architectures, such as Timer and UniTS, combine predictive tasks (forecasting multiple outcomes) and generative tasks into one model \cite{hou2025time}. By leveraging tokenization (i.e., task tokens) or multiple output heads, the models can generate as many outputs as needed from a single input stream.

There are two main benefits of multi-task FMs. First, they can improve forecasting performance. If the FM learns information from one task, it could help the model perform better on related tasks. For example, the patterns that were learnt while forecasting solar power may also be beneficial to forecasting wind power, since both are reliant on weather predictions \cite{cheng2023imbalanced, deng2021mvmt}. Second, multi-task FMs improve efficiency and scalability. Rather than having several individual models memorizing and/or calculating the same input for several forecasting objectives, operators only need to maintain and operate a single FM across individual forecasting tasks \cite{li2024water}. Multi-task training can also provide a form of regularization and provide better fidelity of the training data by leveraging the independent "tasks" within the multi-task FM, leading to improved accuracy on individual tasks \cite{nikentari2022narmax}.

Nevertheless, even if it is relatively easier to train a multi-task FM, combining tasks is still not easy. Careful selection of tasks to ensure they complement each other (e.g., another task should help the FM learn) is essential, and may also require the use of other techniques such as uncertainty-weighted loss balancing or curriculum learning to ensure the training can proceed in a stable manner \cite{mendis2024multivariate}. In conclusion, hybrid and multi-task FM frameworks go beyond classic single-task framework approaches. Hybrid FMs leverage existing knowledge defined by domain-specific knowledge, whereas multi-task approaches incorporate multiple forecasting objectives into the model under a single problem framework. Both these trends are aimed at improving FM frameworks towards addressing the unique and extraordinarily complex challenges of clean energy forecasting.

\section{Data Perspective: Advances in Data and Representation for Energy Forecasting}
\label{sec:data}

\subsection{Integrating Diverse Data Sources}
Clean energy forecasting relies on many different types of data, including weather measurements and grid operations data. Recent FMs aim to integrate these varied data sources. Advances in sensors and IoT technologies provide extensive datasets beyond traditional historical data. Modern forecasting models can incorporate the following:

\begin{itemize}
\item \textbf{Meteorological data}: Numerical Weather Prediction results, weather station measurements, radar and satellite cloud images, all important for wind and solar forecasts \cite{patil2024hybrid, bellinguer2021assessment}.

\item \textbf{Satellite and remote sensing data}: Geostationary satellite images for cloud movement, satellite-based solar irradiance, LiDAR data, or sky camera images used to track local cloud conditions at solar farms \cite{bright2021statistical, tournadre2022heliosat, gregor2021nowcasting, rodriguez2021nowcasting, logothetis2021forecasting}.

\item \textbf{IoT sensor and smart meter data}: High-resolution sensor measurements, such as building thermostat and occupancy sensor data for load forecasts, SCADA data from wind turbines, and EV charging data for grid load forecasting \cite{suheb2022iot, kaur2023iot, alhaddad2024forecasting, mlynek2022smartmeters, chen2023smartmeter}.

\item \textbf{Grid and market data}: Operational grid data (voltage levels, power flows), electricity market prices, demand response signals, and contextual information like grid events or holidays \cite{yuan2021demand, wang2024demand, rawal2022comparative, cui2023game, shaquor2021dayahead}.
\end{itemize}

\begin{figure}[t]
\centering
\includegraphics[width=\columnwidth]{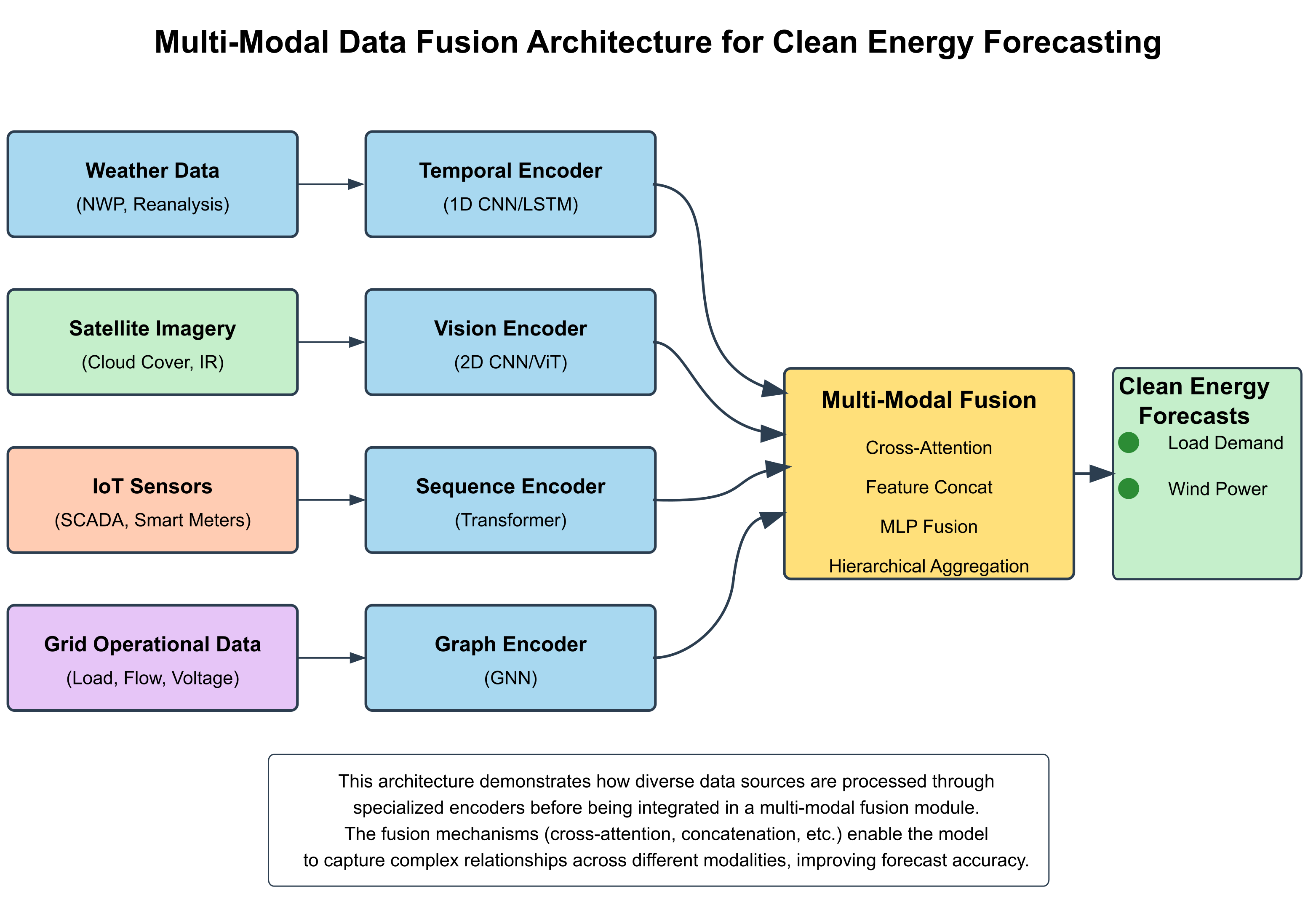}
\caption{Multi-Modal Data Fusion Architecture for Clean Energy Forecasting. The diagram illustrates how diverse data sources (weather, satellite imagery, IoT sensors, and grid data) are processed through specialized encoders before being integrated in a fusion module to generate accurate clean energy forecasts.}
\label{fig:data_fusion}
\end{figure}

FMs are very adept at integrating multiple data sources, mainly due to their established size and training on different datasets. A single model can recognize many types of inputs (images, sequences, spatial grids) to help shape a full representation of the energy system. IBM, for example, recently created climate-oriented FMs trained on satellite hyperspectral images in conjunction with ground-based LiDAR data to estimate environmental variables \cite{schmude2024prithvi}. Similarly, FMs used for weather and energy forecasting more frequently incorporate data that is of varying spatial and temporal resolutions. An example of this is Microsoft's Aurora model, trained on over a million hours of simulated weather data from several datasets of variable resolution and variables \cite{bodnar2024aurora}. As Aurora learned jointly from these many data inputs, it understood atmospheric phenomena across multiple scales and operated reliably within limited data availability \cite{bodnar2024aurora}. These capabilities are particularly useful when doing renewable energy forecasting, especially when looking at extreme or rare events.

When looking at the energy industry, there is indeed increased effort to combine data streams to enhance forecasts. For example, SolarAnywhere's production basics combines data from satellite imagery, satellite-based modeling, and ground-based sensor data into solar forecasting models that are based on deep learning \cite{paletta2023advances}. This integration provides more accurate irradiance predictions by leveraging several datasets of detailed cloud coverage from the satellite imagery combined with local sensor measurements. In the same way, in data-sparse locations, researchers have used nighttime satellite imagery along with infrastructure characteristics and very sparse data on electricity use to improve their demand forecasts \cite{lee2023multimodal}. More importantly, these examples of combining data in different ways show the improved reliability of forecasts: e.g., satellite imagery offers more coverage, IoT sensors are localized real-time, and weather models offer a level of physicality.

Recently, new FMs were also explicitly constructed to integrate multiple data sources with incorporation of specialized architecture and training that will deal with many types of varying inputs at once. The emergence of FMs explicitly designed to incorporate multiple data sources, illustrates a shift from traditional single-source datasets to using integrated data altogether. Clean energy forecasting increasingly combines data from weather, grid, and sensors to better understand and predict energy system dynamics.

\subsection{Multi-Modal Data Fusion Techniques for Forecasting}
Having a wide range of data doesn’t answer an important problem, namely how to combine different data modalities in a uniquely defined Forecasting Model. State-of-the-art studies about FMs have defined new data integration approaches using recent FMs that are designed to analyze images and time series at the same time. FMs usually require a separate encoder or sub-network for each type of data and then merge them at a later time in the process, for an example in multi-modal solar forecasting, CNN encoders might be used to process satellite cloud images and a Transformer processes historical solar production time series data separately.

After extracting features independently, the model can use the features extracted from each modality and learn how to relate them to each other. Tortora et al. uses this structure in MATNet that is a multi-level fusion Transformer for photovoltaic (PV) based Forecasting Models. MATNet uses the historical PV generation data and weather data (historical and forecasted) separately by having dedicated self-attention modules in MATNet. At several stages in MATNet the weather and PV streams were integrated by combining the information in a separate self-attention fusion layer, allowing the model to both learn about short-term effects associated with changes in weather (cloud passing) and also the longer-term patterns we associate with different weather types through having multiple levels of fusions. MATNet incorporated in its architecture this deep level of fusing and this was found to enhance day-ahead forecasts of total PV generation accuracy \cite{tortora2023matnet}. The figure \ref{fig:data_fusion} provides an example of a typical multi-modal fusion architecture and shows how different data sources are potentially processed through encoders that are tailored to individual modalities, before being fused to make accurate clean energy prediction.

Another way of approaching data fusion is to use "cross-modal attention", where the features of one modality can determine how the features of another modality are processed. As an example, in wind power prediction, the wind field maps may help the model identify which parts of its turbine sensor data particular attention should be paid to \cite{bharti2022attention, meng2022hybrid}. Through cross-attention, the model can be dynamically guided to emphasis information coming from one data source when patterns identified in another \cite{zhang2022crossattention}. This is quite similar to visual attention in image captioning, where the model uses textual inputs to perform analysis on the image. In this case, the weather patterns could mediate attention over the power output time-series data. While research on the use of cross-attention for energy FMs is still in its infancy, it represents a viable alternative to simpler fusion methods, such as concatenation \cite{chen2021ultrashortterm, sendanayake2023attention}.

Spatiotemporal alignment is also an essential aspect of data fusion. Often times, energy datasets will be aligned in time and space, but will subsequently have different spatial resolution and temporal resolutions. For example, it may be that satellite imagery is taken every five minutes, while solar plant measurements are taken by a specific sensor every minute. More sophisticated models will be programmed to account for these differences using interpolation, specific encodings of the spatial coordinates, or hierarchical pooling to accommodate the differing modalities. For example, the Aurora model with a 3D Swin Transformer learns using climate and model data with multiple resolutions and many different pressure levels at once. By jointly optimizing the information provided from these datasets, it can create a common representation of that data \cite{bodnar2024aurora}. It is worth noting that Aurora also implements a Perceiver-based encoder to manage multiple input formats and map them into a common latent space.

Simultaneously, researchers are identifying more universal methods of encoding various types of data that can be represented as sequences of tokens that are compatible with Transformers. \cite{Hamann2024} notes that it is important to consider what tokens will be understandable to a model when building FMs for a new domain. This may include splitting a time-series into patches, or representing graphs in a token like manner \cite{Hamann2024}. In the multi-modal instance, an image can be separated into image patches and similarly, a time series can be represented as segments of temporal patches. Although these distinctions remain distributionally differentiated in most modeling methods, both the images and the time series can be treated by a single model developed as a Transformer model. This token based fusion approach makes the multi-modal focus more straightforward while allowing optimal use of the Transformer style sub-sequential processing. For example, Time-LLM incorporated textual information with time based events, coding time-based events as text tokens interpretable to a language model \cite{hou2025time} The Time-LLM paper develops specifically a process for where external information used in the decision making process is based on textual context (for example news reports) to be incorporated in the forecasting. However, the mechanism is fundamentally similar: each mode of information is coded into equivalent sequences of tokens, and then the FM to simultaneously process the combined tokens.

\begin{table}[htbp]
  \caption{Data Sources and Representations for Foundation Models in Clean Energy Forecasting}
  \label{tab:data_sources}
  \centering
  %------ style tweaks ------
  \scriptsize
  \setlength{\tabcolsep}{4pt}     % horizontal padding
  \renewcommand{\arraystretch}{1.2} % vertical padding
  %--------------------------
  \begin{tabularx}{\columnwidth}{@{}
      >{\raggedright\arraybackslash}p{0.19\columnwidth}
      >{\raggedright\arraybackslash}p{0.21\columnwidth}
      >{\raggedright\arraybackslash}p{0.15\columnwidth}
      >{\raggedright\arraybackslash}p{0.19\columnwidth}
      >{\raggedright\arraybackslash}X@{}}
    \toprule
    \textbf{Data Category} & \textbf{Sources} & \textbf{Temporal Res.} &
    \textbf{Integration Methods} & \textbf{FM Applications} \\
    \midrule
    Meteorological               & NWP outputs; weather stations; atmospheric reanalysis
                                  & 1--6\,h (forecast); 5\,min--hourly (obs.)
                                  & Feature embedding; specialised encoders; cross-attention
                                  & Wind forecasting; solar irradiance; load prediction \\
    \addlinespace
    Satellite / Remote Sensing   & GEO satellites; MODIS; Landsat; LiDAR
                                  & 10–30\,min (GEO); daily (polar)
                                  & CNN pre-processing; multi-modal fusion; tokenisation
                                  & Cloud tracking; solar forecasting; environmental monitoring \\
    \addlinespace
    IoT / Smart Meters           & SCADA; smart meters; building sensors; EV chargers
                                  & Real-time–15\,min
                                  & Time-series patching; edge processing; sequential encoding
                                  & Demand response; building load; prosumer behaviour \\
    \addlinespace
    Grid Operations              & System operators; energy markets; substations
                                  & Real-time–hourly
                                  & Graph representations; temporal attention; multi-view modelling
                                  & Grid stability; price prediction; congestion management \\
    \addlinespace
    Behavioural / Contextual     & Calendar events; social media; user patterns
                                  & Daily–weekly
                                  & Feature augmentation; auxiliary inputs; context embedding
                                  & Demand prediction; anomaly detection; peak identification \\
    \addlinespace
    Synthetic                    & GAN-generated data; physics simulations; augmentation
                                  & Variable
                                  & Pre-training enhancement; domain adaptation; regularisation
                                  & Rare-event training; uncertainty modelling; robustness \\
    \addlinespace
    Multimodal                   & Cross-domain composites
                                  & Multi-scale
                                  & Multi-stream Transformers; cross-attention; token-based fusion
                                  & Comprehensive forecasting; extreme-event prediction \\
    \bottomrule
  \end{tabularx}
\end{table}

Multi-modal data fusion in FMs, which can occur via either late fusion (i.e., features are processed separately and combined into a model) or early fusion (i.e., data are fused into a single representation), for example, through using multi-stream Transformers, cross-attention between modalities, or tokenizing the text and imagery to make the model learn about activities in the weather images with the sensor data \cite{wang2022transformer, cui2023temporal}. The benefits are crystal clear from empirical projects in this area. Models can exploit, or differentiate, between complementary data sources that add low error or uncertainty to model forecasts. For example, sky images reduced overall uncertainty by providing an explicit measurement of cloud movement, which could be jointly predicted with irradiance time series data from ground sensor data into some sort of 'winning' method to provide a short-term solar forecast \cite{morelli2023multimodal}. Another example is with smart building meter data, where contextual data like weather or calendar / event information led to better prediction of building load due to the capture of variability that could not be captured through traditional time series \cite{morelli2023multimodal}. FMs represent an evolution beyond single time series sequence predictors, towards some form of universal model drawing from different streams of data. As energy systems produce more sensor data, they will have more information \cite{jia2024geminifusion, pellegrain2021streaming}. Table \ref{tab:data_sources} summarizes the varying data sources that can be used by foundation models to forecast clean energy. The table distinguishes different sources of data based on time resolution, class of data, where and how they are integrated, and what they have been used to forecast. This organization shows the differing sources of data integrate different and unique information designed to improve forecasting across renewable energy contexts, including meteorological data from ground stations, satellite data, IoT sensor data, or disparate grid operational data.

\subsection{Improved Data Representations: Impact and Case Studies}
Innovations in data integration and representation enhance forecasting accuracy and reliability. Several case studies illustrate improved outcomes from richer data and better representations of data sources in clean energy forecasting:

\textbf{Solar Forecasting Using Satellite Data}: The inclusion of satellite cloud data substantially improved the accuracy of solar forecasts. A recent survey indicated that deep learning predictive models that used multiple Earth observation data sources (satellite imagery, sky cameras) have higher accuracy than predictive models that only used ground sensor data sources \cite{paletta2023advances}. For example, SolarAnywhere's operational forecasting model combines cloud data from satellite images with radiative transfer modeling, enabling much more accurate solar irradiance nowcast forecasts than traditional methods using ground sensors \cite{paletta2023advances}. The satellite imagery allows clouds to be detected and tracked, enabling models to forecast perturbations in solar power output in the short term that historical PV generating data alone cannot indicate. This advantage of using both satellite imagery and historical PV generation data is especially advantageous for nowcasting, or predicting the near-term (sub-hourly) variability in solar generation output, where knowledge of current sky conditions is necessary. Utilities report that solar forecasts using satellite imagery improve their ability to predict solar ramp events, enabling them to carry out load management and better stabilize the grid in regions with high solar penetration.

\textbf{Wind Power Forecasting with Weather Features}: Wind power forecasting has recently improved through the better usage of weather information. Current forecasting models use, in different proportions, outputs of numerical weather prediction (such as wind speeds at various heights, atmospheric pressure fields) and turbine SCADA measurements together. The combined data of historical and probabilistic information has reduced forecast errors, improved forecasting of extreme wind events (which cause changes in production) and many forecasting systems are in flux due to the domestic influence of the EU. A case study from a grid operator showed that a hybrid machine learning model, utilizing weather forecasts, in conjunction with historical wind farm data, could reduce the day ahead forecasting mean absolute error (MAE) by 20\% compared to purely statistical methods available at the time \cite{tortora2023matnet}. The improvements in the accuracy simply comes from the fact that the model has access to larger weather context (e.g., storm fronts), and not just what the turbine did the previous day. FMs further this concept, by including technology from large-scale global weather reanalysis datasets, allowing them to understand complex weather patterns, and thus better map that information into wind power predictions.

\textbf{Electric Load Forecasting Using Behavioral and IoT Data}: The problem of electricity demand forecasting now incorporates behavioral data and IoT sensor data, like occupancy data, human activities data, and appliance usage data. Up to this point, load forecasts had largely been based solely on temperature and calendar data. Utility companies now have smart meters and IoT sensors that allow them to directly see the drivers of demand \cite{palety2022residential, rajendiran2022nilm}. In one example, a utility implemented a FM for load forecasting using smart thermostat data and occupancy sensor data from some of their buildings. The model was able to recognize patterns, such as pre-cooling activities or spikes in demand associated with special events. This model was able to have lower forecasting errors on unusual days (like holidays or large sports events) because it was able to take into account the ways that demand variations occurred based on human activity that we would miss capturing in weather only based models \cite{avordeh2021demand}. While industry reports and use case studies are often kept confidential, academic literature is equally demonstrating that load forecast accuracy is significantly improved and flexible when detailed information defining or indicating behavior is incorporated to the load forecast models. \cite{vasuprada2021smartmeter, bashawyah2021ml}.

\textbf{Use of Synthetic Data for Better Generalization}: Data representation can be improved not only with real sensor data, but with synthetic data as well. The Chronos model example illustrates this. Integrating synthetically-generated time series data when training the Chronos model tended to improve its average zero-shot forecasting performance on previously unseen datasets \cite{ansari2024chronos}. In this case, the synthetic data was also considered to be a form of data augmentation by providing clear examples of behaviours and patterns that would rarely occur together in nature, for instance, atypical seasonal trends or extreme events. This utility is useful in practice: a FM trained on a larger set of simulated scenarios is less likely to find itself in unexpected conditions during forecast tasks in the real world. In other words, synthetic data helped the model learn from a wider variety of foreseeable scenarios and conditions, making it more resilient when it encounters novel scenarios during issues such as wind farm outages or rare demand pattern behaviour in buildings.

\textbf{Building Energy Forecasting - Data Issues and Solutions}: A recent study on building energy forecasting (BEF) suggested that simply fine-tuning a FM from a very general model generated very poor initial performance \cite{liang2024enabling}. The first problem arose from the difference in the training data of the FM and the distinctive characteristics of building energy data; in particular, noisy, tenant-specific, and uniquely patterned weekly (i.e., occupancy) data. To overcome this issue of the mismatch, the researchers suggested a second approach using contrastive curriculum learning. They organized the fine-tuning data in stages, from what they called easy patterns to more complex patterns, while also using contrastive objectives that would encourage the model to find signals that are specific to buildings \cite{liang2024enabling}. Effectively, this approach improved zero/few-shot forecasting from building energy loads to an average accuracy (improvement) of 14.6\% compared to basic fine-tuning \cite{liang2024enabling}. This example emphasizes the significance of data selection and organization during model adaptation; essentially improving data representation during training.

In all these examples, we tend to see a similar conclusion across studies: better data consistently lead to better forecasts. Whether combining data sources (having access to more data), claiming to encode data more appropriately, or augmenting the training data (to have many examples of even more represented experiences), will likely lead to better - and potentially reliable - forecasts. Since FMs are well-suited to this type of acquisition and experience owing to their ability to ingest and work with robust, larger, and potentially more varied datasets, we also concede they can pose challenges (such as ensuring quality of data and interoperability between the different sources of data); meaning that bad data may also be equally bad for a strong FM. Better practices are being undertaken at standardizing data formats and creating data spaces for energy data that support these phenomena; allowing data sharing and interpreting data compatible actions through study \cite{ArxivGrid2024}. As such, further data experiments and real-world use cases of FMs will be introduce; notably grid operators/a Grid operator, is starting to try large FMs that are speculated to have been ingesting meteorological data for weather outcome guidance and usage data for allocation/final demand prediction. These operators are hopeful to find these models trained across the two data sources produce better forecasts and fall at a trust threshold compared to smaller separate specialized models of usage, a unified FM concept. These early practical successes will begin to create trust internally with data representation and standards, increasing confidence that consideration of investments in better data representations and FMs improves accuracy (performance or acceptable accuracy), reliability, and improves energy system efficiencies.

\section{Methodology Perspective: Innovations in Modeling Techniques}
\label{sec:methods}

\subsection{Advances in Transformer-Based Models and Alternatives}
The Transformer architecture is commonly used in FMs for clean energy forecasting, but ongoing research aims to better adapt it to time-series tasks or explore alternatives when necessary. A key area of innovation is improving efficiency in handling long temporal sequences and high-dimensional inputs. Standard Transformers have quadratic complexity with respect to sequence length, making them inefficient for long historical windows or high-frequency data. To solve this, researchers introduced methods like segmenting or patching the input data before it enters the Transformer. For instance, TimesFM applied a patching approach, grouping consecutive time steps into "patches," reducing the effective input length for self-attention and significantly speeding up computation (approximately by the factor of patch size) while also improving prediction accuracy \cite{das2023decoder}.

% \begin{figure}[t]
% \centering
% \includegraphics[width=\columnwidth]{fig6_transformer_architectures.pdf}
% \caption{Advanced Transformer Architectures for Time Series Foundation Models. The diagram illustrates key components of modern transformer-based foundation models for energy forecasting, including specialized modules for time-frequency representation, patching, and self-attention mechanisms optimized for temporal data.}
% \label{fig:transformer_architectures}
% \end{figure}

The model can handle long sequences by compressing multiple consecutive time steps (e.g., 32 points) into a single token. Patch-based Transformers, similar to Vision Transformers, have made it indicative that even at long horizons (e.g. hourly predictions for one week), the sequence does not have to be large if we reduce effective length. Furthermore there were also specific attention mechanisms, such as Informer or log-sparse attention, developed in the past to address long sequences that are becoming frequently incorporated into FMs, setting the stage for these methods to handle longer historic time periods.

Another recent development were the efforts to design Transformers which explicitly model seasonal and frequency patterns common with energy data. Some methods are embedding Fourier or wavelet transforms in the network to detect periodic signals. For example, the FreqMixer framework (2025) processes frequency domain and attention signals in parallel by creating a branch designed to catch seasonal patterns. The output from this branch was then merged with the main Transformer path\cite{hou2025time}. This is an explicit design feature to model regular structure like day to day to week to week cycles, and rely less alone on the activity of attention alone. Such architectural modifications will especially help energy forecasting tasks where patterns are clear with something like difference between weekday and weekend energy use or daily solar generation cycle.

Graph Neural Networks (GNNs) are becoming a favored choice for forecasting tasks requiring spatial structure, like predicting power flows in a grid or renewable generation sites. GNNs are not widely used in FM applications yet (compared to, say, Transformers), but a few research teams are investigating GNN-based approaches to grid forecasting, called GridFM. In a grid context, nodes (e.g., substations or generating sources) can be regarded as vertices of a graph and edges will exist depending on physical connections, such that a GNN can naturally represent the power flows and their dependencies. Hamann et al. propose that graph-based FMs can learn patterns of a diverse set of grid topologies and operational configurations, so that the inherent dynamics of a network can be represented \cite{Hamann2024}. This would be particularly useful for problems such as distributed load forecasting or grid stability assessment, as these applications could use learnings derived from a single model regardless of varying grid configurations. Other works have begun to integrate recent advancements for a GNN-based approach to improve forecasts, such as using Graph Neural Transformers (self-attention combined with graph convolutions) for forecasting on a regional load scale, and applying GNNs to downscale forecasts from larger spatial resolutions down to each grid node.

Diffusion models and other generative architectures are also coming about as alternatives to deterministic forecasting methods. Diffusion models were first successful in image generation tasks; researchers are now working on diffusion models as probabilistic time-series forecasting models. For example, the TimeDiT model combines a diffusion process (KL divergence) with a transformer-style architecture where it acts as a probabilistic FM for time-series forecasting \cite{ansari2024chronos}. In their approach, the user prepares a noise prediction and the model iteratively transforms that prediction into a time series, while generating a distribution of possible, believable futures. For this reason, diffusion models are particularly valuable for quantifying uncertainty (see section 6.3). Though diffusion-based models for forecasting are still in the early innings, they might supplement or even supplant current models. This is especially true for multi-facetted forecasting problems like wind power predictions in variable weather with conflicting factors regarding power and energy versus independence for a wide range of acceptable operational conditions.

In summary, recent advances in model architectures include:

\begin{itemize}
\item \textbf{Enhanced Transformers}: incorporating techniques such as patch-based segmentation, frequency-domain components, and improved handling of long time sequences.

\item \textbf{Graph-based models}: leveraging relational information for spatial forecasting tasks, such as grid node predictions.

\item \textbf{Hybrid architectures}: combining CNNs (for local patterns or image-based data) with Transformers, or integrating multiple Transformer variants for multi-modal forecasting.

\item \textbf{Generative models}: including diffusion-based models and autoregressive language models adapted for probabilistic forecasting or prompt-driven prediction tasks.
\end{itemize}

Each of these innovations addresses specific challenges, such as handling long sequences, capturing correlations across multiple locations, or producing full probability distributions. In practice, FMs often combine or cascade multiple techniques, for example, using graph-based modules alongside Transformers, to benefit from their different strengths \cite{phyo2021hybrid, jalali2022hybrid}. As a result, the current toolbox for energy forecasting is more diverse. It now includes specialized designs instead of relying on general-purpose architectures \cite{mendis2024multivariate, thaker2024solar, usmani2024omme}.

\subsection{Pre-Training Strategies, Fine-Tuning, and Task Adaptation}
In recent work, training and adaptation procedures of FMs have jumped forward, particularly in the practicality of pre-training effectively, and the evolving capabilities to adapt FMs to new tasks. An example of a development has been the addition of multi-task learning and multi-objective learning using a pre-training method. Using earlier generations of FM methods, the methods would only use pre-training for one task (e.g. predicting the next data point). In contrast, newer methods in the FM community are using multiple-objectives in the pre-training process simultaneously to train and learn multiple objectives. For example, \cite{bodnar2024aurora} pre-trained the Aurora model with datasets made of datasets containing various atmospheric variables (e.g. CO2, CH4, radiation, temperature, and so on) and rather than using a multi-task learning approach, optimized one single combined loss using multiple datasets simultaneously. Using multi-objective pre-training for the FM, helps the FM to learn a more generalized representation, and the upshot is an FM that performs well on various prediction tasks and various types of datasets, both short-term and long-term predictions. Further, certain time-series methodologies are using the task adaptation of forecasting, interpolation (e.g. filling in missing values, future predictions, or trend classifications), and share exactly the same objectives of using context to help with multiple prediction objectives in order to develop generalization.

Another advancement in model training and adaptation of FMs, is the use of curriculum learning as part of pre-training, where often pre-training usually starts with an easier task (e.g. simplified or deseasonalized data) to harder tasks (raw noisey data). The potential of curriculum learning has shown advancements relating to the applicability of FMs to building energy data, to where they improved the performance significantly \cite{liang2024enabling}.

When fine-tuning FM models to new tasks, ideally efficiency and flexibility are two desirable aspects. The PEFT is becoming known framework for FM fine-tuning paradigms since it allows the FM model to perform flexible fine-tuning procedures without having to update any model parameters. PEFT approaches typically only fine-tune a small number of model parameters and the overwhelming majority fo the model parameters are not finetuned to data. Furthermore, fine-tuning all the parameters in a model can take time, and be subject to overfitting for the small datasets. A few PEFT approaches that we have found useful are as follows:

\begin{itemize}
\item \textbf{Adapters}: small neural network modules placed between the layers of a pre-trained model. Only these adapter parameters are trained during fine-tuning, while the original weights remain largely unchanged. Adapters, initially used in NLP, have been adapted for time series—for example, Hou et al. proposed adapters that process frequency-domain features alongside the main model \cite{hou2025time}.

\item \textbf{LoRA (Low-Rank Adaptation)}: introduces small, trainable low-rank matrices into the Transformer's attention mechanism, enabling efficient adaptation without retraining the entire model \cite{hou2025time}. LoRA has been successfully applied in climate and weather forecasting models to quickly adapt for different prediction horizons \cite{bodnar2024aurora}.

\item \textbf{BitFit and Bias-Tuning}: fine-tuning only the bias parameters of a model's neurons. This simple method has been effective in certain scenarios \cite{hou2025time}.

\item \textbf{Prompt Tuning}: instead of adjusting model parameters, this method trains a small set of virtual "prompt vectors" added to the input data to guide the model toward the desired task \cite{liu2023pttuning}. This approach has been explored in time-series classification and could also be extended to forecasting tasks \cite{jia2024gpt4mts, xue2022promptcast}.
\end{itemize}

Using PEFT allows adapting a large FM to multiple tasks without significant computational expense. For instance, a study applied a SideTuning method (a form of PEFT) to an energy forecasting model, showing it effectively retained the model's general capabilities while quickly adapting to task-specific features \cite{hou2025time}. Researchers generally agree that fine-tuning only a small, targeted set of parameters is usually enough because FMs already contain rich, general-purpose representations.

Another methodological improvement is adapting FMs using transfer learning and continual learning. Energy systems often change over time: seasonal patterns shift, new solar plants begin operation, and demand changes due to events like COVID-19 or increased EV usage. FMs need continuous adaptation to remain effective. Continual learning methods help FMs learn new data without losing previously learned knowledge (avoiding catastrophic forgetting). Researchers are exploring regular updates or online training methods, where models are lightly fine-tuned with recent data on a daily or weekly basis. Hamann et al. highlight that continual learning is essential for FMs to remain relevant as power grids and operational patterns evolve \cite{Hamann2024}. Techniques such as elastic weight consolidation (which limits significant changes to important parameters) or rehearsal (training on a mix of old and new data) can help models maintain their broad forecasting abilities while adjusting to new trends.

Finally, adaptation through meta-learning is another promising direction. Meta-learning aims to train FMs so that they quickly adapt to new tasks with minimal additional training. Rather than pre-training only on general tasks, models learn how to efficiently adapt to new scenarios \cite{xiao2022metalearning}. In energy forecasting, an FM could be trained across many small forecasting tasks (each representing different locations or energy variables). This would enable the model to quickly adapt to new tasks—such as forecasting for a new site—with only a few gradient updates (similar to methods like MAML) \cite{jawed2023forecasting}. Although still relatively unexplored in the energy field, meta-learning is particularly relevant for applications like microgrid forecasting, where each microgrid has distinct characteristics, but the goal is a model that rapidly adapts to each new scenario \cite{shen2024microgrid}.

In summary, methods for training and adapting FMs are becoming more flexible and efficient. Pre-training methods now include multi-task learning and curriculum strategies, while fine-tuning methods increasingly rely on lightweight approaches like adapters, LoRA, and prompt-tuning \cite{german2024lliama, paischer2024eva}. These developments make FMs practical for energy forecasting tasks, allowing large pre-trained models to be customized for specific applications without excessive computational costs \cite{gupta2024lora, khanna2024explora, meo2024bayesianlora}. As a result, energy companies can effectively use FMs with modest resources, significantly lowering barriers to adoption beyond specialized research labs.

\subsection{Uncertainty Quantification and Risk-Aware Forecasting}
Accurate point forecasts are useful, but understanding forecast uncertainty is equally important in the energy sector. Grid operators and traders need information on the range of likely outcomes (for example, knowing there is a 90\% chance wind generation will be between 50 and 70 MW) to manage risk effectively. Therefore, a key methodological goal is developing FMs that produce probabilistic forecasts and quantify their uncertainty \cite{chaouch2023probabilistic, arpino2021gaussian}. Traditional methods for uncertainty quantification include prediction intervals, quantile regression, and ensemble forecasts. FMs are now using and extending these traditional methods in new ways \cite{liu2022probabilistic, li2024diffusion, lyu2024probabilistic, ye2024tadnet}.

\begin{figure*}[t]
\centering
\includegraphics[width=0.88\textwidth]{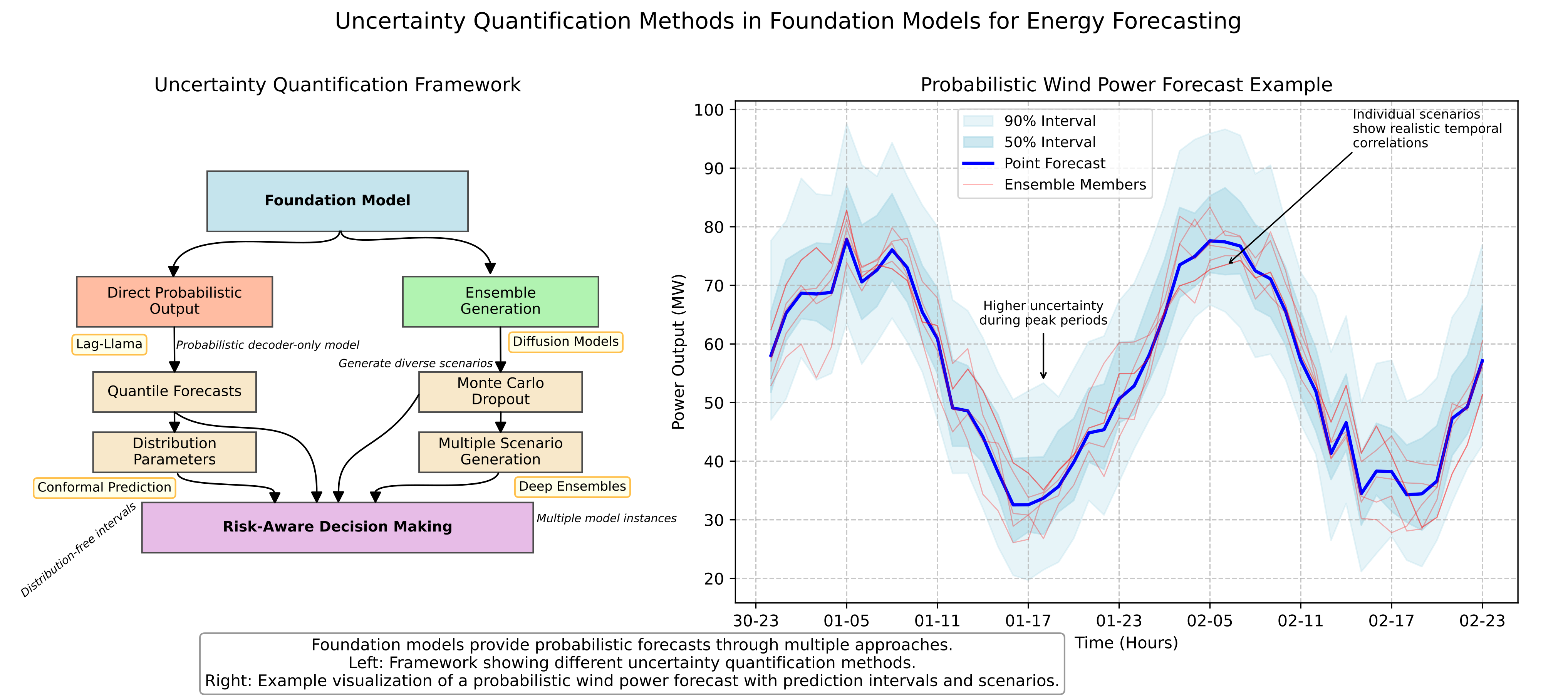}
\caption{Uncertainty Quantification Methods in FMs for Energy Forecasting. The left panel presents a framework for different approaches to uncertainty estimation, while the right panel demonstrates a probabilistic wind power forecast with prediction intervals and ensemble scenarios.}
\label{fig:uncertainty}
\end{figure*}

One approach FMs use to handle uncertainty is directly predicting the probability distribution of future values. For instance, Lag-Llama is designed as a probabilistic forecasting FM that outputs a full distribution for each forecast step rather than just one predicted value \cite{rasul2023lag}. It uses a decoder-only Transformer architecture trained either with quantile loss (to predict specific quantiles) or by predicting distribution parameters such as mean and variance (e.g., assuming a Gaussian distribution). As a result, Lag-Llama can provide median predictions, prediction intervals, or even multiple scenario samples for future points. Such probabilistic models enable users to assess forecast confidence, clearly indicating higher uncertainty (e.g., wide intervals when storms approach solar plants) or greater certainty (tight intervals during clear weather).

Another approach uses the generative capability of FMs to produce multiple plausible scenarios. Many FMs (such as Chronos or GPT-based models) can generate diverse outcomes by introducing noise or using different sampling methods during generation \cite{liu2021deep, tang2021probabilistic}. Similar to how language models produce many realistic sentences, these FMs can generate multiple realistic future scenarios. The variation among these scenarios provides a measure of uncertainty \cite{capel2022diffusion}.

While smaller, specialized models typically could not do this effectively, FMs can easily generate multiple forecasts because they learn the overall data distribution. For instance, a FM could generate 100 different wind power forecasts by slightly altering initial conditions. These forecasts can then be used to calculate risk metrics, such as the likelihood of significantly low wind power production \cite{rasul2023lagllama, wang2024generative}.

Ensemble methods remain important. An option is to combine multiple FMs or integrate a FM with traditional forecasting models \cite{sha2024ensemble, vercosa2021hybrid}. Since FMs are data-driven, they can complement physics-based ensemble forecasts, such as numerical weather prediction (NWP) ensembles. A hybrid ensemble could start from a FM prediction and then adjust it based on uncertainty estimates from a physical ensemble \cite{skoien2021postprocessing}. Several studies have examined FMs for post-processing ensemble forecasts: each ensemble member serves as an input feature, allowing the FM to produce calibrated probabilistic forecasts by correcting biases and refining uncertainty intervals \cite{jeworrek2022analog, baran2023postprocessing}.

Recent advances in uncertainty quantification also include new loss functions and evaluation metrics during training. Rather than optimizing solely for MAPE or MSE, some FMs are trained with quantile loss (Pinball loss) to directly improve quantile forecasts \cite{lopez2022quantile, li2024quantile}. Others use the Continuous Ranked Probability Score (CRPS), a proper scoring rule designed explicitly for probabilistic forecasts \cite{zheng2024mvgcrps}. These approaches encourage the model to accurately represent uncertainty, beyond merely predicting the mean correctly. Additionally, there is interest in distributional consistency, which ensures that predicted distributions remain realistic over multiple forecast horizons. Diffusion-based models naturally produce coherent time-series samples, providing a notable advantage for uncertainty quantification tasks \cite{koo2024vaeneu, wang2024diffusion}. Figure \ref{fig:uncertainty} presents both a framework for different approaches to uncertainty estimation (left panel) and a visualization of a probabilistic wind power forecast with prediction intervals and ensemble scenarios (right panel)

Real-world forecasting applications have already benefited from these methodological developments:

\begin{itemize}
\item In power grid operations, probabilistic load forecasts from FMs help operators decide how much spinning reserve to maintain. A narrow predicted distribution allows operators to reduce reserve requirements and save costs, while a wide distribution signals greater uncertainty, prompting increased reserve allocation \cite{Hamann2024}.

\item For renewable energy trading, accurately estimating the probability of over- or under-producing relative to contract commitments is essential. FMs capable of generating probabilistic scenarios help traders manage risks effectively—for instance, by informing hedging strategies such as option purchases or demand response agreements.

\item In extreme event prediction (such as infrequent but severe wind ramp-downs), reliable uncertainty quantification is critical. FMs have demonstrated capability in capturing tail risks. For example, a properly trained weather FM may identify a low but meaningful probability of an extreme event that deterministic forecasts would miss entirely.
\end{itemize}

Uncertainty quantification methods also face several challenges. One issue is calibration—ensuring predicted probabilities align with observed frequencies. Researchers often address this by applying calibration methods such as Platt scaling or isotonic regression to FM outputs \cite{borgohain2022bayesian, bicici2023calibrating}. Another challenge is ensuring uncertainty estimates appropriately increase with forecasting horizon. This can be addressed by using recursive predictions (feeding back the model's own forecasts) or injecting noise during multi-step forecasting \cite{qian2022uncertainty}. Some FMs incorporate Bayesian neural network approaches, such as Monte Carlo dropout or deep ensembles, to capture both model uncertainty and data uncertainty, providing a more complete picture of forecast risk \cite{hekler2023posthoc, lin2022bayesian}.

In conclusion, FMs are shifting from point forecasting toward full probabilistic forecasting. By training these models using distribution-based objectives and utilizing their generative capabilities, forecasts now include explicit uncertainty measures \cite{capel2022denoising, lara2022stochastic}. This shift significantly impacts decision-making, moving from deterministic scheduling toward probabilistic risk management \cite{draz2024risk}. As these techniques continue to mature, risk-aware forecasts will likely become standard practice. FMs will then provide not just single-value predictions, but detailed probability distributions to support reliable and economically sound decisions in energy systems \cite{ye2023deep, wang2024bayesian}.

\subsection{Computational Efficiency and Scalability of Models}
FMs, despite their capabilities, require substantial computational resources for training and inference. Improving their efficiency and scalability, thus enabling timely and cost-effective deployment, has become a key research area. Several strategies have emerged to address this challenge:

\textbf{Model Compression and Distillation}: Approaches such as knowledge distillation, where a smaller "student" model replicates outputs from a large FM, as well as parameter pruning and quantization, are being explored to reduce model complexity \cite{huang2023uncertaintydriven, kim2023quantization}. For example, a 1-billion-parameter weather FM could be distilled into a 100-million-parameter version, significantly reducing computational costs while retaining accuracy. Although literature specifically addressing energy-focused FMs is limited, research from the broader machine learning domain demonstrates that large Transformer models can often be compressed by 50\% or more with minimal performance degradation \cite{feng2023modelcompression}. Adapting these methods to energy models would facilitate deployment in resource-constrained environments, such as edge devices or control centers with limited GPU resources \cite{mnif2024compression, zhu2023compression}.

\textbf{Patching and Sparse Computation}: Patching techniques reduce sequence length, directly decreasing computational requirements \cite{das2023decoder}. Sparse attention mechanisms similarly reduce computations by focusing only on relevant interactions and omitting calculations for distant or less influential time steps \cite{li2023bidformer, chen2023resformer}. This approach is particularly effective for extremely long sequences, such as decade-long climate datasets, where short-term interactions typically have greater importance. Methods such as local attention or learnable sparsity patterns enable FMs to process longer sequences efficiently without proportional increases in computation \cite{feng2024halveformer}. Long-sequence Transformer variants (e.g., Longformer, Informer) introduced these concepts, and recent work is applying them to time-series FMs, enabling the use of extensive historical data (e.g., sequences of 10,000 steps) \cite{wen2024crossscale, ding2024drformer}.

\textbf{Scalability through Parallelism}: Distributed computing significantly accelerates the training of large FMs used in energy forecasting. Recent studies have successfully trained climate FMs using model and data parallelism on high-performance computing clusters with hundreds of GPUs \cite{bodnar2024aurora}. During inference, these trained models typically run much faster than traditional physics-based simulations. For example, IBM researchers demonstrated that a tuned FM provides forecasts more rapidly and cost-effectively compared to a numerical weather prediction model at similar resolution \cite{schmude2024prithvi}. This speed advantage greatly enhances scalability, allowing forecasts that previously required hours of computation to be generated in milliseconds. Such improvements support real-time decision-making, such as a solar farm controller making rapid operational adjustments based on immediate forecasts. Although the initial training cost is substantial, the operational cost per forecast can become very low when spread over many uses, driving interest in FMs to reduce overall computational demands in operational settings \cite{Hamann2024}.

\textbf{Scaling Laws and Model Size vs. Data Trade-offs}: Empirical studies on FMs for time-series forecasting reveal scaling laws similar to those observed in NLP. Generally, larger models and larger datasets improve forecasting performance, although benefits diminish at very large scales. For instance, experiments with TimesFM compared models of 17M, 70M, and 200M parameters, showing increased forecast accuracy with larger models, but with diminishing gains at higher scales \cite{das2023decoder}. This finding allows researchers to select model sizes that align well with their specific forecasting problems. A regional load forecasting model may perform adequately with 100 million parameters rather than a billion, especially if training data availability is limited. In contrast, global-scale weather models might benefit from significantly larger parameter counts. Current practice typically involves increasing model size to the threshold of diminishing returns and then applying compression or efficient fine-tuning techniques. This approach enables scalability during training by leveraging extensive datasets, while also ensuring practical scalability at deployment.

\textbf{Hardware-aware Optimizations}: Practical efforts are underway to optimize FM implementations for modern hardware, including GPUs and TPUs. Techniques such as mixed-precision training (FP16/BF16), memory-efficient GPU utilization through checkpointing, and optimized tensor computations are being explored \cite{peng2023fp8lm, liu2024neuralprocessor}. Some time-series FMs leverage highly efficient libraries originally developed for large language models, benefiting directly from established hardware optimizations \cite{santos2022gpu}. Cloud providers have introduced optimized inference platforms, such as NVIDIA's Triton Inference Server, enabling efficient deployment of large FMs with low latency to many simultaneous users \cite{xie2022bfloat16, jain2022timeseriesgpu}.

\textbf{Benchmarking and Efficiency Metrics}: Recent methodological advances include rigorous benchmarking of FMs, emphasizing efficiency in addition to accuracy. Meyer et al. (2024) compared inference time and required input context length between FMs and traditional models in short-term load forecasting \cite{meyer2024benchmarking}. They observed that while FMs typically require more computational resources initially, they can process longer historical sequences without retraining. This capability simplifies data pipelines by eliminating extensive manual feature engineering, such as creating lag features. Studies like this clarify the trade-offs involved: in scenarios requiring frequent retraining of numerous small-scale models (e.g., daily forecasting for thousands of households), deploying a single FM trained once may prove more efficient overall.

In summary, scalability challenges are being addressed at multiple levels. Researchers are developing methods to efficiently train large FMs, as well as techniques to reduce or optimize these models for faster inference. Consequently, FMs are becoming increasingly accessible. In the near future, even mid-sized grid operators may be able to fine-tune pre-trained FMs on their own datasets without extensive computational resources, enabled by parameter-efficient fine-tuning (PEFT) techniques and cloud-based services offering FMs as a service. In operational settings, a well-designed FM could significantly reduce computational requirements by replacing many specialized models and computationally intensive simulations \cite{schmude2024prithvi}. Continued research into scaling and optimization is essential for realizing these benefits in practical energy systems.

\subsection{Interpretability and Transparency of Models}
As FMs become increasingly common in clean energy forecasting, their interpretability and transparency are receiving greater attention. Energy stakeholders—such as grid operators, engineers, and regulators—traditionally prefer models with understandable decision processes, including physics-based models or simpler regression methods. Trusting opaque AI models with critical decisions related to grid stability or market participation is difficult without insight into their reasoning. Consequently, a key methodological effort involves developing techniques to interpret FM behaviors, thereby ensuring their reliability and acceptance in operational contexts \cite{Hamann2024}.

\begin{figure*}[t]
\centering
\includegraphics[width=0.88\textwidth]{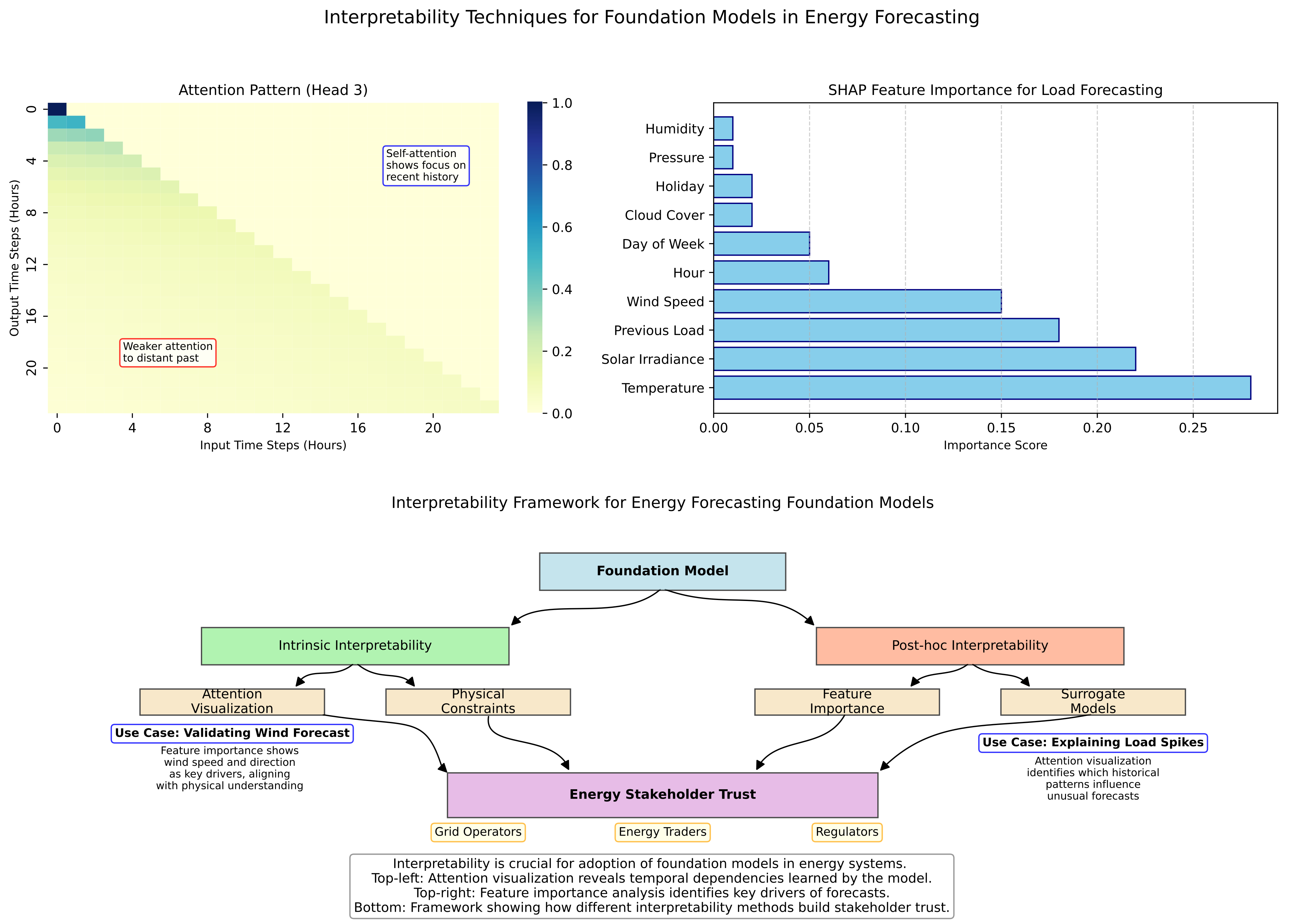}
\caption{Interpretability Techniques for FMs in Energy Forecasting. The figure illustrates attention visualization (top-left), feature importance analysis (top-right), and a comprehensive framework showing how different interpretability methods build stakeholder trust in FMs for energy applications.}
\label{fig:interpretability}
\end{figure*}

One interpretability strategy is to integrate physical knowledge or known constraints directly into FMs, as done in hybrid modeling approaches. A FM designed to comply with established physical laws inherently becomes easier to trust—for instance, unrealistic wind speed predictions can be constrained by a physics-based module \cite{lermer2021hybrid, nahid2021hybrid}. Although this approach does not fully explain the internal workings of the model, it narrows the model's uncertainty by limiting outputs to physically plausible scenarios. A hybrid FM might explicitly enforce energy conservation or operational limits as constraints, thereby improving reliability and contextual interpretability ("the model respects constraints X, Y, and Z by design") \cite{nadeem2024interpretable, zhou2024physicsai, watermeyer2023hybrid}.

For purely data-driven FMs, post-hoc interpretability methods are actively explored:

\textbf{Attention Weight Analysis}: Transformer models provide attention scores that can indicate which input time steps or features contributed most significantly to a prediction. By visualizing attention patterns, analysts can determine, for example, whether a load forecasting model primarily considered temperatures from two days prior or the load pattern from the previous week when predicting tomorrow's demand \cite{niu2024attention, wu2024revisiting}. Such insights can align with human intuition—recognizing familiar patterns—and thus increase confidence in the model. However, interpreting attention scores requires caution, since not all attention heads correspond directly to meaningful features, and low attention scores do not necessarily imply irrelevance \cite{aguilera2024local, li2024linwa, baric2021benchmarking}.

\textbf{Feature Importance and Sensitivity}: Techniques such as SHAP or Integrated Gradients can be applied to FMs to quantify how each input feature influences predictions \cite{cremades2024shap, ioannou2023baselines}. In multi-modal FMs, these methods can clarify, for example, the relative importance of satellite imagery versus historical power data in forecasting solar output. Perturbing inputs allows evaluation of model sensitivity \cite{raykar2023tsshap}. If the model behaves consistently, such analyses may yield intuitive insights—for instance, revealing that today's solar forecast depends primarily on current cloud cover (80\%) rather than previous day's output (20\%) \cite{huang2024shap, zhuo2024ig2}.

\textbf{Surrogate Modeling}: Another interpretability approach involves approximating a FM locally using simpler models. Around a specific operating condition, for example, one could fit a local linear model or decision tree to replicate the FM's predictions \cite{heidari2023interpretable, kleinlein2022local}. Such surrogate models offer direct interpretability, allowing straightforward analysis of relationships between inputs and outputs within that local region. This method parallels surrogate approaches used for neural network interpretability in other fields \cite{sigut2023surrogate, ranjbar2022lime}.

\textbf{Concept Analysis}: Interpretability can also be assessed through the model's response to known patterns or "concepts," such as seasonality or ramp events. Researchers may create synthetic test scenarios—such as a week of steadily rising temperatures—to observe how the FM responds \cite{ozyegen2021evaluation, kuvshinova2024foundation}. If the model's predictions align with expected causal relationships (e.g., increased air conditioning load as temperatures rise), this strengthens interpretability and trust in the model \cite{potosnak2024reasoning}. Conversely, unexpected responses could highlight areas requiring model adjustments or additional caution in application \cite{sprang2024interpretability, kjaernli2024robustness}.

The trustworthiness of FMs is increasingly established through validation against known benchmarks and physical principles. For instance, AI-based weather emulators such as GraphCast have shown the ability to capture atmospheric physics accurately, even surpassing traditional physics-based models in weather forecasting tasks \cite{Hamann2024}. This indicates that FMs may indeed encode underlying system dynamics, rather than simply identifying correlations, although the exact mechanisms remain opaque. Such evidence helps build confidence that critical factors are not overlooked by these models. In power systems, initial comparisons between FM predictions and traditional methods like load-flow or stability analyses are underway to verify consistency. Early findings suggest that, for specific applications, well-validated AI models can achieve trust levels comparable to traditional first-principles models \cite{Hamann2024}. Figure \ref{fig:interpretability} illustrates the various interpretability techniques that support this trustworthiness, showing how attention visualization can reveal temporal dependencies learned by the model (top-left) and how feature importance analysis helps identify key drivers of predictions (top-right), all within a framework for building stakeholder trust in energy forecasting applications.

Regulatory bodies and standards organizations are increasingly addressing interpretability requirements. For example, the European Union's draft AI Act is expected to mandate explainability for high-stakes AI systems, such as those controlling critical infrastructure \cite{Hamann2024}. In response, researchers are developing standardized testing protocols and "experimentation sandboxes" to systematically evaluate FMs across diverse scenarios and document their behaviors. Additionally, formal verification methods—traditionally used in software—are being explored to confirm key properties of FMs, such as stability, ensuring minor input variations do not lead to significant or unpredictable changes in model outputs \cite{Hamann2024}.

\begin{table}[htbp]
  \caption{Future Research Directions and Challenges for Foundation Models in Clean Energy Forecasting}
  \label{tab:future_directions}
  \centering
  \scriptsize
  \setlength{\tabcolsep}{4pt}
  \renewcommand{\arraystretch}{1.2}
  \begin{tabularx}{\columnwidth}{@{}
      >{\raggedright\arraybackslash}p{0.20\columnwidth}
      >{\raggedright\arraybackslash}p{0.25\columnwidth}
      >{\raggedright\arraybackslash}p{0.25\columnwidth}
      >{\raggedright\arraybackslash}X@{}}
    \toprule
    \textbf{Research Area} & \textbf{Current Limitations} & \textbf{Promising Approaches} & \textbf{Expected Impact} \\
    \midrule
    \multirow{2}{*}{Interpretability} 
    & Black-box nature of FMs & Attention visualization, feature attribution & Increased trust, regulatory compliance \\
    & Lack of physical consistency & Physics-informed constraints, hybrid modeling & Better alignment with domain knowledge \\
    \addlinespace
    \multirow{2}{0.2\columnwidth}{Computational Efficiency} 
    & High training costs & Knowledge distillation, parameter-efficient tuning & Wider adoption, edge deployment \\
    & Inference latency & Model compression, sparse computation & Real-time applications, resource optimization \\
    \addlinespace
    \multirow{2}{*}{Data Challenges} 
    & Limited historical data & Synthetic data generation, transfer learning & Improved performance for new assets \\
    & Data quality issues & Robust pre-training, anomaly detection & Resilience to noisy measurements \\
    \addlinespace
    \multirow{2}{0.2\columnwidth}{Integration with Energy Systems} 
    & Operational integration gaps & Decision support frameworks, API standardization & Seamless deployment in control systems \\
    & Multi-scale forecasting needs & Hierarchical models, reconciliation techniques & Consistent forecasts across time horizons \\
    \bottomrule
  \end{tabularx}
\end{table}

In practice, interpretability typically involves combining multiple methods. For example, an ISO deploying a FM for demand forecasting might initially run it alongside the traditional model for a full year, carefully examining cases of disagreement and using interpretability techniques to diagnose differences \cite{airlangga2024xai, gou2023interpretability}. If the FM occasionally emphasizes a previously unused feature (such as wind speed influencing temperature-driven load), operators can either gain new insights or reconsider feature selection strategies \cite{harel2022featureimportance}. Thus, FMs can reveal subtle correlations or unexpected relationships, but extracting these insights requires clear and systematic interpretation approaches \cite{samimi2022boosting, kaneko2023feature}.

To summarize, interpretability remains a critical and active research area for FMs in clean energy. Methods being explored include internal structural analyses (attention mechanisms), external interpretability techniques (SHAP, surrogate models), and systematic validation against physical principles and historical data \cite{zheng2023tcn, airlangga2024xai}. The ultimate goal is for power system operators and decision-makers to trust these models not as opaque "black boxes," but as reliable tools whose predictions align clearly with domain knowledge \cite{zeng2024shap}. Achieving this will likely require iterative feedback: real-world deployments will expose limitations and errors, driving further interpretability research. Enhanced interpretability will then improve safety, transparency, and acceptance, promoting further use in operations \cite{raykar2023tsshap}. Ultimately, combining high predictive performance with strong explainability will firmly establish FMs as standard tools in clean energy forecasting \cite{toubeau2022interpretable}. Table \ref{tab:future_directions} summarizes these interpretability challenges alongside other critical areas for future research in foundation models for clean energy forecasting, including computational efficiency, data quality issues, and integration with existing energy systems. These challenges represent key focus areas for advancing the practical implementation of foundation models in renewable energy applications.

\section{Conclusion and Future Directions}
\label{sec:conclusion}

This comprehensive review has examined the rapid emergence and evolution of FMs for clean energy forecasting. These large-scale pre-trained models, typically built on transformer architectures, are revolutionizing how we predict renewable energy generation and load demand by leveraging transfer learning and domain adaptation techniques. Beginning with statistical and physics-based methods, the field has progressed through machine learning approaches to today's FMs that can generalize across different forecasting tasks with minimal fine-tuning.

The key advantages of FMs in clean energy forecasting include their ability to: (1) capture complex temporal dependencies through self-attention mechanisms, (2) transfer knowledge across different energy domains and geographical regions, (3) integrate diverse data sources including meteorological, satellite, and IoT sensor data, and (4) provide probabilistic forecasts with uncertainty quantification. These capabilities address critical challenges in renewable energy integration, particularly the intermittent nature of wind and solar resources.

Our analysis of the current state of research reveals several important developments:

\begin{itemize}
\item Specialized transformer architectures for time series, such as patching techniques and frequency-domain integration, have significantly improved computational efficiency and forecasting accuracy.
\item Multi-modal data fusion approaches now effectively combine diverse data sources, enhancing the models' ability to capture complex dependencies in energy systems.
\item Parameter-efficient fine-tuning methods like LoRA and adapters enable adaptation to specific energy applications without extensive retraining.
\item Uncertainty quantification techniques, including probabilistic outputs and scenario generation, provide essential risk assessments for grid operations.
\item Interpretability tools, though still developing, increasingly bridge the gap between black-box models and operational requirements in critical energy infrastructure.
\end{itemize}

Despite these advances, several challenges remain. Computational costs continue to be prohibitive for some organizations, data availability (especially standardized, high-quality energy data) remains limited, and domain-specific biases must be carefully addressed. Regulatory and interpretability constraints further complicate widespread deployment in mission-critical energy applications.

Looking ahead, we identify promising research directions for the field:

\begin{itemize}
\item Development of energy-specific pre-training objectives that incorporate physical laws and operational constraints
\item Expansion of hybrid approaches combining data-driven FMs with physics-based simulations
\item Creation of standardized benchmarks for evaluating FMs across diverse energy forecasting tasks
\item Advancement of continual learning methods to keep models current as energy systems evolve
\item Further research on model compression and distillation to enable deployment on resource-constrained devices
\item Enhanced interpretability techniques specifically designed for time series FMs in energy applications
\end{itemize}

The transition to clean energy requires increasingly sophisticated forecasting tools to manage the variability and uncertainty inherent in renewable resources. FMs represent a significant step forward in this domain, offering a unified approach that can adapt to diverse energy forecasting needs while providing the accuracy and reliability required for operational decision-making. As research continues to address current limitations, these models will likely play an increasingly central role in enabling the reliable integration of higher proportions of renewable energy into our power systems.

\section*{CRediT authorship contribution statement}
\textbf{Author One:} Conceptualization, Methodology, Writing - original draft, Writing - review \& editing, Supervision. \textbf{Author Two:} Investigation, Data curation, Formal analysis, Visualization, Writing - original draft. \textbf{Author Three:} Investigation, Validation, Writing - original draft, Writing - review \& editing.

\section*{Declaration of competing interest}
The authors declare that they have no known competing financial interests or personal relationships that could have appeared to influence the work reported in this paper.

\section*{Data availability}
No data was used for the research described in the article.

\section*{Acknowledgments}
This research did not receive any specific grant from funding agencies in the public, commercial, or not-for-profit sectors.

\bibliographystyle{elsarticle-num}
\bibliography{cas-refs}

\begin{thebibliography}{100}
\expandafter\ifx\csname url\endcsname\relax
  \def\url#1{\texttt{#1}}\fi
\expandafter\ifx\csname urlprefix\endcsname\relax\def\urlprefix{URL }\fi
\expandafter\ifx\csname href\endcsname\relax
  \def\href#1#2{#2} \def\path#1{#1}\fi

\bibitem{mane2024forecasting}
O.~Mane, A.~Zagade, S.~Sonpatki, S.~Chavan, K.~Nimbalkar, Forecasting renewable energy production using ai-based weather prediction models, International Journal For Multidisciplinary Research (2024).
\newblock \href {https://doi.org/10.36948/ijfmr.2024.v06i03.21917} {\path{doi:10.36948/ijfmr.2024.v06i03.21917}}.

\bibitem{zhou2024universal}
X.~Zhou, J.~Ye, S.~Zhao, M.~Jin, Z.~Hou, C.-S. Yang, Z.~Li, Y.~Wen, X.~Yuan, Towards universal large-scale foundational model for natural gas demand forecasting, ArXiv abs/2409.15794 (2024).
\newblock \href {https://doi.org/10.48550/arXiv.2409.15794} {\path{doi:10.48550/arXiv.2409.15794}}.

\bibitem{li2024foundts}
Z.~Li, X.~Qiu, P.~Chen, Y.~Wang, H.~Cheng, Y.~Shu, J.~Hu, C.~Guo, A.~Zhou, Q.~Wen, C.~S. Jensen, B.~Yang, Foundts: Comprehensive and unified benchmarking of foundation models for time series forecasting (2024).

\bibitem{sergeev2023review}
N.~Sergeev, P.~Matrenin, A review of international experience in forecasting renewable energy generation using machine learning methods, iPolytech Journal (2023).
\newblock \href {https://doi.org/10.21285/1814-3520-2023-2-354-369} {\path{doi:10.21285/1814-3520-2023-2-354-369}}.

\bibitem{jung2022photovoltaic}
A.-H. Jung, D.-H. Lee, J.-Y. Kim, C.~K. Kim, H.-G. Kim, Y.-S. Lee, Regional photovoltaic power forecasting using vector autoregression model in south korea, Energies (2022).
\newblock \href {https://doi.org/10.3390/en15217853} {\path{doi:10.3390/en15217853}}.

\bibitem{zou2023survey}
Y.~Zou, Survey of wind power output power forecasting technology, Highlights in Science, Engineering and Technology (2023).
\newblock \href {https://doi.org/10.54097/hset.v50i.8490} {\path{doi:10.54097/hset.v50i.8490}}.

\bibitem{alpackaya2024renewable}
I.~Alpackaya, M.~H. Fallah, N.~Mounika, S.~Sood, S.~Rajvanshi, S.~Lakhanpal, P.~Cajla, A.~Sharma, Y.~S. Lalitha, Renewable energy forecasting using deep learning techniques, E3S Web of Conferences (2024).
\newblock \href {https://doi.org/10.1051/e3sconf/202458101011} {\path{doi:10.1051/e3sconf/202458101011}}.

\bibitem{benti2023forecasting}
N.~E. Benti, M.~D. Chaka, A.~Semie, Forecasting renewable energy generation with machine learning and deep learning: Current advances and future prospects, Sustainability (2023).
\newblock \href {https://doi.org/10.3390/su15097087} {\path{doi:10.3390/su15097087}}.

\bibitem{pham2023wind}
T.~S. Pham, D.~Nguyen, N.~D.~M. Phan, Wind energy forecasting: A comparative analysis of machine learning approaches, in: 2023 8th International Scientific Conference on Applying New Technology in Green Buildings (ATiGB), 2023, pp. 195--200.
\newblock \href {https://doi.org/10.1109/ATiGB59969.2023.10364510} {\path{doi:10.1109/ATiGB59969.2023.10364510}}.

\bibitem{bassey2023solar}
K.~E. Bassey, Solar energy forecasting with deep learning technique, Engineering Science \& Technology Journal (2023).
\newblock \href {https://doi.org/10.51594/estj.v4i2.1286} {\path{doi:10.51594/estj.v4i2.1286}}.

\bibitem{mohan2023evaluation}
L.~Mohan, R.~Durga, Evaluation of renewable energy forecasting (ref) technique using by machine learning, in: 2023 International Conference on New Frontiers in Communication, Automation, Management and Security (ICCAMS), Vol.~1, 2023, pp. 1--8.
\newblock \href {https://doi.org/10.1109/ICCAMS60113.2023.10526008} {\path{doi:10.1109/ICCAMS60113.2023.10526008}}.

\bibitem{saeed2024adaptive}
F.~Saeed, S.~Aldera, Adaptive renewable energy forecasting utilizing a data-driven pca-transformer architecture, IEEE Access 12 (2024) 109269--109280.
\newblock \href {https://doi.org/10.1109/ACCESS.2024.3440226} {\path{doi:10.1109/ACCESS.2024.3440226}}.

\bibitem{walczewski2024prediction}
M.~J. Walczewski, H.~Wöhrle, Prediction of electricity generation using onshore wind and solar energy in germany, Energies (2024).
\newblock \href {https://doi.org/10.3390/en17040844} {\path{doi:10.3390/en17040844}}.

\bibitem{ghazaei2024wind}
E.~Ghazaei, O.~Feizi, A.~Ghavifekr, M.~Salim, A.~Hassanzadeh, Wind farm power prediction with transformer encoder, in: 2024 9th International Conference on Technology and Energy Management (ICTEM), 2024, pp. 1--5.
\newblock \href {https://doi.org/10.1109/ICTEM60690.2024.10631900} {\path{doi:10.1109/ICTEM60690.2024.10631900}}.

\bibitem{dai2023learning}
X.~Dai, G.~Liu, W.~Hu, Z.~Lei, H.~Zhou, Learning from chatgpt: A transformer-based model for wind power forecasting, in: 2023 IEEE International Conference on Environment and Electrical Engineering and 2023 IEEE Industrial and Commercial Power Systems Europe (EEEIC / I\&CPS Europe), 2023, pp. 1--6.
\newblock \href {https://doi.org/10.1109/EEEIC/ICPSEurope57605.2023.10194807} {\path{doi:10.1109/EEEIC/ICPSEurope57605.2023.10194807}}.

\bibitem{han2024multi}
H.~Han, X.~Liang, K.~Zhu, Multi-modal transformer-based wind power prediction by utilizing heterogeneous sources 13159 (2024) 131591H -- 131591H--6.
\newblock \href {https://doi.org/10.1117/12.3024579} {\path{doi:10.1117/12.3024579}}.

\bibitem{ghasemi2023harnessing}
A.~Ghasemi, M.~Hashemi, Harnessing the power of transformer learning with long-term memory for renewable energy forecasting, 2023 IEEE Virtual Conference on Communications (VCC) (2023) 171--176\href {https://doi.org/10.1109/VCC60689.2023.10475044} {\path{doi:10.1109/VCC60689.2023.10475044}}.

\bibitem{zeng2022transformers}
A.~Zeng, M.-H. Chen, L.~Zhang, Q.~Xu, Are transformers effective for time series forecasting?, ArXiv abs/2205.13504 (2022).
\newblock \href {https://doi.org/10.48550/arXiv.2205.13504} {\path{doi:10.48550/arXiv.2205.13504}}.

\bibitem{joseph2024transformers}
S.~Joseph, A.~A. Jo, E.~D. Raj, Improving time series forecasting accuracy with transformers: A comprehensive analysis with explainability, in: 2024 Third International Conference on Electrical, Electronics, Information and Communication Technologies (ICEEICT), 2024, pp. 1--7.
\newblock \href {https://doi.org/10.1109/ICEEICT61591.2024.10718609} {\path{doi:10.1109/ICEEICT61591.2024.10718609}}.

\bibitem{rasul2023lagllama}
K.~Rasul, A.~Ashok, A.~R. Williams, A.~Khorasani, G.~Adamopoulos, R.~Bhagwatkar, M.~Bilovs, H.~Ghonia, N.~Hassen, A.~Schneider, S.~Garg, A.~Drouin, N.~Chapados, Y.~Nevmyvaka, I.~Rish, Lag-llama: Towards foundation models for probabilistic time series forecasting, ArXiv abs/2310.08278 (2023).
\newblock \href {https://doi.org/10.48550/arXiv.2310.08278} {\path{doi:10.48550/arXiv.2310.08278}}.

\bibitem{das2023decoder}
A.~Das, W.~Kong, R.~Sen, Y.~Zhou, A decoder-only foundation model for time-series forecasting, arXiv preprint arXiv:2310.10688 (2023).

\bibitem{potosnak2024implicit}
W.~Potosnak, C.~Challu, M.~Goswami, M.~Wiliński, N.~Zukowska, Implicit reasoning in deep time series forecasting, ArXiv abs/2409.10840 (2024).
\newblock \href {https://doi.org/10.48550/arXiv.2409.10840} {\path{doi:10.48550/arXiv.2409.10840}}.

\bibitem{bharti2024transformer}
M.~K. Bharti, R.~Wadhvani, M.~Gyanchandani, M.~Gupta, Transformer-based multivariate time series forecasting, in: 2024 IEEE International Students' Conference on Electrical, Electronics and Computer Science (SCEECS), 2024, pp. 1--6.
\newblock \href {https://doi.org/10.1109/SCEECS61402.2024.10482217} {\path{doi:10.1109/SCEECS61402.2024.10482217}}.

\bibitem{uremović2023framework}
N.~Uremović, M.~Bizjak, P.~Sukič, G.~Štumberger, B.~Žalik, N.~Lukač, A new framework for multivariate time series forecasting in energy management system, IEEE Transactions on Smart Grid 14 (2023) 2934--2947.
\newblock \href {https://doi.org/10.1109/TSG.2022.3224559} {\path{doi:10.1109/TSG.2022.3224559}}.

\bibitem{conti2022physics_adaptation}
Z.~X. Conti, R.~Choudhary, L.~Magri, A physics-based domain adaptation framework for modeling and forecasting building energy systems, ArXiv abs/2208.09456 (2022).
\newblock \href {https://doi.org/10.48550/arXiv.2208.09456} {\path{doi:10.48550/arXiv.2208.09456}}.

\bibitem{gupta2024lora_timeseries}
D.~Gupta, A.~Bhatti, S.~Parmar, C.~Dan, Y.~Liu, B.~Shen, S.~Lee, Low-rank adaptation of time series foundational models for out-of-domain modality forecasting, ArXiv abs/2405.10216 (2024).
\newblock \href {https://doi.org/10.48550/arXiv.2405.10216} {\path{doi:10.48550/arXiv.2405.10216}}.

\bibitem{mundotiya2022wind_uncertainty}
P.~Mundotiya, P.~Mathuria, H.~Tiwari, Comprehensive review on uncertainty in wind power forecasting-models and challenges, in: 2022 2nd International Conference on Innovative Sustainable Computational Technologies (CISCT), 2022, pp. 1--6.
\newblock \href {https://doi.org/10.1109/CISCT55310.2022.10046605} {\path{doi:10.1109/CISCT55310.2022.10046605}}.

\bibitem{borrotti2024uncertainty}
M.~Borrotti, Quantifying uncertainty with conformal prediction for heating and cooling load forecasting in building performance simulation, Energies (2024).
\newblock \href {https://doi.org/10.3390/en17174348} {\path{doi:10.3390/en17174348}}.

\bibitem{bommasani2021opportunities}
R.~Bommasani, D.~A. Hudson, E.~Adeli, R.~Altman, S.~Arora, S.~von Arx, M.~S. Bernstein, J.~Bohg, A.~Bosselut, E.~Brunskill, et~al., On the opportunities and risks of foundation models, arXiv preprint arXiv:2108.07258 (2021).

\bibitem{german2024transfer}
M.~Germ{\'a}n-Morales, A.~Rivera-Rivas, M.~D{\'\i}az, C.~Carmona, Transfer learning with foundational models for time series forecasting using low-rank adaptations, arXiv preprint arXiv:2410.11539 (2024).

\bibitem{liang2024foundation}
Y.~Liang, H.~Wen, Y.~Nie, Y.~Jiang, M.~Jin, D.~Song, S.~Pan, Q.~Wen, Foundation models for time series analysis: A tutorial and survey, in: Proceedings of the 30th ACM SIGKDD conference on knowledge discovery and data mining, 2024, pp. 6555--6565.

\bibitem{meyer2024benchmarking}
M.~Meyer, D.~Zapata, S.~Kaltenpoth, O.~Mueller, Benchmarking time series foundation models for short-term household electricity load forecasting, arXiv preprint arXiv:2410.09487 (2024).

\bibitem{ansari2024chronos}
A.~F. Ansari, L.~Stella, C.~Turkmen, X.~Zhang, P.~Mercado, H.~Shen, O.~Shchur, S.~S. Rangapuram, S.~P. Arango, S.~Kapoor, et~al., Chronos: Learning the language of time series, arXiv preprint arXiv:2403.07815 (2024).

\bibitem{garza2023timegpt}
A.~Garza, C.~Challu, M.~Mergenthaler-Canseco, Timegpt-1, ArXiv (2023).

\bibitem{dalal2024beyond}
L.~Dalal, R.~Verma, R.~Chahar, Beyond words: Adapting nlp methodologies for time series forecasting challenges, in: 2023 4th International Conference on Intelligent Technologies (CONIT), 2024, pp. 1--8.
\newblock \href {https://doi.org/10.1109/CONIT61985.2024.10627058} {\path{doi:10.1109/CONIT61985.2024.10627058}}.

\bibitem{yang2024vitime}
L.~Yang, Y.~Wang, X.~Fan, I.~Cohen, Y.~Zhao, Z.~Zhang, Vitime: A visual intelligence-based foundation model for time series forecasting, ArXiv abs/2407.07311 (2024).
\newblock \href {https://doi.org/10.48550/arXiv.2407.07311} {\path{doi:10.48550/arXiv.2407.07311}}.

\bibitem{chai2024fomo}
H.~Chai, S.~Zhang, X.~Qi, Y.~Li, Fomo: A foundation model for mobile traffic forecasting with diffusion model, ArXiv (2024).

\bibitem{makridakis1997arma}
S.~Makridakis, M.~Hibon, Arma models and the box--jenkins methodology, Journal of forecasting 16~(3) (1997) 147--163.

\bibitem{stock2001vector}
J.~H. Stock, M.~W. Watson, Vector autoregressions, Journal of Economic perspectives 15~(4) (2001) 101--115.

\bibitem{lerch2021postprocessing}
S.~Lerch, B.~Schulz, M.~E. Ayari, S.~Baran, Post-processing numerical weather prediction ensembles for probabilistic solar irradiance forecasting, Solar Energy (2021).
\newblock \href {https://doi.org/10.1016/J.SOLENER.2021.03.023} {\path{doi:10.1016/J.SOLENER.2021.03.023}}.

\bibitem{wu2021probabilistic}
Y.-K. Wu, Y.-C. Wu, J.-S. Hong, L.~T. Phan, Q.~Phan, Probabilistic forecast of wind power generation with data processing and numerical weather predictions, IEEE Transactions on Industry Applications 57 (2021) 36--45.
\newblock \href {https://doi.org/10.1109/TIA.2020.3037264} {\path{doi:10.1109/TIA.2020.3037264}}.

\bibitem{vennila2024predicting}
A.~Vennila, S.~Balambigai, P.~V, S.~U. P, V.~Kavyashree, M.~Jinisha, Predicting solar and wind power production with the weather map data approach, in: 2024 International Conference on Signal Processing, Computation, Electronics, Power and Telecommunication (IConSCEPT), 2024, pp. 1--6.
\newblock \href {https://doi.org/10.1109/IConSCEPT61884.2024.10627791} {\path{doi:10.1109/IConSCEPT61884.2024.10627791}}.

\bibitem{kumhar2023threat}
B.~Kumhar, A.~Saxena, M.~Nagar, Advanced threat intelligence forecasting using machine learning algorithms, in: 2023 IEEE International Conference on ICT in Business Industry \& Government (ICTBIG), 2023, pp. 1--7.
\newblock \href {https://doi.org/10.1109/ICTBIG59752.2023.10456004} {\path{doi:10.1109/ICTBIG59752.2023.10456004}}.

\bibitem{andrews2022forecasting}
J.~Andrews, O.~Gkountouna, E.~Blaisten-Barojas, Forecasting molecular dynamics energetics of polymers in solution from supervised machine learning, Chemical Science 13 (2022) 7021--7033.
\newblock \href {https://doi.org/10.1039/d2sc01216b} {\path{doi:10.1039/d2sc01216b}}.

\bibitem{ayyildiz2023stock}
N.~Ayyıldız, Prediction of stock market index movements with machine learning (2023).
\newblock \href {https://doi.org/10.58830/ozgur.pub354} {\path{doi:10.58830/ozgur.pub354}}.

\bibitem{moraitis2024path}
N.~Moraitis, L.~Tsipi, D.~Vouyioukas, Machine-learning-based path loss prediction for in-cabin wireless networks, in: 2024 IEEE International Conference on Machine Learning for Communication and Networking (ICMLCN), 2024, pp. 393--398.
\newblock \href {https://doi.org/10.1109/ICMLCN59089.2024.10624765} {\path{doi:10.1109/ICMLCN59089.2024.10624765}}.

\bibitem{deep2024financial}
A.~Deep, Advanced financial market forecasting: integrating monte carlo simulations with ensemble machine learning models, Quantitative Finance and Economics (2024).
\newblock \href {https://doi.org/10.3934/qfe.2024011} {\path{doi:10.3934/qfe.2024011}}.

\bibitem{hochreiter1997long}
S.~Hochreiter, Long short-term memory, Neural Computation MIT-Press (1997).

\bibitem{muzaffar2019short}
S.~Muzaffar, A.~Afshari, Short-term load forecasts using lstm networks, Energy Procedia 158 (2019) 2922--2927.

\bibitem{eua2021enhancing}
B.~Eua-Arporn, S.-L. Huang, E.~Kuruoglu, Enhancing neural network based hybrid learning with empirical wavelet transform for time series forecasting, in: 2021 IEEE 33rd International Conference on Tools with Artificial Intelligence (ICTAI), 2021, pp. 386--390.
\newblock \href {https://doi.org/10.1109/ICTAI52525.2021.00063} {\path{doi:10.1109/ICTAI52525.2021.00063}}.

\bibitem{ghide2022wavelet}
L.~Ghide, S.~Wei, Y.~Ding, Comparative study of wavelet-sarima and emd-sarima for forecasting daily temperature series, International Journal of Analysis and Applications (2022).
\newblock \href {https://doi.org/10.28924/2291-8639-20-2022-17} {\path{doi:10.28924/2291-8639-20-2022-17}}.

\bibitem{wang2023ssa}
J.~Wang, X.~Peng, J.~Wu, Y.~Ding, B.~Ali, Y.~Luo, Y.~Hu, K.~Zhang, Singular spectrum analysis-based hybrid models for emergency ambulance demand time series forecasting, IMA Journal of Management Mathematics (2023).
\newblock \href {https://doi.org/10.1093/imaman/dpad019} {\path{doi:10.1093/imaman/dpad019}}.

\bibitem{li2021arima}
T.-L. Li, C.-A. Tsai, A new strategy of hybrid models using arima, ann, and dwt in time series modelling, in: Journal of Statistics: Advances in Theory and Applications, 2021.
\newblock \href {https://doi.org/doi.org/10.18642/jsata_7100122182} {\path{doi:doi.org/10.18642/jsata_7100122182}}.

\bibitem{darmawan2022ssa}
G.~Darmawan, D.~Rosadi, B.~N. Ruchjana, Hybrid model of singular spectrum analysis and arima for seasonal time series data, CAUCHY (2022).
\newblock \href {https://doi.org/10.18860/ca.v7i2.14136} {\path{doi:10.18860/ca.v7i2.14136}}.

\bibitem{weerasinghe2023abc}
C.~Weerasinghe, R.~Loaiza-Maya, G.~Martin, D.~T. Frazier, Abc-based forecasting in state space models (2023).

\bibitem{ahmad2023ensemble}
T.~Ahmad, N.~Zhou, Ensemble methods for probabilistic solar power forecasting: A comparative study, in: 2023 IEEE Power \& Energy Society General Meeting (PESGM), 2023, pp. 1--5.
\newblock \href {https://doi.org/10.1109/PESGM52003.2023.10253133} {\path{doi:10.1109/PESGM52003.2023.10253133}}.

\bibitem{jensen2022encqr}
V.~Jensen, F.~Bianchi, S.~Anfinsen, Ensemble conformalized quantile regression for probabilistic time series forecasting, IEEE Transactions on Neural Networks and Learning Systems PP (2022).
\newblock \href {https://doi.org/10.1109/TNNLS.2022.3217694} {\path{doi:10.1109/TNNLS.2022.3217694}}.

\bibitem{doubleday2021bma}
K.~Doubleday, S.~Jascourt, W.~Kleiber, B.~Hodge, Probabilistic solar power forecasting using bayesian model averaging, IEEE Transactions on Sustainable Energy 12 (2021) 325--337.
\newblock \href {https://doi.org/10.1109/TSTE.2020.2993524} {\path{doi:10.1109/TSTE.2020.2993524}}.

\bibitem{althoff2023conformal}
S.~Althoff, J.~H. Szabadv'ary, J.~Anderson, L.~Carlsson, Evaluation of conformal-based probabilistic forecasting methods for short-term wind speed forecasting, 2023, pp. 100--115.

\bibitem{wu2021autoformer}
H.~Wu, J.~Xu, J.~Wang, M.~Long, Autoformer: Decomposition transformers with auto-correlation for long-term series forecasting, ArXiv abs/2106.13008 (2021).

\bibitem{laborda2023tft}
J.~Laborda, S.~Ruano, I.~Zamanillo, Multi-country and multi-horizon gdp forecasting using temporal fusion transformers, SSRN Electronic Journal (2023).
\newblock \href {https://doi.org/10.2139/ssrn.4329721} {\path{doi:10.2139/ssrn.4329721}}.

\bibitem{ying2024tfeformer}
C.~Ying, J.~Lu, Tfeformer: Temporal feature enhanced transformer for multivariate time series forecasting, IEEE Access 12 (2024) 153694--153708.
\newblock \href {https://doi.org/10.1109/ACCESS.2024.3480953} {\path{doi:10.1109/ACCESS.2024.3480953}}.

\bibitem{oliveira2024evaluating}
J.~Oliveira, P.~Ramos, Evaluating the effectiveness of time series transformers for demand forecasting in retail, Mathematics (2024).
\newblock \href {https://doi.org/10.3390/math12172728} {\path{doi:10.3390/math12172728}}.

\bibitem{shabani2022scaleformer}
A.~Shabani, A.~Abdi, L.~Meng, T.~Sylvain, Scaleformer: Iterative multi-scale refining transformers for time series forecasting, ArXiv abs/2206.04038 (2022).
\newblock \href {https://doi.org/10.48550/arXiv.2206.04038} {\path{doi:10.48550/arXiv.2206.04038}}.

\bibitem{hou2025time}
Y.~Hou, C.~Ma, X.~Li, Y.~Sun, H.~Yu, Z.~Fang, Time series foundation model for improved transformer load forecasting and overload detection, Energies 18~(3) (2025) 660.

\bibitem{shi2024timemoe}
X.~Shi, S.~Wang, Y.~Nie, D.~Li, Z.~Ye, Q.~Wen, M.~Jin, Time-moe: Billion-scale time series foundation models with mixture of experts, ArXiv abs/2409.16040 (2024).
\newblock \href {https://doi.org/10.48550/arXiv.2409.16040} {\path{doi:10.48550/arXiv.2409.16040}}.

\bibitem{panja2022parnn}
M.~Panja, T.~Chakraborty, U.~Kumar, A.~Hadid, Parnn: A probabilistic autoregressive neural network framework for accurate forecasting (2022).

\bibitem{evseenkov2021hybrid}
A.~Evseenkov, D.~K. Kuchkildin, K.~I. Krechetov, S.~A. Ospishchev, V.~Kotezhekov, E.~Yudin, Short-term forecasting of well production based on a hybrid probabilistic approach, Day 2 Wed, October 13, 2021 (2021).
\newblock \href {https://doi.org/10.2118/206519-ms} {\path{doi:10.2118/206519-ms}}.

\bibitem{garg2022machine}
R.~Garg, S.~Barpanda, S.~GirishRaoSalankeN, S.~Ramya, Machine learning algorithms for time series analysis and forecasting, ArXiv abs/2211.14387 (2022).
\newblock \href {https://doi.org/10.48550/arXiv.2211.14387} {\path{doi:10.48550/arXiv.2211.14387}}.

\bibitem{wang2024sttransnet}
J.~Wang, X.~Wang, J.~Guan, L.~Zhang, T.~Chang, W.~Yu, St-transnet: A spatiotemporal transformer network for uncertainty estimation from a single deterministic precipitation forecast, Monthly Weather Review (2024).
\newblock \href {https://doi.org/10.1175/mwr-d-23-0097.1} {\path{doi:10.1175/mwr-d-23-0097.1}}.

\bibitem{phan2021hybrid}
Q.-T. Phan, Y.-K. Wu, Q.~Phan, A hybrid wind power forecasting model with xgboost, data preprocessing considering different nwps, Applied Sciences (2021).
\newblock \href {https://doi.org/10.3390/APP11031100} {\path{doi:10.3390/APP11031100}}.

\bibitem{deng2023universal}
B.~Deng, K.~Jia, Universal domain adaptation from foundation models, ArXiv abs/2305.11092 (2023).
\newblock \href {https://doi.org/10.48550/arXiv.2305.11092} {\path{doi:10.48550/arXiv.2305.11092}}.

\bibitem{conti2022physics}
Z.~X. Conti, R.~Choudhary, L.~Magri, A physics-based domain adaptation framework for modeling and forecasting building energy systems, ArXiv abs/2208.09456 (2022).
\newblock \href {https://doi.org/10.48550/arXiv.2208.09456} {\path{doi:10.48550/arXiv.2208.09456}}.

\bibitem{fang2021hybrid}
X.~Fang, G.~Gong, G.~Li, L.~Chun, W.~Li, P.~Peng, A hybrid deep transfer learning strategy for short term cross-building energy prediction, Energy 215 (2021) 119208.
\newblock \href {https://doi.org/10.1016/j.energy.2020.119208} {\path{doi:10.1016/j.energy.2020.119208}}.

\bibitem{gao2022tgdlf2}
J.~Gao, W.~Hu, D.~Zhang, Y.~Chen, Tgdlf2.0: Theory-guided deep-learning for electrical load forecasting via transformer and transfer learning, ArXiv abs/2210.02448 (2022).
\newblock \href {https://doi.org/10.48550/arXiv.2210.02448} {\path{doi:10.48550/arXiv.2210.02448}}.

\bibitem{gao2023improving}
Y.~hong Gao, Y.~Wang, Q.~Wang, Improving the transferability of time series forecasting with decomposition adaptation, ArXiv abs/2307.00066 (2023).
\newblock \href {https://doi.org/10.48550/arXiv.2307.00066} {\path{doi:10.48550/arXiv.2307.00066}}.

\bibitem{spencer2024transfer}
R.~Spencer, S.~Ranathunga, M.~Boulic, A.~van Heerden, T.~Susnjak, Transfer learning on transformers for building energy consumption forecasting--a comparative study, arXiv preprint arXiv:2410.14107 (2024).

\bibitem{rasul2023lag}
K.~Rasul, A.~Ashok, A.~R. Williams, A.~Khorasani, G.~Adamopoulos, R.~Bhagwatkar, M.~Bilo{\v{s}}, H.~Ghonia, N.~Hassen, A.~Schneider, et~al., Lag-llama: Towards foundation models for time series forecasting, in: R0-FoMo: Robustness of Few-shot and Zero-shot Learning in Large Foundation Models, 2023.

\bibitem{bodnar2024aurora}
C.~Bodnar, W.~P. Bruinsma, A.~Lucic, M.~Stanley, J.~Brandstetter, P.~Garvan, M.~Riechert, J.~Weyn, H.~Dong, A.~Vaughan, et~al., Aurora: A foundation model of the atmosphere, arXiv preprint arXiv:2405.13063 (2024).

\bibitem{Hamann2024}
H.~F. Hamann, T.~Brunschwiler, B.~Gjorgiev, L.~S.~A. Martins, A.~Puech, A.~Varbella, J.~Weiss, J.~Bernabé-Moreno, A.~B. Mass'e, S.~Choi, I.~Foster, B.~Hodge, R.~Jain, K.~Kim, V.~Mai, F.~Miralles, M.~D. Montigny, O.~Ramos-Leaños, H.~Suprême, L.~Xie, E.-N.~S. Youssef, A.~Zinflou, A.~J. Belvi, R.~J. Bessa, B.~P. Bhattari, J.~Schmude, S.~Sobolevsky, A perspective on foundation models for the electric power grid, ArXiv abs/2407.09434 (2024).
\newblock \href {https://doi.org/10.48550/arXiv.2407.09434} {\path{doi:10.48550/arXiv.2407.09434}}.

\bibitem{farhadloo2025towards}
M.~Farhadloo, A.~Sharma, M.~Yang, B.~Jayaprakash, W.~Northrop, S.~Shekhar, Towards physics-guided foundation models, arXiv preprint arXiv:2502.15013 (2025).

\bibitem{huang2022adversarial}
M.~Huang, J.~Yin, Research on adversarial domain adaptation method and its application in power load forecasting, Mathematics (2022).
\newblock \href {https://doi.org/10.3390/math10183223} {\path{doi:10.3390/math10183223}}.

\bibitem{schreiber2022model}
J.~Schreiber, B.~Sick, Model selection, adaptation, and combination for transfer learning in wind and photovoltaic power forecasts, Energy and AI (2022).
\newblock \href {https://doi.org/10.1016/j.egyai.2023.100249} {\path{doi:10.1016/j.egyai.2023.100249}}.

\bibitem{tortora2023matnet}
M.~Tortora, F.~Conte, G.~Natrella, P.~Soda, Matnet: Multi-level fusion and self-attention transformer-based model for multivariate multi-step day-ahead pv generation forecasting, arXiv preprint arXiv:2306.10356 (2023).

\bibitem{cheng2023imbalanced}
J.~Cheng, K.~Huang, Z.~Zheng, Fitting imbalanced uncertainties in multi-output time series forecasting, ACM Transactions on Knowledge Discovery from Data (2023).
\newblock \href {https://doi.org/10.1145/3584704} {\path{doi:10.1145/3584704}}.

\bibitem{deng2021mvmt}
J.~Deng, X.~Chen, R.~Jiang, X.~Song, I.~Tsang, A multi-view multi-task learning framework for multi-variate time series forecasting, ArXiv abs/2109.01657 (2021).
\newblock \href {https://doi.org/10.1109/tkde.2022.3218803} {\path{doi:10.1109/tkde.2022.3218803}}.

\bibitem{li2024water}
L.~Li, Y.~Dai, Z.~Wei, S.~Wei, Y.~Zhang, N.~Wei, Q.~Li, Enforcing water balance in multitask deep learning models for hydrological forecasting, Journal of Hydrometeorology (2024).
\newblock \href {https://doi.org/10.1175/jhm-d-23-0073.1} {\path{doi:10.1175/jhm-d-23-0073.1}}.

\bibitem{nikentari2022narmax}
N.~Nikentari, H.-L. Wei, Multi-task learning for time series forecasting using narmax-lstm, 2022 27th International Conference on Automation and Computing (ICAC) (2022) 1--6\href {https://doi.org/10.1109/ICAC55051.2022.9911071} {\path{doi:10.1109/ICAC55051.2022.9911071}}.

\bibitem{mendis2024multivariate}
K.~Mendis, M.~Wickramasinghe, P.~Marasinghe, Multivariate time series forecasting: A review, Proceedings of the 2024 2nd Asia Conference on Computer Vision, Image Processing and Pattern Recognition (2024).
\newblock \href {https://doi.org/10.1145/3663976.3664241} {\path{doi:10.1145/3663976.3664241}}.

\bibitem{patil2024hybrid}
A.~Patil, K.~Kulkarni, A hybrid machine learning - numerical weather prediction approach for day ahead solar irradiance prediction, in: 2024 IEEE 4th International Conference on Sustainable Energy and Future Electric Transportation (SEFET), 2024, pp. 1--6.
\newblock \href {https://doi.org/10.1109/SEFET61574.2024.10718271} {\path{doi:10.1109/SEFET61574.2024.10718271}}.

\bibitem{bellinguer2021assessment}
K.~Bellinguer, R.~Girard, G.~Bontron, G.~Kariniotakis, Assessment of alternative ways to integrate weather predictions in photovoltaic generation forecasting, EGU General Assembly (2021).
\newblock \href {https://doi.org/10.5194/EGUSPHERE-EGU21-16091} {\path{doi:10.5194/EGUSPHERE-EGU21-16091}}.

\bibitem{bright2021statistical}
J.~M. Bright, X.~Sun, Statistical satellite-derived irradiance estimation: A case study in singapore, in: 2021 IEEE 48th Photovoltaic Specialists Conference (PVSC), 2021, pp. 0835--0842.
\newblock \href {https://doi.org/10.1109/PVSC43889.2021.9518398} {\path{doi:10.1109/PVSC43889.2021.9518398}}.

\bibitem{tournadre2022heliosat}
B.~Tournadre, B.~Gschwind, Y.~Saint-Drenan, X.~Chen, R.~A. e~Silva, P.~Blanc, An alternative cloud index for estimating downwelling surface solar irradiance from various satellite imagers in the framework of a heliosat-v method, Atmospheric Measurement Techniques (2022).
\newblock \href {https://doi.org/10.5194/amt-15-3683-2022} {\path{doi:10.5194/amt-15-3683-2022}}.

\bibitem{gregor2021nowcasting}
P.~Gregor, B.~Mayer, T.~Zinner, J.~Schreder, L.~Bugliaro, Towards seamless irradiance nowcasting on intrahour to intraday scales using all-sky and satellite imagery, Tech. rep., Copernicus Meetings (2021).

\bibitem{rodriguez2021nowcasting}
F.~J. Rodr{\'i}guez-Ben{\'i}tez, M.~L{\'o}pez-Cuesta, C.~Arbizu-Barrena, M.~M. Fern{\'a}ndez-Le{\'o}n, M.~{\'A}. Pamos-Ure{\~n}a, J.~Tovar-Pescador, F.~Santos-Alamillos, D.~Pozo-V{\'a}zquez, Assessment of new solar radiation nowcasting methods based on sky-camera and satellite imagery, Applied Energy 292 (2021) 116838.
\newblock \href {https://doi.org/10.1016/J.APENERGY.2021.116838} {\path{doi:10.1016/J.APENERGY.2021.116838}}.

\bibitem{logothetis2021forecasting}
S.-A. Logothetis, V.~Salamalikis, S.~Wilbert, J.~Remund, L.~Zarzalejo, Y.~Xie, B.~Nouri, E.~Ntavelis, J.~Nou, L.~Visser, M.~Sengupta, M.~Pó, R.~Chauvin, S.~Grieu, W.~van Sark, A.~Kazantzidis, Forecasting of solar irradiance and ramp events with all-sky imagers (2021).
\newblock \href {https://doi.org/10.5194/ems2021-75} {\path{doi:10.5194/ems2021-75}}.

\bibitem{suheb2022iot}
Suheb, J.~V. N, Iot enabled smart metering in smart energy grid, in: 2022 IEEE North Karnataka Subsection Flagship International Conference (NKCon), 2022, pp. 1--8.
\newblock \href {https://doi.org/10.1109/NKCon56289.2022.10126752} {\path{doi:10.1109/NKCon56289.2022.10126752}}.

\bibitem{kaur2023iot}
A.~P. Kaur, M.~Singh, Iot-based net meter for vehicle-to-grid application with time-of-use, in: 2023 Second International Conference On Smart Technologies For Smart Nation (SmartTechCon), 2023, pp. 153--158.
\newblock \href {https://doi.org/10.1109/SmartTechCon57526.2023.10391799} {\path{doi:10.1109/SmartTechCon57526.2023.10391799}}.

\bibitem{alhaddad2024forecasting}
Y.~Al-Haddad, A.~A. Ibrahim, R.~Naeem, Forecasting energy consumption in smart grids: A comparative analysis of recurrent neural networks, Iraqi Journal for Computers and Informatics (2024).
\newblock \href {https://doi.org/10.25195/ijci.v50i1.492} {\path{doi:10.25195/ijci.v50i1.492}}.

\bibitem{mlynek2022smartmeters}
P.~Mlynek, V.~Uher, J.~Misurec, Forecasting of smart meters energy consumption for data analytics and grid monitoring, 2022 22nd International Scientific Conference on Electric Power Engineering (EPE) (2022) 1--5\href {https://doi.org/10.1109/EPE54603.2022.9814101} {\path{doi:10.1109/EPE54603.2022.9814101}}.

\bibitem{chen2023smartmeter}
Z.~Chen, A.~Amani, X.~Yu, M.~Jalili, Control and optimisation of power grids using smart meter data: A review, Sensors (Basel, Switzerland) 23 (2023).
\newblock \href {https://doi.org/10.3390/s23042118} {\path{doi:10.3390/s23042118}}.

\bibitem{yuan2021demand}
L.~Yuan, C.~Ying, S.~Yongfu, Z.~Xuenan, L.~Yucai, L.~Yang, Research on short-term load forecasting under demand response of multi-type power grid connection based on dynamic electricity price, in: 2021 IEEE International Conference on Power, Intelligent Computing and Systems (ICPICS), 2021, pp. 141--144.
\newblock \href {https://doi.org/10.1109/ICPICS52425.2021.9524096} {\path{doi:10.1109/ICPICS52425.2021.9524096}}.

\bibitem{wang2024demand}
Z.~Wang, Demand forecasting and resource scheduling of independent energy storage market in power grid with deep learning, Journal of Electrical Systems (2024).
\newblock \href {https://doi.org/10.52783/jes.2609} {\path{doi:10.52783/jes.2609}}.

\bibitem{rawal2022comparative}
K.~Rawal, A.~Ahmad, A comparative analysis of supervised machine learning algorithms for electricity demand forecasting, in: 2022 Second International Conference on Power, Control and Computing Technologies (ICPC2T), 2022, pp. 1--6.
\newblock \href {https://doi.org/10.1109/ICPC2T53885.2022.9776960} {\path{doi:10.1109/ICPC2T53885.2022.9776960}}.

\bibitem{cui2023game}
F.~Cui, D.~An, G.~Zhang, A game strategy for demand response based on load monitoring in smart grid, Frontiers in Energy Research (2023).
\newblock \href {https://doi.org/10.3389/fenrg.2023.1240542} {\path{doi:10.3389/fenrg.2023.1240542}}.

\bibitem{shaquor2021dayahead}
A.~Shaqour, H.~Farzaneh, H.~Almogdady, Day-ahead residential electricity demand response model based on deep neural networks for peak demand reduction in the jordanian power sector, Applied Sciences (2021).
\newblock \href {https://doi.org/10.3390/APP11146626} {\path{doi:10.3390/APP11146626}}.

\bibitem{schmude2024prithvi}
J.~Schmude, S.~Roy, W.~Trojak, J.~Jakubik, D.~S. Civitarese, S.~Singh, J.~Kuehnert, K.~Ankur, A.~Gupta, C.~E. Phillips, et~al., Prithvi wxc: Foundation model for weather and climate, arXiv preprint arXiv:2409.13598 (2024).

\bibitem{paletta2023advances}
Q.~Paletta, G.~Terr{\'e}n-Serrano, Y.~Nie, B.~Li, J.~Bieker, W.~Zhang, L.~Dubus, S.~Dev, C.~Feng, Advances in solar forecasting: Computer vision with deep learning, Advances in Applied Energy (2023) 100150.

\bibitem{lee2023multimodal}
S.~J. Lee, Multimodal data fusion for estimating electricity access and demand, Ph.D. thesis, Massachusetts Institute of Technology (2023).

\bibitem{bharti2022attention}
S.~Bharti, V.~Saini, R.~Kumar, A.~Vijayvargiya, Attention mechanism in deep learning for wind power forecasting, in: 2022 IEEE International Conference on Power Electronics, Drives and Energy Systems (PEDES), 2022, pp. 1--6.
\newblock \href {https://doi.org/10.1109/PEDES56012.2022.10080136} {\path{doi:10.1109/PEDES56012.2022.10080136}}.

\bibitem{meng2022hybrid}
A.~Meng, S.~Chen, Z.~Ou, W.~Ding, H.~Zhou, J.~Fan, H.~Yin, A hybrid deep learning architecture for wind power prediction based on bi-attention mechanism and crisscross optimization, Energy 238 (2022) 121795.
\newblock \href {https://doi.org/10.1016/J.ENERGY.2021.121795} {\path{doi:10.1016/J.ENERGY.2021.121795}}.

\bibitem{zhang2022crossattention}
Y.~Zhang, C.~Peng, A hybrid model based on deep learning and cross-attention for short-term wind power prediction, in: 2022 5th International Conference on Renewable Energy and Power Engineering (REPE), 2022, pp. 351--355.
\newblock \href {https://doi.org/10.1109/REPE55559.2022.9948810} {\path{doi:10.1109/REPE55559.2022.9948810}}.

\bibitem{chen2021ultrashortterm}
F.~Chen, Y.~Zhang, J.~Yan, J.~You, Y.~Liu, M.~B. Paskyabi, Ultra-short-term wind power forecasting based on attention mechanism, in: The 10th Renewable Power Generation Conference (RPG 2021), Vol. 2021, 2021, pp. 186--192.
\newblock \href {https://doi.org/10.1049/icp.2021.2387} {\path{doi:10.1049/icp.2021.2387}}.

\bibitem{sendanayake2023attention}
C.~M. Sendanayake, M.~A. Juman, T.~W. Shan, T.~C. Pin, Attention mechanisms in deep learning models for short-term energy load forecasting, 2023 IEEE 21st Student Conference on Research and Development (SCOReD) (2023) 87--92\href {https://doi.org/10.1109/SCOReD60679.2023.10563696} {\path{doi:10.1109/SCOReD60679.2023.10563696}}.

\bibitem{wang2022transformer}
Z.~Wang, Z.~Zhu, G.~Xiao, B.~Bai, Y.~Zhang, A transformer-based multi-entity load forecasting method for integrated energy systems, Frontiers in Energy Research 10 (2022).
\newblock \href {https://doi.org/10.3389/fenrg.2022.952420} {\path{doi:10.3389/fenrg.2022.952420}}.

\bibitem{cui2023temporal}
D.~Cui, W.~Xiang, Z.~Zang, H.~Yu, Z.~Ou, Y.~Mao, Z.~He, Temporal fusion transformer with non-intrusive attention for data-driven electricity load forecasting, 2023 7th International Conference on Power and Energy Engineering (ICPEE) (2023) 275--281\href {https://doi.org/10.1109/ICPEE60001.2023.10453838} {\path{doi:10.1109/ICPEE60001.2023.10453838}}.

\bibitem{morelli2023multimodal}
V.~G. Morelli, M.~P. Barbato, F.~Piccoli, P.~Napoletano, Multimodal fusion methods with vision transformers for remote sensing semantic segmentation, in: 2023 13th Workshop on Hyperspectral Imaging and Signal Processing: Evolution in Remote Sensing (WHISPERS), 2023, pp. 1--5.
\newblock \href {https://doi.org/10.1109/WHISPERS61460.2023.10430788} {\path{doi:10.1109/WHISPERS61460.2023.10430788}}.

\bibitem{jia2024geminifusion}
D.~Jia, J.~Guo, K.~Han, H.~Wu, C.~Zhang, C.~Xu, X.~Chen, Geminifusion: Efficient pixel-wise multimodal fusion for vision transformer, ArXiv abs/2406.01210 (2024).
\newblock \href {https://doi.org/10.48550/arXiv.2406.01210} {\path{doi:10.48550/arXiv.2406.01210}}.

\bibitem{pellegrain2021streaming}
V.~Pellegrain, M.~Tami, M.~Batteux, C.~Hudelot, Streamult: Streaming multimodal transformer for heterogeneous and arbitrary long sequential data, ArXiv abs/2110.08021 (2021).

\bibitem{palety2022residential}
B.~N. Palety, C.~Mahalakshmi, Analysing the residential electricity consumption using smart meter, International Journal on Recent and Innovation Trends in Computing and Communication (2022).
\newblock \href {https://doi.org/10.17762/ijritcc.v10i2s.5910} {\path{doi:10.17762/ijritcc.v10i2s.5910}}.

\bibitem{rajendiran2022nilm}
G.~Rajendiran, Non-intrusive load monitoring: a promising path to the society for responsible energy utilization and sustainability, Proceedings of the Thirteenth ACM International Conference on Future Energy Systems (2022).
\newblock \href {https://doi.org/10.1145/3538637.3539635} {\path{doi:10.1145/3538637.3539635}}.

\bibitem{avordeh2021demand}
T.~K. Avordeh, S.~Gyamfi, A.~Opoku, The role of demand response in residential electricity load reduction using appliance shifting techniques, International Journal of Energy Sector Management (2021).
\newblock \href {https://doi.org/10.1108/ijesm-05-2020-0014} {\path{doi:10.1108/ijesm-05-2020-0014}}.

\bibitem{vasuprada2021smartmeter}
B.~M.~R. Vasuprada, D.~A. Nasreen, Forecasting peak and appliance level demand using smart meter data, 2021.

\bibitem{bashawyah2021ml}
D.~A. Bashawyah, S.~Qaisar, Machine learning based short-term load forecasting for smart meter energy consumption data in london households, in: 2021 IEEE 12th International Conference on Electronics and Information Technologies (ELIT), 2021, pp. 99--102.
\newblock \href {https://doi.org/10.1109/ELIT53502.2021.9501104} {\path{doi:10.1109/ELIT53502.2021.9501104}}.

\bibitem{liang2024enabling}
R.~Liang, Y.~Deng, D.~Xie, F.~He, D.~Wang, Enabling time-series foundation model for building energy forecasting via contrastive curriculum learning, arXiv preprint arXiv:2412.17285 (2024).

\bibitem{ArxivGrid2024}
H.~F. Hamann, et~al., \href{https://arxiv.org/html/2407.09434v1}{A perspective on foundation models for the electric power grid}, arXiv preprint (2024).
\newblock \href {http://arxiv.org/abs/2407.09434} {\path{arXiv:2407.09434}}.
\newline\urlprefix\url{https://arxiv.org/html/2407.09434v1}

\bibitem{phyo2021hybrid}
P.~Phyo, Y.~Byun, Hybrid ensemble deep learning-based approach for time series energy prediction, Symmetry 13 (2021) 1942.
\newblock \href {https://doi.org/10.3390/sym13101942} {\path{doi:10.3390/sym13101942}}.

\bibitem{jalali2022hybrid}
S.~M. Jalali, G.~Osório, S.~Ahmadian, M.~Lotfi, V.~A. Campos, M.~Shafie‐khah, A.~Khosravi, J.~Catalão, New hybrid deep neural architectural search-based ensemble reinforcement learning strategy for wind power forecasting, IEEE Transactions on Industry Applications 58 (2022) 15--27.
\newblock \href {https://doi.org/10.1109/TIA.2021.3126272} {\path{doi:10.1109/TIA.2021.3126272}}.

\bibitem{thaker2024solar}
J.~A. Thaker, R.~Höller, M.~Kapasi, Short-term solar irradiance prediction with a hybrid ensemble model using eumetsat satellite images, Energies (2024).
\newblock \href {https://doi.org/10.3390/en17020329} {\path{doi:10.3390/en17020329}}.

\bibitem{usmani2024omme}
M.~Usmani, Z.~Memon, K.~U. Danyaro, R.~Qureshi, Optimized multi-level multi-type ensemble (omme) forecasting model for univariate time series, IEEE Access 12 (2024) 35700--35715.
\newblock \href {https://doi.org/10.1109/ACCESS.2024.3370679} {\path{doi:10.1109/ACCESS.2024.3370679}}.

\bibitem{liu2023pttuning}
H.~Liu, J.~Gan, X.~Fan, Y.~Zhang, C.~Luo, J.~Zhang, G.~Jiang, Y.~Qian, C.~Zhao, H.~Ma, Z.~Guo, Pt-tuning: Bridging the gap between time series masked reconstruction and forecasting via prompt token tuning, ArXiv abs/2311.03768 (2023).
\newblock \href {https://doi.org/10.48550/arXiv.2311.03768} {\path{doi:10.48550/arXiv.2311.03768}}.

\bibitem{jia2024gpt4mts}
F.~Jia, K.~Wang, Y.~Zheng, D.~Cao, Y.~Liu, Gpt4mts: Prompt-based large language model for multimodal time-series forecasting (2024) 23343--23351\href {https://doi.org/10.1609/aaai.v38i21.30383} {\path{doi:10.1609/aaai.v38i21.30383}}.

\bibitem{xue2022promptcast}
H.~Xue, F.~D. Salim, Promptcast: A new prompt-based learning paradigm for time series forecasting, IEEE Transactions on Knowledge and Data Engineering (2022).
\newblock \href {https://doi.org/10.1109/tkde.2023.3342137} {\path{doi:10.1109/tkde.2023.3342137}}.

\bibitem{xiao2022metalearning}
F.~Xiao, L.~Liu, J.~Han, D.~Guo, S.~Wang, H.~Cui, T.~Peng, Meta-learning for few-shot time series forecasting, J. Intell. Fuzzy Syst. 43 (2022) 325--341.
\newblock \href {https://doi.org/10.3233/jifs-212228} {\path{doi:10.3233/jifs-212228}}.

\bibitem{jawed2023forecasting}
S.~Jawed, K.~Madhusudhanan, V.~K. Yalavarthi, L.~Schmidt-Thieme, Forecasting early with meta learning, 2023 International Joint Conference on Neural Networks (IJCNN) (2023) 1--8\href {https://doi.org/10.1109/IJCNN54540.2023.10191229} {\path{doi:10.1109/IJCNN54540.2023.10191229}}.

\bibitem{shen2024microgrid}
H.~Shen, X.~Shen, Y.~Chen, Real-time microgrid energy scheduling using meta-reinforcement learning, Energies (2024).
\newblock \href {https://doi.org/10.3390/en17102367} {\path{doi:10.3390/en17102367}}.

\bibitem{german2024lliama}
M.~Germ´an-Morales, A.~J. Rivera-Rivas, M.~J. del, J.~D´ıaz, C.~J. Carmona, Transfer learning with foundational models for time series forecasting using low-rank adaptations (2024).

\bibitem{paischer2024eva}
F.~Paischer, L.~Hauzenberger, T.~Schmied, B.~Alkin, M.~Deisenroth, S.~Hochreiter, One initialization to rule them all: Fine-tuning via explained variance adaptation (2024).

\bibitem{gupta2024lora}
D.~Gupta, A.~Bhatti, S.~Parmar, C.~Dan, Y.~Liu, B.~Shen, S.~Lee, Low-rank adaptation of time series foundational models for out-of-domain modality forecasting, ArXiv abs/2405.10216 (2024).
\newblock \href {https://doi.org/10.48550/arXiv.2405.10216} {\path{doi:10.48550/arXiv.2405.10216}}.

\bibitem{khanna2024explora}
S.~Khanna, M.~Irgau, D.~B. Lobell, S.~Ermon, Explora: Parameter-efficient extended pre-training to adapt vision transformers under domain shifts, ArXiv abs/2406.10973 (2024).
\newblock \href {https://doi.org/10.48550/arXiv.2406.10973} {\path{doi:10.48550/arXiv.2406.10973}}.

\bibitem{meo2024bayesianlora}
C.~Meo, K.~Sycheva, A.~Goyal, J.~Dauwels, Bayesian-lora: Lora based parameter efficient fine-tuning using optimal quantization levels and rank values through differentiable bayesian gates, ArXiv abs/2406.13046 (2024).
\newblock \href {https://doi.org/10.48550/arXiv.2406.13046} {\path{doi:10.48550/arXiv.2406.13046}}.

\bibitem{chaouch2023probabilistic}
M.~Chaouch, Probabilistic wind speed forecasting for wind turbine allocation in the power grid, Energies (2023).
\newblock \href {https://doi.org/10.3390/en16227615} {\path{doi:10.3390/en16227615}}.

\bibitem{arpino2021gaussian}
G.~Arpino, H.~Shih, W.~Bruinsma, E.~P. Martins, Gaussian processes for probabilistic electricity price forecasting (2021).

\bibitem{liu2022probabilistic}
Y.~Liu, M.~Zhu, J.~Bai, Y.~Qin, Y.~Zhang, Short-term probabilistic forecasting of renewable energy generation with direct multistep-ahead strategy and recurrent neural network, 2022 4th International Conference on Power and Energy Technology (ICPET) (2022) 975--980\href {https://doi.org/10.1109/ICPET55165.2022.9918497} {\path{doi:10.1109/ICPET55165.2022.9918497}}.

\bibitem{li2024diffusion}
L.~Li, R.~Carver, I.~Lopez‐Gomez, F.~Sha, J.~Anderson, Generative emulation of weather forecast ensembles with diffusion models, Science Advances 10 (2024).
\newblock \href {https://doi.org/10.1126/sciadv.adk4489} {\path{doi:10.1126/sciadv.adk4489}}.

\bibitem{lyu2024probabilistic}
C.~Lyu, S.~Eftekharnejad, Probabilistic solar generation forecasting for rapidly changing weather conditions, IEEE Access 12 (2024) 79091--79103.
\newblock \href {https://doi.org/10.1109/ACCESS.2024.3407778} {\path{doi:10.1109/ACCESS.2024.3407778}}.

\bibitem{ye2024tadnet}
J.~Ye, B.~Zhao, D.~Liu, Q.~Wei, Y.~Wang, Tadnet: Temporal attention decomposition networks for probabilistic energy forecasting, IEEE Transactions on Power Systems 39 (2024) 7190--7202.
\newblock \href {https://doi.org/10.1109/TPWRS.2024.3380388} {\path{doi:10.1109/TPWRS.2024.3380388}}.

\bibitem{liu2021deep}
H.~Liu, C.~Liu, X.~Jiang, X.~Chen, S.~Yang, X.~Wang, Deep probabilistic time series forecasting using augmented recurrent input for dynamic systems, ArXiv abs/2106.05848 (2021).

\bibitem{tang2021probabilistic}
B.~Tang, D.~S. Matteson, Probabilistic transformer for time series analysis (2021) 23592--23608.

\bibitem{capel2022diffusion}
E.~H. Capel, J.~Dumas, Denoising diffusion probabilistic models for probabilistic energy forecasting, ArXiv abs/2212.02977 (2022).
\newblock \href {https://doi.org/10.48550/arXiv.2212.02977} {\path{doi:10.48550/arXiv.2212.02977}}.

\bibitem{wang2024generative}
X.~Wang, L.~Tong, Q.~Zhao, Generative probabilistic time series forecasting and applications in grid operations, 2024 58th Annual Conference on Information Sciences and Systems (CISS) (2024) 1--6\href {https://doi.org/10.1109/CISS59072.2024.10480214} {\path{doi:10.1109/CISS59072.2024.10480214}}.

\bibitem{sha2024ensemble}
Y.~Sha, R.~Sobash, D.~J. Gagne, Improving ensemble extreme precipitation forecasts using generative artificial intelligence, ArXiv abs/2407.04882 (2024).
\newblock \href {https://doi.org/10.48550/arXiv.2407.04882} {\path{doi:10.48550/arXiv.2407.04882}}.

\bibitem{vercosa2021hybrid}
L.~Verçosa, V.~R. Borba, P.~H. E.~S. Lima, L.~A.~G. Malagón, B.~Giublin, J.~F.~L. Oliveira, An ensemble based hybrid system for residual forecasting in industrial data, Anais do 15. Congresso Brasileiro de Inteligência Computacional (2021).
\newblock \href {https://doi.org/10.21528/cbic2021-31} {\path{doi:10.21528/cbic2021-31}}.

\bibitem{skoien2021postprocessing}
J.~Skøien, K.~Bogner, P.~Salamon, F.~Wetterhall, On the implementation of post-processing of runoff forecast ensembles, Journal of Hydrometeorology (2021).
\newblock \href {https://doi.org/10.1175/jhm-d-21-0008.1} {\path{doi:10.1175/jhm-d-21-0008.1}}.

\bibitem{jeworrek2022analog}
J.~Jeworrek, G.~West, R.~Stull, Optimizing analog ensembles for sub-daily precipitation forecasts, Atmosphere (2022).
\newblock \href {https://doi.org/10.3390/atmos13101662} {\path{doi:10.3390/atmos13101662}}.

\bibitem{baran2023postprocessing}
Ágnes Baran, S.~Baran, Parametric model for post-processing visibility ensemble forecasts, Advances in Statistical Climatology, Meteorology and Oceanography (2023).
\newblock \href {https://doi.org/10.5194/ascmo-10-105-2024} {\path{doi:10.5194/ascmo-10-105-2024}}.

\bibitem{lopez2022quantile}
S.~López-Ruiz, C.~I.~H. Castellanos, K.~Rodríguez-Vázquez, Multi-objective framework for quantile forecasting in financial time series using transformers, in: Proceedings of the Genetic and Evolutionary Computation Conference, 2022.
\newblock \href {https://doi.org/10.1145/3512290.3528740} {\path{doi:10.1145/3512290.3528740}}.

\bibitem{li2024quantile}
Y.-H. Li, Estimating uncertainty with implicit quantile network, ArXiv abs/2408.14525 (2024).
\newblock \href {https://doi.org/10.48550/arXiv.2408.14525} {\path{doi:10.48550/arXiv.2408.14525}}.

\bibitem{zheng2024mvgcrps}
V.~Z. Zheng, L.~Sun, Mvg-crps: A robust loss function for multivariate probabilistic forecasting (2024).

\bibitem{koo2024vaeneu}
A.~Koochali, E.~Tahaei, A.~Dengel, S.~Ahmed, Vaeneu: A new avenue for vae application on probabilistic forecasting, ArXiv abs/2405.04252 (2024).
\newblock \href {https://doi.org/10.48550/arXiv.2405.04252} {\path{doi:10.48550/arXiv.2405.04252}}.

\bibitem{wang2024diffusion}
K.~Wang, Y.~Zhang, F.~Lin, J.~Wang, M.~Zhu, Nonparametric probabilistic forecasting for wind power generation using quadratic spline quantile function and autoregressive recurrent neural network, IEEE Transactions on Sustainable Energy 13 (2022) 1930--1943.
\newblock \href {https://doi.org/10.1109/TSTE.2022.3175916} {\path{doi:10.1109/TSTE.2022.3175916}}.

\bibitem{borgohain2022bayesian}
S.~Borgohain, K.~Ackermann, R.~Loaiza-Maya, Bayesian neural network versus ex-post calibration for prediction uncertainty, ArXiv abs/2209.14594 (2022).
\newblock \href {https://doi.org/10.48550/arXiv.2209.14594} {\path{doi:10.48550/arXiv.2209.14594}}.

\bibitem{bicici2023calibrating}
E.~Biçici, H.~Saribas, Calibrating neural networks for ctr prediction, 2023 31st Signal Processing and Communications Applications Conference (SIU) (2023) 1--4\href {https://doi.org/10.1109/SIU59756.2023.10223867} {\path{doi:10.1109/SIU59756.2023.10223867}}.

\bibitem{qian2022uncertainty}
W.~Qian, D.~Zhang, Y.~Zhao, K.~Zheng, J.~J.~Q. Yu, Uncertainty quantification for traffic forecasting: A unified approach, ArXiv abs/2208.05875 (2022).
\newblock \href {https://doi.org/10.48550/arXiv.2208.05875} {\path{doi:10.48550/arXiv.2208.05875}}.

\bibitem{hekler2023posthoc}
A.~Hekler, T.~Brinker, F.~Buettner, Test time augmentation meets post-hoc calibration: Uncertainty quantification under real-world conditions, ArXiv abs/2303.02644 (2023).
\newblock \href {https://doi.org/10.1609/aaai.v37i12.26735} {\path{doi:10.1609/aaai.v37i12.26735}}.

\bibitem{lin2022bayesian}
Y.-H. Lin, G.~Li, A bayesian deep learning framework for rul prediction incorporating uncertainty quantification and calibration, IEEE Transactions on Industrial Informatics 18 (2022) 7274--7284.
\newblock \href {https://doi.org/10.1109/TII.2022.3156965} {\path{doi:10.1109/TII.2022.3156965}}.

\bibitem{capel2022denoising}
E.~H. Capel, J.~Dumas, Denoising diffusion probabilistic models for probabilistic energy forecasting, ArXiv abs/2212.02977 (2022).
\newblock \href {https://doi.org/10.48550/arXiv.2212.02977} {\path{doi:10.48550/arXiv.2212.02977}}.

\bibitem{lara2022stochastic}
J.~D. Lara, O.~Dowson, K.~Doubleday, B.~Hodge, D.~S. Callaway, A multi-stage stochastic risk assessment with markovian representation of renewable power, IEEE Transactions on Sustainable Energy 13 (2022) 414--426.
\newblock \href {https://doi.org/10.1109/tste.2021.3114615} {\path{doi:10.1109/tste.2021.3114615}}.

\bibitem{draz2024risk}
M.~Draz, F.~Pagel, S.~Albayrak, Probabilistic risk assessment in power systems with high wind energy penetration, IEEE Access 12 (2024) 140097--140111.
\newblock \href {https://doi.org/10.1109/ACCESS.2024.3463882} {\path{doi:10.1109/ACCESS.2024.3463882}}.

\bibitem{ye2023deep}
Z.~Ye, Y.~He, F.~Luo, Evaluation study of deep learning-based models on probabilistic power load forecasting, 2023 IEEE 7th Conference on Energy Internet and Energy System Integration (EI2) (2023) 4312--4317\href {https://doi.org/10.1109/EI259745.2023.10512431} {\path{doi:10.1109/EI259745.2023.10512431}}.

\bibitem{wang2024bayesian}
C.~Wang, Y.~Wang, Z.~Ding, K.~Zhang, Probabilistic multi-energy load forecasting for integrated energy system based on bayesian transformer network, IEEE Transactions on Smart Grid 15 (2024) 1495--1508.
\newblock \href {https://doi.org/10.1109/TSG.2023.3296647} {\path{doi:10.1109/TSG.2023.3296647}}.

\bibitem{huang2023uncertaintydriven}
T.~Huang, W.~Dong, F.~Wu, X.~Li, G.~Shi, Uncertainty-driven knowledge distillation for language model compression, IEEE/ACM Transactions on Audio, Speech, and Language Processing 31 (2023) 2850--2858.
\newblock \href {https://doi.org/10.1109/TASLP.2023.3289303} {\path{doi:10.1109/TASLP.2023.3289303}}.

\bibitem{kim2023quantization}
J.~Kim, Quantization robust pruning with knowledge distillation, IEEE Access 11 (2023) 26419--26426.
\newblock \href {https://doi.org/10.1109/ACCESS.2023.3257864} {\path{doi:10.1109/ACCESS.2023.3257864}}.

\bibitem{feng2023modelcompression}
K.~Feng, L.~Xu, D.~Zhao, S.~Liu, X.~Huang, Toward model compression for a deep learning–based solar flare forecast on satellites, The Astrophysical Journal Supplement Series 268 (2023).
\newblock \href {https://doi.org/10.3847/1538-4365/ace96a} {\path{doi:10.3847/1538-4365/ace96a}}.

\bibitem{mnif2024compression}
M.~Mnif, S.~Sahnoun, M.~Djemaa, A.~Fakhfakh, O.~Kanoun, Exploring model compression techniques for efficient 1d cnn-based hand gesture recognition on resource-constrained edge devices, 2024 IEEE 7th International Conference on Advanced Technologies, Signal and Image Processing (ATSIP) 1 (2024) 659--664.
\newblock \href {https://doi.org/10.1109/ATSIP62566.2024.10638925} {\path{doi:10.1109/ATSIP62566.2024.10638925}}.

\bibitem{zhu2023compression}
X.~Zhu, J.~Li, Y.~Liu, C.~Ma, W.~Wang, A survey on model compression for large language models, ArXiv abs/2308.07633 (2023).
\newblock \href {https://doi.org/10.48550/arXiv.2308.07633} {\path{doi:10.48550/arXiv.2308.07633}}.

\bibitem{li2023bidformer}
W.~Li, X.~Meng, C.~Chen, H.~Mi, H.~Wang, Bidformer: A transformer-based model via bidirectional sparse self-attention mechanism for long sequence time-series forecasting, in: 2023 IEEE International Conference on Systems, Man, and Cybernetics (SMC), 2023, pp. 4076--4082.
\newblock \href {https://doi.org/10.1109/SMC53992.2023.10394310} {\path{doi:10.1109/SMC53992.2023.10394310}}.

\bibitem{chen2023resformer}
G.~Chen, H.~Wang, Y.~Liu, M.~Zhang, F.~Zhang, Resformer: Combine quadratic linear transformation with efficient sparse transformer for long-term series forecasting, Intelligent Data Analysis (2023).
\newblock \href {https://doi.org/10.3233/ida-227006} {\path{doi:10.3233/ida-227006}}.

\bibitem{feng2024halveformer}
Y.~Feng, K.~Xia, X.~Qu, X.~Hu, L.~Sun, Z.~Zhang, Halveformer: A novel architecture combined with linear models for long sequences time series forecasting, in: 2024 International Joint Conference on Neural Networks (IJCNN), 2024, pp. 1--8.
\newblock \href {https://doi.org/10.1109/IJCNN60899.2024.10651438} {\path{doi:10.1109/IJCNN60899.2024.10651438}}.

\bibitem{wen2024crossscale}
L.~Wen, Q.~Hu, C.~Guo, A.~Hu, M.~Zhang, Cross-scale attention for long-term time series forecasting, IEEE Signal Processing Letters 31 (2024) 2675--2679.
\newblock \href {https://doi.org/10.1109/LSP.2024.3439103} {\path{doi:10.1109/LSP.2024.3439103}}.

\bibitem{ding2024drformer}
R.~Ding, Y.~Chen, Y.-T. Lan, W.~Zhang, Drformer: Multi-scale transformer utilizing diverse receptive fields for long time-series forecasting, ArXiv abs/2408.02279 (2024).
\newblock \href {https://doi.org/10.1145/3627673.3679724} {\path{doi:10.1145/3627673.3679724}}.

\bibitem{peng2023fp8lm}
H.~Peng, K.~Wu, Y.~Wei, G.~Zhao, Y.~Yang, Z.~Liu, Y.~Xiong, Z.~Yang, B.~Ni, J.~Hu, R.~Li, M.~Zhang, C.~Li, J.~Ning, R.~Wang, Z.~Zhang, S.~Liu, J.~Chau, H.~Hu, P.~Cheng, Fp8-lm: Training fp8 large language models, ArXiv abs/2310.18313 (2023).
\newblock \href {https://doi.org/10.48550/arXiv.2310.18313} {\path{doi:10.48550/arXiv.2310.18313}}.

\bibitem{liu2024neuralprocessor}
R.~Liu, C.~Wei, Y.~Yang, W.~Wang, B.~Yuan, H.~Yang, Y.~Liu, A dynamic execution neural network processor for fine-grained mixed-precision model training based on online quantization sensitivity analysis, IEEE Journal of Solid-State Circuits 59 (2024) 3082--3093.
\newblock \href {https://doi.org/10.1109/JSSC.2024.3377292} {\path{doi:10.1109/JSSC.2024.3377292}}.

\bibitem{santos2022gpu}
F.~Santos, P.~Rech, A.~Kritikakou, O.~Sentieys, Evaluating the impact of mixed-precision on fault propagation for deep neural networks on gpus, 2022 IEEE Computer Society Annual Symposium on VLSI (ISVLSI) (2022) 327--327\href {https://doi.org/10.1109/ISVLSI54635.2022.00071} {\path{doi:10.1109/ISVLSI54635.2022.00071}}.

\bibitem{xie2022bfloat16}
Z.~Xie, S.~Raskar, M.~Emani, Throughput-oriented and accuracy-aware dnn training with bfloat16 on gpu, 2022 IEEE International Parallel and Distributed Processing Symposium Workshops (IPDPSW) (2022) 1084--1087\href {https://doi.org/10.1109/IPDPSW55747.2022.00176} {\path{doi:10.1109/IPDPSW55747.2022.00176}}.

\bibitem{jain2022timeseriesgpu}
M.~Jain, S.~Ghosh, S.~Nandanoori, Workload characterization of a time-series prediction system for spatio-temporal data, Proceedings of the 19th ACM International Conference on Computing Frontiers (2022).
\newblock \href {https://doi.org/10.1145/3528416.3530242} {\path{doi:10.1145/3528416.3530242}}.

\bibitem{lermer2021hybrid}
M.~Lermer, C.~Reich, D.~Abdeslam, Hybrid ai improves energy forecasts by combining fuzzy rules, evolutionary strategies and neural networks, in: IECON 2021 – 47th Annual Conference of the IEEE Industrial Electronics Society, 2021, pp. 1--6.
\newblock \href {https://doi.org/10.1109/IECON48115.2021.9589186} {\path{doi:10.1109/IECON48115.2021.9589186}}.

\bibitem{nahid2021hybrid}
F.~A. Nahid, W.~Ongsakul, N.~M. M., T.~Laopaiboon, Hybrid neural networks for renewable energy forecasting, Advances in Computer and Electrical Engineering (2021).
\newblock \href {https://doi.org/10.4018/978-1-7998-3970-5.ch011} {\path{doi:10.4018/978-1-7998-3970-5.ch011}}.

\bibitem{nadeem2024interpretable}
A.~Nadeem, M.~F. Hanif, M.~S. Naveed, M.~T. Hassan, M.~Gul, N.~Husnain, J.~Mi, Ai-driven precision in solar forecasting: Breakthroughs in machine learning and deep learning, AIMS Geosciences (2024).
\newblock \href {https://doi.org/10.3934/geosci.2024035} {\path{doi:10.3934/geosci.2024035}}.

\bibitem{zhou2024physicsai}
Y.~Zhou, C.~MacPhee, T.~Zhou, B.~Jalali, Nonlinear schrödinger network, ArXiv abs/2407.14504 (2024).
\newblock \href {https://doi.org/10.48550/arXiv.2407.14504} {\path{doi:10.48550/arXiv.2407.14504}}.

\bibitem{watermeyer2023hybrid}
M.~Watermeyer, T.~Mobius, O.~Grothe, F.~Musgens, A hybrid model for day-ahead electricity price forecasting: Combining fundamental and stochastic modelling (2023).

\bibitem{niu2024attention}
P.~Niu, T.~Zhou, X.~Wang, L.~Sun, R.~Jin, Attention as robust representation for time series forecasting, ArXiv abs/2402.05370 (2024).
\newblock \href {https://doi.org/10.48550/arXiv.2402.05370} {\path{doi:10.48550/arXiv.2402.05370}}.

\bibitem{wu2024revisiting}
H.~Wu, Revisiting attention for multivariate time series forecasting, ArXiv abs/2407.13806 (2024).
\newblock \href {https://doi.org/10.48550/arXiv.2407.13806} {\path{doi:10.48550/arXiv.2407.13806}}.

\bibitem{aguilera2024local}
I.~Aguilera-Martos, A.~Herrera-Poyatos, J.~Luengo, F.~Herrera, Local attention: Enhancing the transformer architecture for efficient time series forecasting, 2024 International Joint Conference on Neural Networks (IJCNN) (2024) 1--8\href {https://doi.org/10.1109/IJCNN60899.2024.10650762} {\path{doi:10.1109/IJCNN60899.2024.10650762}}.

\bibitem{li2024linwa}
Q.~Li, J.~Qin, D.~Sun, F.~Shi, D.~Cui, J.~Xie, Linwa: Linear weights attention for time series forecasting, 2024 International Joint Conference on Neural Networks (IJCNN) (2024) 1--8\href {https://doi.org/10.1109/IJCNN60899.2024.10650033} {\path{doi:10.1109/IJCNN60899.2024.10650033}}.

\bibitem{baric2021benchmarking}
D.~Baric, P.~Fumić, D.~Horvatic, T.~Lipić, Benchmarking attention-based interpretability of deep learning in multivariate time series predictions, Entropy 23 (2021).
\newblock \href {https://doi.org/10.3390/e23020143} {\path{doi:10.3390/e23020143}}.

\bibitem{cremades2024shap}
A.~Cremades, S.~Hoyas, R.~Vinuesa, Additive-feature-attribution methods: a review on explainable artificial intelligence for fluid dynamics and heat transfer, ArXiv abs/2409.11992 (2024).
\newblock \href {https://doi.org/10.48550/arXiv.2409.11992} {\path{doi:10.48550/arXiv.2409.11992}}.

\bibitem{ioannou2023baselines}
G.~Ioannou, A.~Stafylopatis, The issue of baselines in explainability methods, 2023 IEEE International Conference on Data Mining Workshops (ICDMW) (2023) 958--965\href {https://doi.org/10.1109/ICDMW60847.2023.00127} {\path{doi:10.1109/ICDMW60847.2023.00127}}.

\bibitem{raykar2023tsshap}
V.~Raykar, A.~Jati, S.~Mukherjee, N.~Aggarwal, K.~K. Sarpatwar, G.~Ganapavarapu, R.~Vaculín, Tsshap: Robust model agnostic feature-based explainability for time series forecasting, ArXiv abs/2303.12316 (2023).
\newblock \href {https://doi.org/10.48550/arXiv.2303.12316} {\path{doi:10.48550/arXiv.2303.12316}}.

\bibitem{huang2024shap}
X.~Huang, J.~Marques-Silva, Updates on the complexity of shap scores, ArXiv abs/2405.11766 (2024).
\newblock \href {https://doi.org/10.48550/arXiv.2405.11766} {\path{doi:10.48550/arXiv.2405.11766}}.

\bibitem{zhuo2024ig2}
Y.~Zhuo, Z.~Ge, Ig2: Integrated gradient on iterative gradient path for feature attribution, IEEE Transactions on Pattern Analysis and Machine Intelligence 46 (2024) 7173--7190.
\newblock \href {https://doi.org/10.1109/TPAMI.2024.3388092} {\path{doi:10.1109/TPAMI.2024.3388092}}.

\bibitem{heidari2023interpretable}
F.~Heidari, P.~Taslakian, G.~Rabusseau, Explaining graph neural networks using interpretable local surrogates (2023) 146--155.

\bibitem{kleinlein2022local}
R.~Kleinlein, A.~Hepburn, R.~Santos-Rodríguez, F.~Fernández-Martínez, Sampling based on natural image statistics improves local surrogate explainers (2022) 1083\href {https://doi.org/10.48550/arXiv.2208.03961} {\path{doi:10.48550/arXiv.2208.03961}}.

\bibitem{sigut2023surrogate}
J.~Sigut, F.~Fumero, R.~Arnay, J.~Estévez, T.~Díaz-Alemán, Interpretable surrogate models to approximate the predictions of convolutional neural networks in glaucoma diagnosis, Machine Learning: Science and Technology 4 (2023).
\newblock \href {https://doi.org/10.1088/2632-2153/ad0798} {\path{doi:10.1088/2632-2153/ad0798}}.

\bibitem{ranjbar2022lime}
N.~Ranjbar, R.~Safabakhsh, Using decision tree as local interpretable model in autoencoder-based lime, 2022 27th International Computer Conference, Computer Society of Iran (CSICC) (2022) 1--7\href {https://doi.org/10.48550/arXiv.2204.03321} {\path{doi:10.48550/arXiv.2204.03321}}.

\bibitem{ozyegen2021evaluation}
O.~Ozyegen, I.~Ilic, M.~Cevik, Evaluation of interpretability methods for multivariate time series forecasting, Applied Intelligence (2021) 1--17\href {https://doi.org/10.1007/s10489-021-02662-2} {\path{doi:10.1007/s10489-021-02662-2}}.

\bibitem{kuvshinova2024foundation}
K.~Kuvshinova, O.~Tsymboi, A.~Kostromina, D.~Simakov, E.~Kovtun, Towards foundation time series model: To synthesize or not to synthesize?, ArXiv abs/2403.02534 (2024).
\newblock \href {https://doi.org/10.48550/arXiv.2403.02534} {\path{doi:10.48550/arXiv.2403.02534}}.

\bibitem{potosnak2024reasoning}
W.~Potosnak, C.~Challu, M.~Goswami, M.~Wiliński, N.~Zukowska, Implicit reasoning in deep time series forecasting, ArXiv abs/2409.10840 (2024).
\newblock \href {https://doi.org/10.48550/arXiv.2409.10840} {\path{doi:10.48550/arXiv.2409.10840}}.

\bibitem{sprang2024interpretability}
A.~van Sprang, E.~Acar, W.~Zuidema, Enforcing interpretability in time series transformers: A concept bottleneck framework (2024).

\bibitem{kjaernli2024robustness}
H.~H. Kjaernli, L.~Mas-Ribas, A.~Ashrafi, G.~Sizov, H.~Langseth, O.~E. Gundersen, Probing the robustness of time-series forecasting models with counterfacts, ArXiv abs/2403.03508 (2024).
\newblock \href {https://doi.org/10.48550/arXiv.2403.03508} {\path{doi:10.48550/arXiv.2403.03508}}.

\bibitem{airlangga2024xai}
G.~Airlangga, Decoding energy usage predictions: An application of xai techniques for enhanced model interpretability, Indonesian Journal of Artificial Intelligence and Data Mining (2024).
\newblock \href {https://doi.org/10.24014/ijaidm.v7i2.29041} {\path{doi:10.24014/ijaidm.v7i2.29041}}.

\bibitem{gou2023interpretability}
X.~Gou, L.~Hu, D.~Wang, X.~Zhang, A fundamental model with stable interpretability for traffic forecasting, in: Proceedings of the 1st ACM SIGSPATIAL International Workshop on Geo-Privacy and Data Utility for Smart Societies, 2023.
\newblock \href {https://doi.org/10.1145/3615889.3628510} {\path{doi:10.1145/3615889.3628510}}.

\bibitem{harel2022featureimportance}
N.~Harel, U.~Obolski, R.~Gilad-Bachrach, Inherent inconsistencies of feature importance (2022).

\bibitem{samimi2022boosting}
R.~Samimi, A.~Alyousef, D.~Baranzini, H.~Meer, Boosting interpretability of non-readable deep learning forecasts: the case of buildings' energy consumptions prediction, in: Proceedings of the Thirteenth ACM International Conference on Future Energy Systems, 2022.
\newblock \href {https://doi.org/10.1145/3538637.3538754} {\path{doi:10.1145/3538637.3538754}}.

\bibitem{kaneko2023feature}
H.~Kaneko, Interpretation of machine learning models for data sets with many features using feature importance, ACS Omega 8 (2023) 23218--23225.
\newblock \href {https://doi.org/10.1021/acsomega.3c03722} {\path{doi:10.1021/acsomega.3c03722}}.

\bibitem{zheng2023tcn}
Q.~Zheng, J.~Zheng, F.~Mei, A.~Gao, X.~Zhang, Y.~Xie, Tcn-gat multivariate load forecasting model based on shap value selection strategy in integrated energy system, Frontiers in Energy Research 11 (2023).
\newblock \href {https://doi.org/10.3389/fenrg.2023.1208502} {\path{doi:10.3389/fenrg.2023.1208502}}.

\bibitem{zeng2024shap}
X.~Zeng, Enhancing the interpretability of shap values using large language models, ArXiv abs/2409.00079 (2024).
\newblock \href {https://doi.org/10.48550/arXiv.2409.00079} {\path{doi:10.48550/arXiv.2409.00079}}.

\bibitem{toubeau2022interpretable}
J.~Toubeau, J.~Bottieau, Y.~Wang, F.~Vallée, Interpretable probabilistic forecasting of imbalances in renewable-dominated electricity systems, IEEE Transactions on Sustainable Energy 13 (2022) 1267--1277.
\newblock \href {https://doi.org/10.1109/tste.2021.3092137} {\path{doi:10.1109/tste.2021.3092137}}.

\end{thebibliography}

\end{document}